\documentclass[12pt]{article}
\pdfoutput=1
\usepackage{amsmath,amssymb,amsthm,bm,graphicx,float,array,multirow,multicol,rotfloat,caption,subcaption,hyperref,cleveref,enumerate,geometry,mathdots,adjustbox,booktabs,parskip,mathtools,tikz,tikz-cd,pdflscape,csquotes,lscape,rotating,empheq,mathrsfs,longtable}
\usetikzlibrary{decorations.pathreplacing,calligraphy,intersections}
\usepackage{stackengine} 
\usepackage{authblk}
\usepackage{tcolorbox}
\usepackage{multicol}
\usepackage{lscape} 
\usepackage{xcolor}
\usepackage[para]{threeparttable}
\usepackage[all]{xy}
\usepackage[normalem]{ulem}
\usepackage[numbers,sort&compress]{natbib}
\usepackage[aligntableaux=center]{ytableau}
\usepackage[numbers,sort&compress]{natbib}
\numberwithin{equation}{section}
\numberwithin{figure}{section}
\allowdisplaybreaks
\makeatletter
\usetikzlibrary{arrows,shapes.misc,positioning,decorations.pathmorphing,decorations.markings,decorations.pathreplacing,matrix,patterns,backgrounds}

\tikzset{
scale cd/.style={every label/.append style={scale=#1}, cells={nodes={scale=#1}}}}

\tikzset{gauge/.style={rounded rectangle, draw=black!100, thick, minimum size=5mm},  gaugeD/.style={rounded rectangle, draw=black!100,double,thick,minimum size=5mm},  empty/.style={rounded rectangle, draw=white!100, thick, minimum size=5mm}, flavor/.style={rectangle, draw=black!100, thick, minimum size=5mm},flavorD/.style={rectangle, draw=black!100, double,thick, minimum size=5mm}}

\tikzset{
node/.style={circle, thick, draw=black!100,fill=white!100,  minimum size=2mm, inner sep=0pt},
sqnode/.style={rectangle
, thick, draw=black!100,fill=white!100,  minimum size=2mm, inner sep=0pt
},
sonode/.style={circle, thick, draw=black!100,fill=red!100,  minimum size=3mm, inner sep=0pt},
spnode/.style={circle, thick, draw=black!100,fill=blue!100,  minimum size=3mm, inner sep=0pt},
fnode/.style={rectangle, thick, draw=black!100,fill=white!100,  minimum size=3mm, inner sep=0pt},
tnode/.style={rounded rectangle, outer sep=0pt, thick, minimum size=5mm}
}
\DeclarePairedDelimiter\floor{\lfloor}{\rfloor}

\newcommand\notsoscript{\@setfontsize\notsoscript{9}{7}}

\theoremstyle{plain}
\newtheorem*{thm*}{Theorem}
\newtheorem{thm}{Theorem}[section]

\newtheorem{alg}[thm]{Algorithm}

\theoremstyle{definition}

\newtheorem*{defn*}{Definition}

\makeatother

\begin{document}

\begin{titlepage}
\vspace*{-3cm} 
\begin{flushright}
{\tt DESY-23-199}\\
\end{flushright}
\begin{center}
\vspace{2cm}
{\LARGE\bfseries The Higgs Branch of Heterotic LSTs: \\
\Large Hasse Diagrams and Generalized Symmetries\\} 
\vspace{1.2cm}

{\large
$^{}$Craig Lawrie and $^{}$Lorenzo Mansi\\}
\vspace{.7cm}
$^{}${Deutsches Elektronen-Synchrotron DESY,}\par
{Notkestr.~85, 22607 Hamburg, Germany}\par
\vspace{.2cm}

\vspace{.3cm}

\scalebox{1.0}{\tt $^{}$craig.lawrie1729@gmail.com,
$^{}$lorenzo.mansi@desy.de}\par
\vspace{1.2cm}
\textbf{Abstract}
\end{center}
\noindent
We study the Higgs branches of the 6d $(1,0)$ little string theories that live on the worldvolume of NS5-branes probing an ADE-singularity in the heterotic $E_8 \times E_8$ and $\mathrm{Spin}(32)/\mathbb{Z}_2$ string theories. On the $E_8 \times E_8$ side, such LSTs are obtained via fusion of orbi-instanton SCFTs.
For the $\mathbb{C}^2/\mathbb{Z}_K$ orbifolds, we determine a magnetic quiver for the Higgs branch 
from
the alternative Type IIA brane system engineering the LST; we show that the magnetic quiver obtained in this way is the same as the Coulomb gauging of the 3d mirrors associated with the orbi-instanton building blocks. Using quiver subtraction, we determine the Hasse diagram of Higgs branch RG flows between the LSTs, and we analyze how the structure constants of the generalized global symmetries vary along the edges of the Hasse diagram. From the Hasse diagram of the Higgs branch, we are immediately able to identify LSTs with the same T-duality-invariant properties, and thus propose candidate T-dual pairs. We perform a similar analysis of the Higgs branch Hasse diagram and putative T-dual families for particular $E_6$-orbifold LSTs by taking advantage of a duality between a rank zero orbi-instanton theory and a rank one conformal matter theory.
\vfill 
\end{titlepage}

\tableofcontents
\newpage

\section{Introduction}\label{sec:Introduction}

Non-perturbative or strong-coupling behavior of quantum field theories (QFTs) often gives rise to especially interesting physics, the quintessential example being confinement in QCD. By construction, properties of QFT at strong-coupling cannot be understood via perturbation theory around a Gaussian fixed point, and we are forced to use alternative methods. One powerful tool for exploring non-perturbative behavior is dualities -- when two a priori different QFTs, in fact, describe the same physics; if one description is at weak-coupling, and the other is at strong-coupling, then we can learn about the strong-coupling behavior of the latter via a perturbative analysis of the former. Determining a dual pair is often challenging, it is useful instead to discover ``invariant'' properties of QFTs and compare different descriptions; descriptions with identical invariants are candidate duals.
Another approach to strong-coupling behavior is to determine protected or symmetry-related quantities, such as anomalies. In this paper, we study the moduli space of vacua of supersymmetric quantum field theories. Not only does the moduli space capture important physical information about the theory, such as its renormalization group flows to different effective descriptions, but it also provides a particularly robust invariant, allowing for the identification of potential dualities.

To have sufficient protected and computable quantities of interest, we first focus on superconformal field theories (SCFTs). When considering such theories, the maximal dimension in which they can exist is six \cite{Nahm:1977tg}. Such SCFTs do not have a weakly-coupled or perturbative description, and instead, their construction relies on string theory. While such constructions appear to give rise to a theory of interacting tensionless strings, they have been shown to, in fact, be local quantum field theories \cite{Witten:1995zh,Strominger:1995ac,Seiberg:1996qx}. It is a marked advantage of the string-theoretic origin that many of the strongly-coupled physics are encoded in the geometry of the string theory compactification space.\footnote{This is a prominent example of a general phenomenon, known as ``geometrization'', whereby the non-perturbative physical properties of a theory in $d$ dimensions can be extracted by considering the compactification of a theory in higher dimensions. See \cite{Argyres:2022mnu} for a recent review of the successes of this top-down approach.} 

In the past decade, there has been a wealth of research on six-dimensional SCFTs with eight supercharges, spurred on by the F-theory realizations of such SCFTs in \cite{Heckman:2013pva,Heckman:2015bfa}. Each SCFT is associated with a non-compact elliptically-fibered Calabi--Yau threefold that is required to satisfy a variety of geometric conditions. From the Calabi--Yau geometry, it is possible to extract many properties of the SCFTs, such as the (global form of the) flavor symmetry, the anomalies, and, of particular interest in this paper, the structure of part of the moduli space of vacua known as the Higgs branch. Specifically, understanding the Higgs branch means, physically, knowing the network of Higgs branch renormalization group flows between interacting fixed points on the Higgs branch, or, mathematically, knowing the foliation of the Higgs branch as a symplectic singularity. The Higgs branch of 6d $(1,0)$ SCFTs has been explored in a significant number of publications, a small sample is: \cite{Heckman:2015ola,Heckman:2016ssk,Mekareeya:2016yal,Heckman:2018pqx,DeLuca:2018zbi,Hassler:2019eso,Baume:2021qho,Bourget:2022tmw,Baume:2020ure,Razamat:2019mdt,Bergman:2020bvi,Baume:2022cot,DKL,Distler:2022yse,Distler:2022kjb,LM,Hanany:2018vph,Hanany:2018uhm,Cabrera:2019dob,Cabrera:2019izd,Baume:2023onr}.

Supersymmetric quantum field theories in any dimension encode symmetry-related quantities in their moduli space of vacua, $\mathcal{M}$. Usually, this space splits off in different \textit{branches}, determined by which scalar fields, coming from a certain type of multiplet, parameterize through their VEVs that portion of the manifold.\footnote{When considering non-Lagrangian theories, there is an alternative parametrization of the moduli space that does not involve a discussion of perturbative supermultiplets.} In particular, the Higgs branch $\mathcal{H}$ is the space contained in $\mathcal{M}$ where only scalars in the hypermultiplets can acquire a non-trivial VEV, and from which we can read information related to the global symmetry of the theory. Limiting to theories with eight supercharges, this space mathematically assumes the structure of a hyperk\"ahler cone \cite{Hitchin:1986ea}, which can also be regarded as a symplectic singularity \cite{Beauville_2000}.  On a generic point of $\mathcal{H}$, given that enough matter is available in the theory, the gauge group is completely broken. But in some positive codimension loci, part of the gauge symmetry may be preserved. Hence a partial ordering naturally emerges when considering a Higgs branch RG flow from a theory with gauge group $G$ to a theory with a gauge group $G'\subset G$, allowing one to construct a Hasse diagram for the Higgs branch of a theory. The symplectic singularity structure of $\mathcal{H}$ in theories with eight supercharges is naturally equipped with a \textit{foliation}, that exactly encodes the pattern of partial Higgsing dictated by the (Higgs branch) RG flow, thus the equivalence between studying Higgsing and singularity foliation \cite{Bourget:2019aer}.

One successful approach that provides an understanding of the Higgs branch of a theory with eight supercharges is to determine and study it via the symplectic singularity structure. In order to accomplish this feat, we would like to engineer a 3d $\mathcal{N}=4$ Lagrangian quiver theory \cite{Cabrera:2019izd}, whose Coulomb branch $\mathcal{C}_{3d}$ is known to be a symplectic singularity, and whose physics is under control, such that
\begin{equation}
    \mathcal{H}_{6d} \cong \mathcal{C}_{3d} \,.
\end{equation}
This establishes a moduli space correspondence between the six-dimensional Higgs branch and a three-dimensional Coulomb branch.\footnote{More generally, a $d\ge3$ dimensional supersymmetric theory $\mathcal{T}$ with $8$ supercharges has a Higgs branch $\mathcal{H}_d$ that may be isomorphic to the union of Coulomb branches of a collection $\mathcal{Q}_i$ of 3d $\mathcal{N}=4$ quiver theories:
\begin{equation*}
    \mathcal{H}_{d}(\mathcal{T}) \cong \bigcup_{i} \mathcal{C}_{3d}(\mathcal{Q}_i) \,.
\end{equation*}
Each of the $\mathcal{Q}_i$ constitutes a \textit{magnetic quiver} for the theory $\mathcal{T}$.}. We sometimes refer to the 6d theory as the \textit{electric} theory and the 3d theory as the \textit{magnetic} theory. The source of this nomenclature relies on the resemblance of this correspondence to 3d mirror symmetry that exchanges the Coulomb and Higgs branches of two three-dimensional theories.

Moving beyond the regime of local quantum field theories, we can consider the so-called six-dimensional little string theories (LSTs). These are non-gravitational 6d theories that contain an intrinsic string scale: $M_\text{string}$. Such theories have a UV-completion which is not a standard quantum field theory, however, below the energy scale set by $M_\text{string}$, we can treat the theory as a quantum field theory with some cutoff scale. Such novel non-local theories have been known for a long time, see, for example, \cite{Seiberg:1997zk,Aharony:1999ks}. Following on from the geometric constructions of 6d $(1,0)$ SCFTs in F-theory, the analysis was extended to provide similar constructions of 6d little string theories with $(1,0)$ supersymmetry in \cite{Bhardwaj:2015oru}. LSTs also possess a Higgs branch as part of the moduli space, and in this paper, we explore the Higgs branch and the related renormalization group flows through the construction of the 3d $\mathcal{N}=4$ magnetic theory. The study of the Higgs branches of heterotic orbifold LSTs via the magnetic quiver has been initiated in \cite{DelZotto:2023nrb}. LSTs distinguish themselves from their purely conformal siblings as they are non-local theories and thus a unique definition for an energy-momentum tensor is not available, instead, they present the emergence of T-duality. This phenomenon consists of the equivalence of different LSTs under $S^1$-compactification in the presence of Wilson lines leading to a unique five-dimensional theory. The Hasse diagram of Higgs branch renormalization group flows provides a useful packaging of invariants for the identification of families of T-dual theories.

Although LSTs present a QFT description only at low energies, an essential tool when investigating T-dual models is the comparison of symmetry-related quantities, such as anomalies, or 5d quantities independent from compactification details; these latter can be computed directly from the six-dimensional model, such as Coulomb branch dimension or flavor symmetry rank. Little string theories also realize generalized global symmetries \cite{Gaiotto:2014kfa}. Such a generalized symmetry is associated with an invariant of the quantum field theory, similar to an anomaly, and which therefore must match across a T-dual pair \cite{DelZotto:2020sop}.

We provide a short description of these generalized global symmetries, following \cite{Gaiotto:2014kfa}, with a view to elucidating these invariant quantities. A QFT may exhibit some generic symmetry structure on $p$-dimensional objects whose conserved current $J^{(p+1)}$ is a $(p+1)$-form, in this language the usual symmetries acting on point-like objects ($p=0$) are called $0$-form symmetries, whereas symmetries acting on line objects ($p=1$) are called $1$-form symmetries, etc. Analogously to the introduction of background (one-form) gauge fields for continuous zero-form symmetries, we introduce a background $(p+1)$-form gauge field, $B^{(p+1)}$, for the $p$-form symmetry, and which couples to the conserved current via the following term in the action:
\begin{equation}
    \delta S =\int d^d x \ B^{(p+1)} \wedge *J^{(p+1)} \,.
\end{equation}
A global $p$-form symmetry transformation is then just a background gauge transformation:
\begin{equation}
    B^{(p+1)}\rightarrow B^{(p+1)} + d \Lambda^{(p)} \,. 
\end{equation}
Often, when considering the background $p$-form symmetry transformations, the partition function of the theory is not invariant, but it picks up a so-called ``operator-valued shift'', which leads to an inconsistency. However, this operator-valued shift may be cancelled by considering a simultaneous $p$-form and $p'$-form background gauge transformation. Consider $p' < p$, and the combined background gauge transformation
\begin{equation}\label{eqn:generalised_gauge}
    B_i^{(p'+1)}\rightarrow B_i^{(p'+1)} + d \Lambda_i^{(p')} \text{ and } B^{(p+1)}\rightarrow B^{(p+1)} + d \Lambda^{(p)} + \kappa f^{(p+1)}\left( B_i^{(p'+1)} , \Lambda_i^{(p')}\right) \,.
\end{equation}
Here, the function $f$ encodes the anomaly term induced by the background gauge transformation, and $\kappa f^{(p+1)}\left( B_i^{(p'+1)} , \Lambda_i^{(p')}\right)$ means that we sum with a different coefficient $\kappa$ over each $p$-form that can be constructed from the various $B_i^{(p'+1)}$ and their gauge shifts. The invariants $\kappa$ are typically fixed by the requirement that the partition function is multiplied only by a constant phase under the combined transformation. Due to a similarity with the Green--Schwarz anomaly cancellation mechanism in string theory, this combined background gauge transformation is sometimes referred to as a ``Green--Schwarz symmetry''.\footnote{Mixed $p$-form and $(p+1)$-form generalized global symmetries have been explored in \cite{Cordova:2018cvg,Cordova:2020tij,Benini:2018reh}, where they are referred to as a ``continuous two-group symmetry''. However, the combined transformation in equation \eqref{eqn:generalised_gauge} does not match the expected transformation for a 2-connection on a principal smooth 2-group bundle, except in very special cases; a careful analysis of the distinction between Green--Schwarz symmetry and smooth 2-group structure has appeared in \cite{Kang:2023uvm}.}

In particular, LSTs manifest several mixed $0$-form/$1$-form global symmetries, resulting from the mixing of a continuous $\mathfrak{u}(1)_{LST}^{(1)}$ $1$-form symmetry with the conventional $0$-form symmetries of the theory: the $\mathfrak{su}(2)_R$ R-symmetry, the Poincare symmetry $\mathfrak{p}$, and any flavor symmetry $\mathfrak{f}$ \cite{DelZotto:2020sop,Cordova:2020tij}. Therefore, the global symmetry of an LST can be expressed as:
\begin{equation}\label{eqn:gensym}
    \left(\mathfrak{su}(2)_R \oplus \mathfrak{p} \oplus \mathfrak{f} \right)^{(0)} \times_{\kappa_R ,\kappa_P , \kappa_{F} } \ \mathfrak{u}(1)_{LST}^{(1)} \,.
\end{equation}
The origin of this continuous $1$-form symmetry lies in the presence of a 2-form current $J_{LST}^{(2)}$ depending on the second Chern class of the Yang--Mills field strength $c_2 \left( f^{(2)} \right)$. Importantly, its Hodge dual is conserved and remains unbroken in the infrared regime. The structure constants $\kappa_R$, $\kappa_P$, and $\kappa_F$ describe the mixing under gauge transformations of the background field for the $1$-form symmetry with the background fields for these $0$-form global symmetries. 

A direct computation of the structure constants for the generalized global symmetries in all LST models can be a tedious exercise, which can be optimized via the exploration of other properties of the theory, such as its moduli space of vacua. In this paper, we study the Higgs branch of little string theories, and one interesting consequence is how the structure constants are encoded in the Hasse diagram of fixed points on the Higgs branch.

As we have discussed, LSTs are particularly interesting as they evince T-duality. As a duality, the relevant invariants must match across a T-dual pair. In this paper, we consider two classes of little string theories, which are the heterotic $E_8 \times E_8$ and heterotic $\mathrm{Spin}\left(32\right) / \mathbb{Z}_2$ ALE instantons. These LSTs have received renewed interest in the last year \cite{Ahmed:2023lhj,DelZotto:2023nrb,DelZotto:2023ahf,DelZotto:2022xrh,DelZotto:2022ohj}. As eight-supercharge theories, their Higgs branches are symplectic singularities, and we study the structure of its foliation via the construction of a dual magnetic quiver. From the structure of the Higgs branch, we can track the changes in the structure constants (and other invariants) along renormalization group flows, and thus we obtain a vast landscape of LSTs which have the same invariants and thus are putative T-dual families. This is a consequence of our detailed study of the Higgs branch of such heterotic LSTs, and we leave a careful geometric analysis of these potential T-dualities for future work.

The structure of this paper is as follows. In Section \ref{sec:build}, we review the geometric construction of 6d $(1,0)$ SCFTs and LSTs in F-theory. Next, in Section \ref{sec:Higgs_Branch_OI}, we discuss the Higgs branches of the 6d $(1,0)$ SCFTs known as the Higgsed rank $N$ $(\mathfrak{e}_8, \mathfrak{su}(K))$ orbi-instantons; this provides the tools necessary to understand the Higgs branches of the heterotic LSTs, which we explore in Section \ref{sec:LSTs}. In Section \ref{subsec:MonotonyThm}, we study the change in the structure constants for the generalized global symmetries along the Higgs branch renormalization group flows. Putting everything together, in Section \ref{sec:tduality}, we enumerate a large collection of heterotic ALE instanton theories that have the same invariant properties, and are thus candidate T-dual families. In Section \ref{sec:beyond}, we go beyond the $\mathfrak{su}(K)$ paradigm, and study the Higgs branches (and the resulting T-dual families) for heterotic ALE instantons associated with the exceptional algebra $\mathfrak{e}_6$. We highlight a few intriguing directions for future study in Section \ref{sec:conc}.

\section{How to Build an LST}\label{sec:build}

Six-dimensional little string theories, just like six-dimensional superconformal field theories, lack simple Lagrangian descriptions. Instead, such theories are typically constructed by considering a top-down approach from string theory -- such a string-theoretic origin involves a compactification geometry, and the power of this perspective lies in the geometrization of diverse, and strongly-coupled, physical behavior in terms of this compactification space. In this paper, we will study 6d LSTs that arise from compactifications of F-theory, such as those described in \cite{Bhardwaj:2015oru}. To begin, in this section, we review the F-theory construction of both 6d $(1,0)$ SCFTs and 6d $(1,0)$ LSTs.

\subsection{General Construction}\label{sec:LSTgeneralconstruction}

It has been established \cite{Heckman:2013pva,Heckman:2015bfa} that F-theory compactified on a non-compact elliptically-fibered Calabi--Yau threefold, where all fibers consist of an irreducible, possibly degenerate, curve, with a non-minimal singularity supported over (possibly) singular points in the base,\footnote{The definition of a non-minimal singularity relevant for engineering 6d $(1,0)$ SCFTs is somewhat technical, in particular, it requires the Weierstrass model of the elliptic fibration; we refer to \cite{Heckman:2015bfa} for the precise definition, as it is not pertinent to this paper.} and where the base itself has no complex curves of finite volume, yields an SCFT. Generally, we will focus on the cases where a non-minimal fiber is supported over only a single point, which leads to an SCFT with a single stress-energy tensor; multiple non-minimal fibers indicate that the SCFT is a product of multiple interacting SCFTs.

Such elliptically-fibered Calabi--Yau threefolds are, by construction, highly singular, however, despite this, we have a procedure with which to construct them \cite{Heckman:2013pva,Heckman:2015bfa}. Consider a non-compact elliptically-fibered Calabi--Yau manifold $\pi : \widetilde{Y} \rightarrow \widetilde{B}$, where the base $\widetilde{B}$ is smooth and contains a collection of intersecting complex curves $C_i$ that have a finite volume, and such that all fibers of the elliptic fibration are minimal. There exists a point in the K\"ahler moduli space of $\widetilde{Y}$ that yields an interacting non-product SCFT if all of the curves $C_i$ can be simultaneously shrunk to zero-volume, and if, under the curve contraction map $\rho : \widetilde{B} \rightarrow B$, the point to which all the curves are contracted is singular in $B$, or else the singular fiber above the image of $\rho$ is non-minimal, or both. 

These constraints on the Calabi--Yau threefolds are powerful. Each curve $C_i$ is forced to be a smooth rational curve, and the intersection matrix $C_i \cdot C_j$ must be negative-definite. The curves can only intersect each other pairwise, and the intersection number between any two non-identical curves is required to be either zero or one. In fact, each such non-compact elliptically-fibered Calabi--Yau $\widetilde{Y}$ can be built in the following way. First, we introduce some standard notation. Let 
\begin{equation}
    \overset{\mathfrak{g}}{n} \,,
\end{equation}
denote a curve $C$ in $\widetilde{B}$, with self-intersection number $C \cdot C = - n$, and with a singular fiber over $C$ of Kodaira-type corresponding to the simple Lie algebra $\mathfrak{g}$.\footnote{There is not a one-to-one map between Kodaira fibers and simple Lie algebras, but it is generally unambiguous for our purposes, and where further data is required we introduce it at that moment.} Then, we can write a set of building blocks known as the non-Higgsable clusters \cite{Morrison:1996na,Morrison:1996pp,Morrison:2012np}:
\begin{equation}
      \overset{\mathfrak{su}_3}{3} \,, \quad \overset{\mathfrak{so}_8}{4} \,, \quad \overset{\mathfrak{f}_4}{5} \,, \quad \overset{\mathfrak{e}_6}{6} \,, \quad \overset{\mathfrak{e}_7}{7} \,, \quad \overset{\mathfrak{e}_7}{8} \,, \quad \overset{\mathfrak{e}_{8}}{12} \,, \quad \overset{\mathfrak{su}_{2}}{2}\overset{\mathfrak{g}_{2}}{3} \,, \quad 2\overset{\mathfrak{su}_{2}}{2}\overset{\mathfrak{g}_{2}}{3} \,, \quad \overset{\mathfrak{su}_{2}}{2}\overset{\mathfrak{so}_{7}}{3}\overset{\mathfrak{su}_{2}}{2} \,, 
\end{equation}
together with the ADE Dynkin diagrams formed out of intersecting $(-2)$-curves:
\begin{equation}
    \underbrace{2\cdots 2}_{N-1} \,, \quad 
    \underbrace{2\cdots 2}_{N-3}\overset{\displaystyle 2}{2}2 \,, \quad 
    22\overset{\displaystyle 2}{2}22 \,, \quad 
    222\overset{\displaystyle 2}{2}22 \,, \quad 
    2222\overset{\displaystyle 2}{2}22 \,.
\end{equation}
These are (combinations of) curves with certain prescribed singular fibers over them.\footnote{Curves that are written adjacently are assumed to have mutual intersection number $1$, otherwise $0$.} The non-Higgsable clusters can be tuned, that is, the singular fiber corresponding to the algebra $\mathfrak{g}$ over a curve can be tuned to an algebra $\mathfrak{g}' \supset \mathfrak{g}$. Of course, the enhancement to $\mathfrak{g}'$ is not arbitrary, as it is necessary to retain the minimality of the fibration.\footnote{Gauge-anomaly cancellation of the corresponding field theory on the tensor branch, which we explain anon, requires that there exists prescribed hypermultiplets charged under $\mathfrak{g}'$.}

Building blocks must be glued together, and the mortar with which they are glued are the so-called decorated E-string configurations. These are single curves of self-intersection $(-1)$ with some singular fiber associated to a Lie algebra $\mathfrak{g}$:
\begin{equation}\label{eqn:tunedEstring}
    \overset{\mathfrak{g}}{1} \,.
\end{equation}
Such a configuration as in equation \eqref{eqn:tunedEstring} itself gives rise to an associated SCFT, by shrinking the $(-1)$-curve to zero-volume and compactifying F-theory on the resulting singular Calabi--Yau. Such an SCFT has a continuous zero-form global symmetry $\mathfrak{f}_\mathfrak{g}$.\footnote{In fact, when $\mathfrak{g} = \mathfrak{su}_6$ or $\mathfrak{g} = \mathfrak{so}_{12}$, the configuration as written in equation \eqref{eqn:tunedEstring} does not uniquely specify an SCFT, and thus the geometric configuration needs additional discrete labels, such as $\mathfrak{g} = \mathfrak{su}_6$ and $\mathfrak{g} = \mathfrak{su}_6'$ which describe different configurations, and which, for example, have distinct flavor symmetries.} We can construct a new non-compact Calabi--Yau threefold satisfying the relevant conditions to be associated with an SCFT by starting from two tuned non-Higgsable clusters 
\begin{equation}
    \cdots \overset{\mathfrak{g}_L}{n} \qquad \text{ and } \qquad \overset{\mathfrak{g}_R}{m} \cdots \,,
\end{equation}
and gluing with the tuned E-string in equation \eqref{eqn:tunedEstring}:
\begin{equation}
     \cdots \overset{\mathfrak{g}_L}{n} \quad + \quad \overset{\mathfrak{g}}{1} \quad + \quad \overset{\mathfrak{g}_R}{m} \cdots \qquad = \qquad \cdots \overset{\mathfrak{g}_L}{n} \, \overset{\mathfrak{g}}{1} \, \overset{\mathfrak{g}_R}{m} \cdots \,.
\end{equation}
This leads to a configuration associated with an SCFT if the glued combination of curves has a negative-definite intersection pairing and if 
\begin{equation}\label{eqn:gaugecdn}
    \mathfrak{f}_\mathfrak{g} \supseteq \mathfrak{g}_L \oplus \mathfrak{g}_R \,.
\end{equation}
Physically, the latter condition is associated with the gauging of a subalgebra of the $\mathfrak{f}_\mathfrak{g}$ flavor symmetry of the tuned E-string SCFT. There are occasional subtleties in this construction of the relevant Calabi--Yau threefolds that have not been recapped here, we refer the reader to the original works \cite{Heckman:2013pva,Heckman:2015bfa}, or the review article \cite{Heckman:2018jxk}, for the complete details.\footnote{When considering a product of multiple interacting 6d $(1,0)$ SCFTs, we use the disjoint union $\sqcup$ between their respective curve configurations. For example, $1 \sqcup 1$ denotes two $(-1)$-curves which do not intersect, and thus when shrunk simultaneously to zero-volume lead to a Calabi--Yau threefold with non-minimal fibers over two separate points in the base.}

We have explained how to construct a non-compact elliptically-fibered Calabi--Yau threefold $\widetilde{Y}$ such that there exists a contraction map $\rho: \widetilde{Y} \rightarrow Y$, which leads to another non-compact elliptically-fibered Calabi--Yau threefold $Y$, such that the compactification of F-theory on $Y$ engineers a 6d superconformal field theory. What is obtained by considering F-theory compactifed on $\widetilde{Y}$ itself? This gives rise to the effective six-dimensional quantum field theory that lives on the generic point of the tensor branch of the 6d SCFT associated with $Y$. D3-branes wrap the compact finite-volume curves; in the six-dimensional theory these give rise to string-like excitations, where the tension of the string is proportional to the volume of the compact curve. The tensionful strings have a two-dimensional worldvolume which couples to a two-form potential, $B_{\mu\nu}$. The potential sits inside of a tensor multiplet which also contains a scalar $\phi$, and the vacuum expectation value of $\phi$ is the volume of the associated compact curve. The moduli space of vacua corresponding to the VEVs of these scalars is called the tensor branch, and at the origin of the tensor branch the tension of all of the BPS-strings goes to zero, and the interacting superconformal field theory emerges. At the generic point of the tensor branch, the fibers of the elliptic fibration are all minimal, and thus there is a description in terms of 6d gauge algebras and matter hypermultiplets as is standard in such F-theory compactifications, see, e.g., \cite{Denef:2008wq,Weigand:2010wm}.

As we have summarized, the conditions that the elliptically-fibered Calabi--Yau threefold $Y$ must satisfy to realize a superconformal field theory when F-theory is compactified on $Y$ are well-known. In \cite{Bhardwaj:2015oru}, a similar analysis of the necessary conditions on $Y$ such that the F-theory compactification gives a 6d $(1,0)$ little string theory was carried out. $Y$ is a non-compact elliptically-fibered Calabi--Yau threefold, $\pi : Y \rightarrow B$, where $B$ contains a single compact curve of self-intersection $0$, and non-minimal fibers may be supported over points of this $(0)$-curve. From the Grauert--Artin contractibility theorem \cite{MR137127,MR146182}, there does not exist a contraction map which shrinks the single finite volume curve while retaining the dimensionality of $B$; D3-branes wrapping this curve thus give rise to tensionful strings, which sets the scale of the LST. Exactly as in the SCFT case, it is known how to construct a Calabi--Yau threefold $\widetilde{Y}$, via the gluing of a small set of building blocks, such that there exists a contraction map $\rho: \widetilde{Y} \rightarrow Y$ where $Y$ engineers an LST. We now briefly review the construction of the possible $\widetilde{Y}$, following \cite{Bhardwaj:2015oru}.

The building blocks for non-compact elliptically-fibered Calabi--Yau threefolds associated with LSTs in F-theory are a small superset of those used to construct 6d SCFTs. The additional building blocks are denoted as follows:
\begin{equation}
    0 \,, \quad I_0 \,, \quad I_1 \,, \quad II \,.
\end{equation}
These refer to, respectively, a smooth rational curve of self-intersection $0$, a smooth torus, a nodal curve, and a cuspidal curve. Each of these specifies a single finite-volume curve in the non-compact base $\widetilde{B}$, and the singular fiber above the curve is not associated with any non-Abelian Lie algebra. In addition, one can have building blocks which are simply the affine ADE Dynkin diagrams formed out of intersecting $(-2)$-curves:
\begin{equation}
    \big/\big/\underbrace{\,2\cdots 2\,}_{N}\big/\big/ \,, \quad 
    2\overset{\displaystyle 2}{2}\underbrace{\,2\cdots 2\,}_{N-5}\overset{\displaystyle 2}{2}2 \,, \quad 
    22\overset{\displaystyle \overset{\displaystyle 2}{2}}{2}22 \,, \quad 
    222\overset{\displaystyle 2}{2}222 \,, \quad 
    22222\overset{\displaystyle 2}{2}22 \,.
\end{equation}
The notation $//\cdots//$ indicates that the left-most and right-most curves intersect, with intersection number $1$, as one expects from the affine A-type Dynkin diagram. It is possible to further tune the fiber to place a non-trivial $\mathfrak{g}$ over the curves, subject to the constraint of minimality of the fibration.

The mortar used to glue the building blocks together is again the tuned E-string configurations. For a glued configuration, it is necessary that the intersection pairing after gluing is negative semi-definite; furthermore, every minor of the intersection matrix must have a non-zero determinant. Minimality of the fibration again imposes an analogous condition to that in equation \eqref{eqn:gaugecdn}. This involves a slight generalization: since $(-1)$-curves can now intersect more than two other finite-volume curves, the constraint becomes
\begin{equation}
    \bigoplus_i \mathfrak{g}_i \subseteq \mathfrak{f}_\mathfrak{g} \,,
\end{equation}
where $\mathfrak{g}_i$ runs over the algebras arising from the singular fibers supported on the curves adjacent to the introduced $(-1)$-curve.

We emphasize that, for LSTs, it is possible for several finite-volume curves in the base to intersect at the same point, or for a pair of curves to intersect with an intersection number greater than $1$. In addition, it is possible to glue two together two tuned E-strings directly, without an interpolating tuned non-Higgsable cluster. We refer the reader to \cite{Bhardwaj:2015oru} for the full description of the rules and constraints for constructing non-compact Calabi--Yau geometries associated with 6d LSTs. In this paper, we are interested in a specific class of LSTs, and thus we do not need to explicate the full, general, construction of LSTs from F-theory. 

One of the advantages to the F-theory description is that there exist robust quantities that can be computed from the tensor branch effective field theory, i.e., from the geometry $\widetilde{Y}$, and for which it is known how those quantities change in the limit where the strings become tensionless. In this way, we can study the well-controlled tensor branch geometry to learn about the strongly-coupled physics of the SCFT or LST. In the SCFT case, one of the hallmarks of the geometric engineering in F-theory was not just that an abstract SCFT can be associated with each appropriate non-compact elliptically-fibered Calabi--Yau, but that one can extract the physical 't Hooft anomalies of the continuous global (zero-form) symmetries in terms of the Calabi--Yau $\widetilde{Y}$ \cite{Ohmori:2014kda,Intriligator:2014eaa,Baume:2021qho}. 

Turning to LSTs, one of the results of \cite{DelZotto:2020sop,DelZotto:2023ahf}, is that the structure constants for the generalized symmetries, $\kappa_P$, $\kappa_R$, and $\kappa_F$ as in equation \eqref{eqn:gensym}, are similarly robust and can be determined from the F-theory tensor branch geometry. The intersection matrix, $A^{ij}$ of compact curves in the base of the elliptic fibration $\pi :\widetilde{Y} \rightarrow \widetilde{B}$ is negative semi-definite and has a single zero-eigenvalue. Let $\ell_j$ be the suitably normalized eigenvector associated to the zero-eigenvalue:
\begin{equation}
    A^{ij} \ell_j = 0 \qquad \text{ such that } \qquad \ell_j > 0 \,\,\text{ and }\,\, \operatorname{gcd}(\ell_1, \ell_2,  \cdots) = 1 \,.
\end{equation}
Thus, to each compact curve $C_i$ is associated the integer $\ell_i$. In terms of these quantities, the structure constants describing the mixing of the $\mathfrak{u}(1)^{(1)}_\text{LST}$ with the Poincare symmetry and R-symmetry are
\begin{equation}\label{eqn:kappaPRgeneral}
    \kappa_P = \sum_i \ell_i (A^{ii} + 2) \,, \qquad \kappa_R = \sum_i \ell_i h_{\mathfrak{g}_i}^\vee \,,
\end{equation}
where the sum runs over the compact curves $C_i$.
Here $h_{\mathfrak{g}_i}^\vee$ is the dual Coxeter number of the algebra $\mathfrak{g}_i$ supported over the curve $C_i$; if the elliptic fiber over $C_i$ is not associated to a non-Abelian Lie algebra, then we define $h_{\mathfrak{g}_i}^\vee = 1$. Denote the semi-simple part of the flavor algebra\footnote{Notice that whenever we refer to the flavor algebra of an LST model, we indicate with $\mathfrak{f}$ the flavor algebra of the six-dimensional theory. Compactifying to 5d introduces an instantonic $\mathfrak{u}(1)_I$ global symmetry that contributes to the five-dimensional global symmetry $\mathfrak{f}_{\text{5d}}$, and is such that $\mathrm{rank}(\mathfrak{f})=\mathrm{rank}(\mathfrak{f}_{\text{5d}})-1$.} of the LST as
\begin{equation}
    \mathfrak{f} = \bigoplus_a \mathfrak{f}_a \,,
\end{equation}
where the sum runs over the simple non-Abelian factors $\mathfrak{f}_a$. Geometrically, each $\mathfrak{f}_a$ is supported on a non-compact curve in $\widetilde{B}$, and we let $B^{ia}$ denote the intersection number between this non-compact curve with the compact curve $C_i$. The structure constants $\kappa_{F_a}$ depend on these intersection numbers, to wit:
\begin{equation}\label{eqn:kappaFgeneral}
    \kappa_{F_a} = \sum_i \ell_i B^{ia} \,.
\end{equation}

Now that we have summarized how to extract particular physical properties of the LSTs from their geometric construction in F-theory, we introduce the class of LSTs with which we are concerned in the remainder of this paper.

\subsection{A-type Orbi-instanton SCFTs}\label{sec:AtypeOI}

In Section \ref{sec:LSTgeneralconstruction}, we have explained the general method for constructing 6d $(1,0)$ LSTs from F-theory. However, in this work, we are expressly interested in only a specific subclass of these LSTs, the so-called heterotic orbifold LSTs. These theories can be obtained from a specific class of 6d $(1,0)$ SCFTs, which are known as the rank $N$ $(\mathfrak{e}_8, \mathfrak{g})$ orbi-instanton theories \cite{DelZotto:2014hpa}. We consider $\mathfrak{g}=\mathfrak{su}_K$, and denote the corresponding theories as $\mathcal{O}_{N,K}$; in this section, we introduce these 6d $(1,0)$ SCFTs, and explain some of their relevant properties.

At the generic point of the tensor branch, these orbi-instanton SCFTs have an effective field theory description as
\begin{equation}\label{eqn:AOI_TB}
    \operatorname{TB}\left[\mathcal{O}_{N,K}\right] \,\, = \,\, 1\,2\,\overset{\mathfrak{su}_2}{2}\overset{\mathfrak{su}_3}{2}\cdots \overset{\mathfrak{su}_{K-1}}{2}\overset{\mathfrak{su}_{K}}{2} \underbrace{\,\overset{\mathfrak{su}_{K}}{2} \cdots \overset{\mathfrak{su}_{K}}{2}\,}_{N-1} \,.
\end{equation}
That is, the SCFTs can be engineered in F-theory by compactification on the singular Calabi--Yau threefold that arises when all of the curves appearing in the non-compact Calabi--Yau threefold given by the right-hand side of equation \eqref{eqn:AOI_TB} are simultaneously shrunk to zero volume. These SCFTs generically possess a flavor symmetry which is
\begin{equation}\label{eqn:OIflavor}
    \mathfrak{e}_8 \oplus \mathfrak{su}_K \oplus \mathfrak{u}_1 \,,
\end{equation}
though the symmetry may enhance for particular combinations of $N$ and $K$. The orbi-instanton theories can generally be regarded as parent theories for a whole family of interacting 6d $(1,0)$ SCFTs that are further generated by Higgs branch renormalization group flows. Such flows have been studied extensively, see, for example, \cite{Heckman:2015bfa,Frey:2018vpw,Elvang:2012st,Heckman:2015ola,Heckman:2016ssk,Heckman:2018pqx,Hassler:2019eso,Heckman:2021nwg,Baume:2021qho,Fazzi:2022hal,Fazzi:2022yca,Distler:2022kjb,Cabrera:2019izd}. In particular, for $N$ sufficiently large,\footnote{If $N < K$ then not all such Higgsed theories are interacting SCFTs.} there is a family of interacting SCFTs which can be denoted
\begin{equation}
    \mathcal{O}_{N,K}(\rho, O) \,.
\end{equation}
Here, $O$ is a choice of nilpotent orbit of $\mathfrak{su}_K$; turning on a non-trivial $O$ corresponds to performing a Higgs branch renormalization group flow triggered by giving a nilpotent vacuum expectation value to the moment map of the $\mathfrak{su}_K$ flavor symmetry of the orbi-instanton. The SCFT obtained after the RG flow does not depend on the precise nilpotent element, only on the nilpotent orbit to which it belongs. Unlike $O$, $\rho$ is a choice of homomorphism from $\mathbb{Z}_K$ to $E_8$, not a choice of nilpotent orbit; that this should be the case is most straightforwardly seen by passing to the dual M-theory description of the orbi-instanton theories.

In M-theory, the orbi-instanton SCFT $\mathcal{O}_{N,K}$ can be realized as the worldvolume theory on a stack of $N$ M5-branes probing a $\mathbb{C}^2/\mathbb{Z}_K$ orbifold singularity, and further contained inside of an M9-brane \cite{DelZotto:2014hpa}. From this perspective, it is necessary to fix a choice of boundary conditions, on the $S^3/\mathbb{Z}_K$ boundary of the orbifold $\mathbb{C}^2/\mathbb{Z}_K$, for the $E_8$-bundle associated with the M9-brane. It is a standard result that flat $E_8$ connections on $S^3/\mathbb{Z}_K$ are in one-to-one correspondence with homomorphisms $\rho: \mathbb{Z}_K \rightarrow E_8$.

It is well-known that nilpotent orbits of $\mathfrak{su}_K$ are in one-to-one correspondence with integer partitions of $K$ \cite{Collingwood_1993}, and that homomorphisms $\mathbb{Z}_K \rightarrow E_8$ are in one-to-one correspondence partitions of $K$ weighted by the Dynkin labels of the affine Dynkin diagram of $\mathfrak{e}_8$ \cite{MR739850}. Specifically, in the latter case, we have that each homomorphism $\rho$ is associated to 
\begin{equation}\label{eqn:thems}
    (m_1, m_2, m_3, m_4, m_5, m_6, m_3', m_4', m_2') \in \mathbb{Z}^9_{\geq 0} 
    \,,
\end{equation}
satisfying
\begin{equation}\label{eqn:sumKac}
    m_1 + 2(m_2 + m_2') + 3(m_3 + m_3') + 4(m_4 + m_4') + 5m_5 + 6m_6 = K \,.
\end{equation}
The effective field theory description on the generic point of the tensor branch of any $\mathcal{O}_{N,K}(\rho, O)$ is given in terms of the partitions in  \cite{Mekareeya:2017jgc}. 

The continuous flavor symmetry of $\mathcal{O}_{N,K}(\rho, O)$ is generically fixed by the choice of Higgsing; we find
\begin{equation}\label{eqn:dora}
    \mathfrak{f} = \mathfrak{f}_\rho \oplus \mathfrak{f}_O \oplus \mathfrak{u}_1 \,.
\end{equation}
As explained in \cite{MR739850}, there is a natural semi-simple Lie algebra associated with the homomorphism $\rho$. This is obtained from the affine Dynkin diagram of $\mathfrak{e}_8$:
\begin{equation}
  \begin{gathered}
    \begin{tikzpicture}
      \node[node, label=below:{\footnotesize $m_1$},fill=black] (A1)  {};
      \node[node, label=below:{\footnotesize $m_2$},fill=black] (A2) [right=6mm of A1] {};
      \node[node, label=below:{\footnotesize $m_3$},fill=black] (A3) [right=6mm of A2] {};
      \node[node, label=below:{\footnotesize $m_4$},fill=black] (A4) [right=6mm of A3] {};
      \node[node, label=below:{\footnotesize $m_5$},fill=black] (A5) [right=6mm of A4] {};
      \node[node, label=below:{\footnotesize $m_6$},fill=black] (N3) [right=6mm of A5] {};
      \node[node, label=below:{\footnotesize $m'_4$},fill=black] (B4) [right=6mm of N3] {};
      \node[node, label=below:{\footnotesize $m'_2$},fill=black] (B2) [right=6mm of B4] {};
      \node[node, label=right:{\footnotesize $m'_3$},fill=black] (Nu) [above=5mm of N3] {};
      \draw (A1.east) -- (A2.west);
      \draw (A2.east) -- (A3.west);
      \draw (A3.east) -- (A4.west);
      \draw (A4.east) -- (A5.west);
      \draw (A5.east) -- (N3.west);
      \draw (N3.east) -- (B4.west);
      \draw (B4.east) -- (B2.west);
      \draw (N3.north) -- (Nu.south); 
    \end{tikzpicture}
  \end{gathered} \,.
\end{equation}
First, delete the nodes for which $m_i$, $m_i'$ are non-zero. The result is the Dynkin diagram of a semi-simple Lie algebra; this is precisely the 
non-Abelian part of the flavor factor arising from the choice of homomorphism: $\mathfrak{f}_\rho$. Similarly, a choice of nilpotent orbit of $\mathfrak{g}$ is associated to a homomorphism $\sigma_O : \mathfrak{su}_2 \rightarrow \mathfrak{g}$. The factor in the flavor algebra associated with the nilpotent orbit $O$ is the centralizer of the image of $\sigma_O$:
\begin{equation}\label{eqn:fO}
    \mathfrak{f}_O = \operatorname{Centralizer}(\mathfrak{su}_K, \sigma_O) \,.
\end{equation}
Note that for $N$ sufficiently small, there can be enhancement of the flavor symmetry appearing in equation \eqref{eqn:dora}.

For the purposes of constructing LSTs in this paper, we only consider cases where $O$ is the nilpotent orbit of $\mathfrak{su}_K$ of trivial dimension; using the correspondence between nilpotent orbits and integer partitions, we denote this via $O = [1^K]$. While the effective field theory at the generic point of the tensor branch depends in an intricate way on the $m_i$ and $m_i'$, this intricacy only affects the ramp, going from the $(-1)$-curve to the curve supporting the first $\mathfrak{su}_K$ algebra, in equation \eqref{eqn:AOI_TB}, whereas the plateau consisting of $N-1$ curves of self-intersection $(-2)$ supporting $\mathfrak{su}_K$ algebras is unmolested. Schematically, the tensor branch configuration can be written as 
\begin{equation}\label{eqn:AOIHiggsed}
    \operatorname{TB}\left[\mathcal{O}_{N,K}(\rho, [1^K])\right] \,\, = \,\, \operatorname{TB}\left[\mathcal{O}_{1,K}(\rho, [1^K])\right] \,\, \underbrace{\,\overset{\mathfrak{su}_{K}}{2} \cdots \overset{\mathfrak{su}_{K}}{2}\,}_{N-1} \,.
\end{equation}

We return to the study of these orbi-instanton SCFTs $\mathcal{O}_{N,K}(\rho, O)$ in Section \ref{sec:Higgs_Branch_OI}. In particular, we use the tensor branch description in equation \eqref{eqn:AOIHiggsed}, as well as the brane construction of the theories in M-theory (utilizing their reduction to Type IIA), to analyze the structure of their Higgs branches.

\subsection{Heterotic \texorpdfstring{$E_8 \times E_8$}{E8 x E8} Orbifold LSTs} \label{sec:HeteroticOrbitfoldLST}

The orbi-instanton theories $\mathcal{O}_{N,K}(\rho, [1^K])$ constitute the SCFT building blocks for a class of 6d $(1,0)$ LSTs that we study in this paper. The LSTs that we explore are those that arise as the worldvolume theories on stacks of NS5-branes in the heterotic string theories, either $E_8 \times E_8$ or $\mathrm{Spin}(32)/\mathbb{Z}_2$.\footnote{There has been a resurgence of interest in this class of LSTs in recent years, see \cite{Ahmed:2023lhj,DelZotto:2023nrb,DelZotto:2023ahf,DelZotto:2022xrh,DelZotto:2022ohj,DelZotto:2020sop}.} In particular, we consider heterotic string theory on an ALE orbifold, $\mathbb{C}^2/\Gamma$ where $\Gamma$ is a finite subgroup of $SU(2)$; these LSTs are often referred to as the LSTs of heterotic ALE instantons. For heterotic $\mathrm{Spin}(32)/\mathbb{Z}_2$, these LSTs are well-understood \cite{Witten:1995gx,Blum:1997mm,Intriligator:1997dh,Sagnotti:1987tw,Bianchi:1990yu}, however, the straightforward Lagrangian techniques applicable there do not apply for heterotic $E_8 \times E_8$ \cite{Aspinwall:1997ye,Ganor:1996mu}. Thus, our focus in this section is on the extraction of the physical properties of the latter from their geometric construction in F-theory.\footnote{Our principle interest in this section is $\Gamma = \mathbb{Z}_K$, however see Section \ref{sec:conc} for a discussion of non-Abelian $\Gamma$.}

\begin{figure}[t]
    \centering
\begin{tikzpicture}[decoration=snake]
  \path[save path=\pathA,name path=A] (0,4.5) to [bend left] (4.5,0);
  \path[save path=\pathB,name path=B]
    (0,0) .. controls (1.5,.45) and (1.995,4.05) .. (4.5,4.5);
  \draw[red][use path=\pathA];
  \draw[red] [use path=\pathB];
    \fill[name intersections={of=A and B}] (intersection-1) circle (2pt);
    \node[] at (5,4.5) () {\color{red} $\mathfrak{su}_K$};
    \draw[->,decorate] (2.5,1) node[below] {\footnotesize $\mathcal{O}_{N_L,K}(\rho_L, [1^K])$} -- ([yshift=-0.3cm]intersection-1);
    \node[] at (6.15,2.25) () {$+$};
     \path[save path=\pathA,name path=A] (8,0) to [bend left] (12.5,4.5);
  \path[save path=\pathB,name path=B]
    (8,4.5) .. controls (10,4.15) and (9.995,.45) .. (12.5,0);
  \draw[red][use path=\pathA];
  \draw[red] [use path=\pathB];
    \fill[name intersections={of=A and B}] (intersection-1) circle (2pt);
    \node[] at (7.5, 4.5) () {\color{red} $\mathfrak{su}_K$};
    \draw[->,decorate] (9.7,1) node[below] {\footnotesize $\mathcal{O}_{N_R,K}(\rho_R, [1^K])$} -- ([yshift=-0.3cm]intersection-1);
\end{tikzpicture}

\begin{tikzpicture}[decoration=snake]
\draw[-{Implies},double distance=3pt] (5.25,5.25)--(5.25,4.25);
  \path[save path=\pathA,name path=A] (0,4.5) to [bend left] (4.5,0);
  \path[save path=\pathB,name path=B] (0,2.25)--(9.5,2.25);
  \path[save path=\pathC,name path=C] (6,0) to [bend left] (10.5,4.5);
    \fill[name intersections={of=A and B,by=J}] (intersection-1) circle (2pt);
  \fill[name intersections={of=B and C,by=I}] (intersection-1) circle (2pt);
  \path[save path=\pathD,name path=D]
    (J) .. controls (4.25,2.75) and (6.25,2.75) .. (I);
      \path[save path=\pathE,name path=E]
    (3,2.1) to[bend right] (J);
          \path[save path=\pathF,name path=F]
    (I) to[bend right] (7.5,2.1);
  \draw[red][use path=\pathA];
  \draw[red] [use path=\pathC];
  \draw[blue][use path=\pathD];
  \draw[blue][use path=\pathE];
  \draw[blue][use path=\pathF];
    \fill[name intersections={of=A and B,by=J}] (intersection-1) circle (2pt);
  \fill[name intersections={of=B and C,by=I}] (intersection-1) circle (2pt);
  \node[] at (5.25,2.9) () {\footnotesize \color{blue} $\mathfrak{su}_K$};
  \node[] at (5.25,2.4) () {\footnotesize \color{blue} $0$-curve};

     \draw[->,decorate] (8.0,1) node[below] {\footnotesize $\mathcal{O}_{N_R,K}(\rho_R, [1^K])$} -- ([yshift=-0.2cm,xshift=0.15cm]I);
     \draw[->,decorate] (2.5,1) node[below] {\footnotesize $\mathcal{O}_{N_L,K}(\rho_L, [1^K])$} -- ([yshift=-0.2cm,xshift=-0.15cm]J);
\end{tikzpicture}

    \caption{A description of the fusion of the F-theory geometry associated to orbi-instanton theories $\mathcal{O}_{N_R,K}$ and $\mathcal{O}_{N_R,K}$ which forms a geometry associated to an LST. We depict the curves in the base of the elliptic fibration and, where relevant, we write the Lie algebra associated to the singular fiber supported over the curve. Non-compact curves responsible for flavor symmetries are shown in red, whereas compact curves are blue.}
    \label{fig:E8fromOI}
\end{figure}
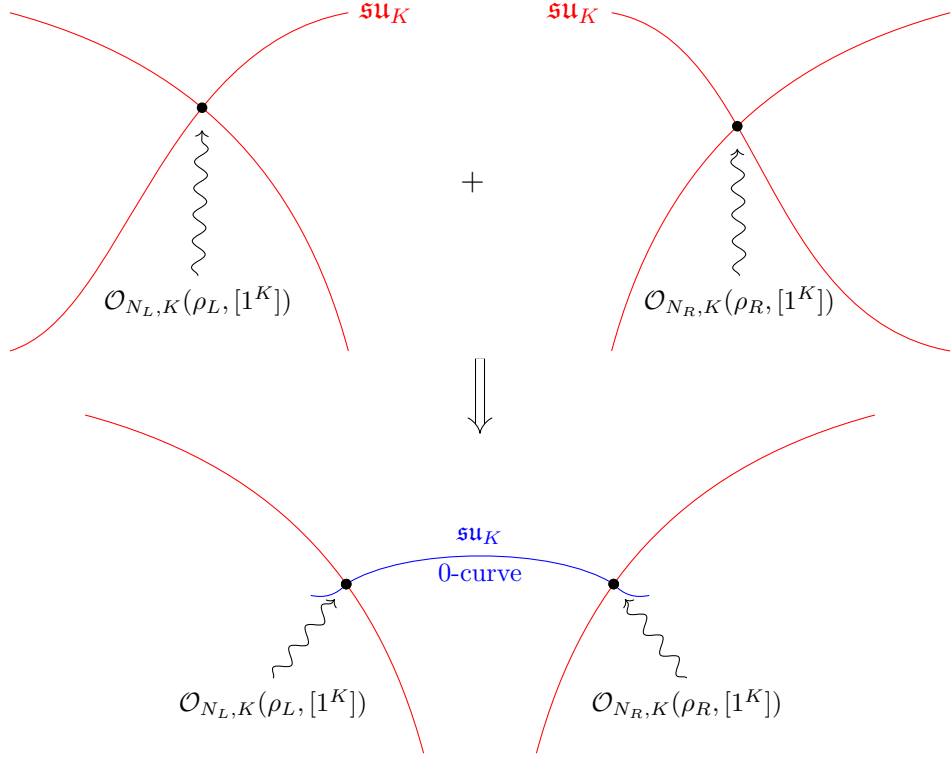

Consider the following non-compact elliptically-fibered Calabi--Yau threefold. Let the base $B$ be a non-compact complex surface containing a single compact curve, which is a smooth rational curve of self-intersection $0$. Above the generic point of this $(0)$-curve, let there be a singular fiber corresponding to the $\mathfrak{su}_K$ algebra. Finally, above two distinct points on the $(0)$-curve we localize the non-minimal singular fibers associated to, respectively, the orbi-instanton SCFTs $\mathcal{O}_{N_L,K}(\rho_L, [1^K])$ and $\mathcal{O}_{N_R,K}(\rho_R, [1^K])$. We depict this construction in Figure \ref{fig:E8fromOI}. This operation is nothing other than diagonal gauging (or, more precisely, fusion) of the $\mathfrak{su}_K$ flavor symmetry factors of the two orbi-instantons.

It is clear that the non-compact Calabi--Yau threefold satisfies all of the necessary conditions referenced in Section \ref{sec:LSTgeneralconstruction}, and thus F-theory compactified on this threefold leads to a 6d $(1,0)$ little string theory. Schematically, we could write this geometry as 
\begin{equation}\label{eqn:swiper}
    \mathcal{O}_{N_L,K}(\rho_L, [1^K]) \,\,\text{---}\,\,
    \overset{\mathfrak{su}_K}{0} \,\,\text{---}\,\, \mathcal{O}_{N_R,K}(\rho_R, [1^K]) \,,
\end{equation}
and we refer to the associated LSTs as
\begin{equation}
    \mathcal{K}_{N_L, N_R, K}(\rho_L, \rho_R) \,.
\end{equation}
From the tensor branch descriptions of the orbi-instanton building blocks that were given in equation \eqref{eqn:AOIHiggsed}, we can see that the effective description of the little string theory associated with the Calabi--Yau in equation \eqref{eqn:swiper} at the generic point of the tensor branch is
\begin{equation}\label{eqn:TBLSTHiggsed}
    \operatorname{TB}\left[\mathcal{K}_{N_L, N_R, K}(\rho_L, \rho_R)\right] \,\, = \,\, \operatorname{TB}\left[\mathcal{O}_{1,K}(\rho_L, [1^K])\right] \,\, \underbrace{\,\overset{\mathfrak{su}_{K}}{2} \cdots \overset{\mathfrak{su}_{K}}{2}\,}_{N_L + N_R -1} \operatorname{TB}\left[\mathcal{O}_{1,K}(\rho_R, [1^K])\right] \,.
\end{equation}
We notice that the tensor branch effective field theory only depends on the sum
\begin{equation}
    N = N_L + N_R \,,
\end{equation}
and not on the individual values of $N_L$ and $N_R$. This highlights an important detail in the F-theoretic construction of 6d LSTs, which is not relevant to the construction of SCFTs. The LST is associated with a tensor branch Calabi--Yau geometry together with a choice of contraction map that shrinks curves in the base under which one obtains the singular Calabi--Yau that engineers the LST itself. The choice of contraction map is not unique, and this explains why distinct LSTs can have the same tensor branch effective field theory, as was already pointed out in \cite{Aspinwall:1997ye}.

There are a variety of physical properties of $\mathcal{K}_{N_L, N_R, K}(\rho_L, \rho_R)$ that can be determined from the tensor branch description.\footnote{In the remainder of this subsection, we assume that $K \geq 2$ to avoid the need to enumerate special cases.} This includes the non-Abelian flavor symmetry algebra, which is generically
\begin{equation}
    \mathfrak{f} = \mathfrak{f}_{\rho_L} \oplus \mathfrak{f}_{\rho_R} \oplus \mathfrak{u}_1 \oplus \mathfrak{u}_1 \,,
\end{equation}
where the $\mathfrak{f}_{\rho}$ are the same as explained around equation \eqref{eqn:dora}. This flavor algebra may enhance; however, any possible enhancement depends on the value of $N$, but not $N_L$ and $N_R$ separately. The structure constants for the generalized symmetries can also be determined using equation \eqref{eqn:kappaPRgeneral}; in particular
\begin{equation}
    \kappa_P = 2 \,,
\end{equation}
and
\begin{equation}\label{eqn:kRcombined}
    \kappa_R = \delta\kappa_R(\mathcal{O}_{1,K}(\rho_L, [1^K])) + \delta\kappa_R(\mathcal{O}_{1,K}(\rho_R, [1^K])) + K(N_L + N_R - 1) \,.
\end{equation}
Here, we have defined $\delta\kappa_R(\mathcal{O}_{N,K}(\rho, O))$ as the contribution to $\kappa_R$ from the Higgsed orbi-instanton. The structure constants describing the mixing between the $\mathfrak{u}(1)^{(1)}_{LST}$ one-form symmetry and the non-Abelian flavor symmetries can also be determined from the tensor branch as in equation \eqref{eqn:kappaFgeneral}; a priori the general result may depend in an intricate way of the $m_i$, $m_i'$ that specify the $E_8$-homomorphism. However, since we have limited our study to the cases where $\mathfrak{g} = \mathfrak{su}_K$, it is straightforward to see that the structure constant for each of the simple factors in the flavor symmetry is 
\begin{equation}
    \kappa_F = 1 \,.
\end{equation}
In particular, the values of $\kappa_F$ for the non-Abelian factors in the flavor symmetry do not depend on $N$.

We also explore the moduli spaces of these LSTs. The first such quantity of interest is the dimension of the Coulomb branch of the five-dimensional theory obtained by circle-reduction. From the tensor branch description, this is simply the total number of curves that contract to zero-volume plus the number of Cartans in each fibral algebra, i.e., 
\begin{equation}
    \operatorname{dim}(\mathcal{C}) = 
    d_C(\mathcal{O}_{1,K}(\rho_L, [1^K])) + d_C(\mathcal{O}_{1,K}(\rho_R, [1^K])) + K(N_L + N_R - 1) \,,
\end{equation}
where we have defined the dimension of the Coulomb branch of the $S^1$ compactification of the arbitrarily Higgsed orbi-instanton SCFT as $d_C(\mathcal{O}_{N,K}(\rho, O))$. Similarly, we can consider the dimension of the Higgs branch of the LST. The dimension of the Higgs branch is related to the gravitational anomaly of the LST, in an obvious generalization of the known formula for either ``very Higgsable'' or ``Higgsable to $(2,0)$'' SCFTs \cite{Ohmori:2015pua,Ohmori:2015pia,Shimizu:2017kzs}. The relevant terms in the anomaly polynomial of the LST are straightforwardly determined from the usual algorithm for extracting the anomaly polynomial of an SCFT from its tensor branch description \cite{Ohmori:2014pca,Ohmori:2014kda,Intriligator:2014eaa,Baume:2021qho}. In the end, the general expression for the Higgs branch dimension is
\begin{equation}
    \operatorname{dim}(\mathcal{H}) = d_H(\mathcal{O}_{1,K}(\rho_L, [1^K])) + d_H(\mathcal{O}_{1,K}(\rho_R, [1^K])) + 30(N_L + N_R - 1) - K^2 - 29 \,,
\end{equation}
where, as usual, we have defined $d_H(\mathcal{O}_{N,K}(\rho, O))$ as the dimension of the Higgs branch of the Higgsed orbi-instanton SCFT $\mathcal{O}_{N,K}(\rho, O)$. Once again, we see that the dependence is only on the linear combination $N_L + N_R$.

Therefore, the invariants of the theories $\mathcal{K}_{N_L, N_R, K}(\rho_L, \rho_R)$ which we have introduced here and denoted as
\begin{equation}\label{eqn:invariants}
    \mathfrak{f} \,,\quad \kappa_P \,,\quad \kappa_R \,,\quad \kappa_F \,,\quad \operatorname{dim}(\mathcal{C}) \,,\quad \operatorname{dim}(\mathcal{H}) \,,
\end{equation}
are identical across the whole family of theories where $N_L + N_R$ is fixed. In this sense, these quantities are somewhat lacking as invariants -- they are exactly equivalent for arbitrarily large numbers of LSTs! We propose that a refined invariant is the Hasse diagram of the Higgs branch, which is sensitive to the different choices of $(N_L, N_R)$, and not just the sum. We return to the Higgs branch Hasse diagrams for $\mathcal{K}_{N_L,N_R,K}(\rho_L, \rho_R)$ in Section \ref{sec:LSTs}. Understanding invariants is particularly important for the study of T-duality among little string theories, the topic to which we now turn.

\subsection{T-duality for LSTs}\label{sec:Tdual}

Dualities provide an important tool for the investigation into non-perturbative behaviors of quantum field theories. Duality has a variety of related definitions, however the key aspect is that there are a priori multiple different theories for which all physical quantities are the same. One example of duality is strong-weak duality, where a theory at strong coupling has a dual description via an alternative theory at weak coupling; such a duality allows us to determine physical observables at strong-coupling in the initial theory in terms of perturbative physical quantities in the dual theory.

An early example \cite{Itzykson:1989sy} of (strong-weak) duality occurred in the context of lattice field theory, where a Hamiltonian $H(g_i)$ dependant on a gauge coupling $g_i$ at strong coupling exhibits the same behavior as a different Hamiltonian depending on a different gauge coupling, $H^*(g^*_i)$, in a weak coupling regime. In the same spirit, one can consider T-duality, a spacetime-based duality, that in its simplest realization connects the behavior of a $d+1$-dimensional theory on $\mathcal{M}_{1,d-1} \times S^1$, where $\mathcal{M}_{1,d-1}$ is a $d$-dimensional spacetime manifold and $S^1$ is a circle of radius $R$, at large spacetime radius with the one at small radius. An elucidating example is the propagation of closed strings in spacetime with one coordinate compactified on a circle: the pure momentum states of energy $E_p=\frac{n}{R}$ can be regarded as pure winding states of energy $E_w=nR$ on a circle of radius $\frac{1}{R}$. A good introductory review on T-duality in string theory is \cite{Alvarez:1994dn}, whereas a more in-depth view of T-duality in toroidal compactifications as an action of elements of a discrete subgroup of the spacetime symmetry is given in \cite{Giveon:1991jj}.

An interesting example of T-duality can be found in toroidal compactifications of the 10-dimensional heterotic string theory with spacetime topology $\mathbb{R}^{10}$ to a $10-d$-non-compact-dimensional model with spacetime topology $\mathbb{R}^{10-d}\times \left(S^1\right)^d$. The $d$-dimensional torus $(S^1)^d$ is obtained by modding the original $d$-dimensional portion of the spacetime via a lattice $\Lambda$, under which vacuum pure winding states are valued. Whereas vacuum pure momenta states are now valued in the dual lattice $\Lambda_{\text{dual}}$ obtained by inverting the metric defined by the generating lattice vector and creating a new orthogonal generator set of vectors.\footnote{For details on the lattice construction and an explicit interpolation of heterotic T-dual models, see \cite{Ginsparg:1986bx}.}
The emergence of T-dual models under this construction \cite{Narain:1985jj} relies on the uniqueness of the self-dual bosonic lattice of the heterotic model: the $E_8$ weight lattice, $\Gamma_8$, and the $\mathrm{Spin}(32)/\mathbb{Z}_2$ weight lattice, $\Gamma_{16}$, also known as the Barnes--Wall lattice, satisfy
\begin{equation}
    \Gamma_8 \oplus \Gamma_8 \oplus \Gamma_{1,1} \cong \Gamma_{16} \oplus \Gamma_{1,1} \,,
\end{equation}
thus implying that a choice of the background gauge field and the antisymmetric tensor field during the compactification is always available to satisfy this equality.

In this context, we can construct heterotic LSTs that, by the argument just given, exhibit T-duality. Here instead of starting directly from the $10$-dimensional model and performing the compactification while tracking down explicitly the values required for the background fields to exhibit the duality, a fruitful approach involves comparing the six-dimensional models directly. Section \ref{sec:LSTgeneralconstruction} summarised how to assemble an LST directly by gluing together some minimal six-dimensional building blocks, whereas Section \ref{sec:HeteroticOrbitfoldLST} dealt with how to realise a heterotic model out of them. In this 6d context, we can regard the duality as a statement of an $S^1$-compactification of those models in the presence of Wilson lines, and try to shed light on dual models without delving into the geometry \cite{DelZotto:2022ohj}.

In this sense, we can consider two 6d LSTs, $\mathcal{T}_1$ and $\mathcal{T}_2$, to be T-dual under the following circumstances. Let $\mathfrak{f}_1$ and $\mathfrak{f}_2$ denote the continuous flavor algebras of the respective LSTs.  We can consider a compactification of each theory on an $S^1$, together with a choice of Wilson lines along each circle valued in the flavor algebra. This way we obtain theories
\begin{equation}
    \mathcal{T}_1 \langle S^1, \operatorname{WL}_1 \rangle \,, \qquad \mathcal{T}_2 \langle S^1, \operatorname{WL}_2 \rangle \,,
\end{equation}
where we have used $\operatorname{WL}_i$ to schematically denote the choice of Wilson lines that were turned on. If there exists a choice of Wilson lines $(\operatorname{WL}_1, \operatorname{WL}_2)$ such that these two 5d theories are identical (at some point on the Coulomb branch), then we say that the two theories are T-dual. To identify potential T-dual pairs, we should then compute invariants of $\mathcal{T}_1$ and $\mathcal{T}_2$ that are unchanged (or modified in a predictable way) under the $S^1$-compactification with Wilson lines. Thus one is interested in quantities that are T-dual invariants, such as \cite{DelZotto:2020sop}: 
    \begin{equation}\label{eqn:T-dual-invariants}
    \quad \kappa_P \,,\quad \kappa_R \,,\quad \kappa_F \,,\quad \operatorname{dim}(\mathcal{C}) \,,\quad \operatorname{rank}(\mathfrak{f}) \,.
\end{equation}
Notice that this set of invariants is a subset of those in equation \eqref{eqn:invariants}, as here the theories under investigation can be taken also in the $\mathrm{Spin}(32)/\mathbb{Z}_2$ model and not only in the $E_8 \times E_8$ one. The first three quantities are anomaly coefficients, and as such must match among dual theories,\footnote{The structure constant $\kappa_F$ requires some care, as it may be multiplied by an embedding index when the Wilson lines break the 6d flavor symmetry to a subalgebra in 5d; thus, it is this scaled version of $\kappa_F$ that must match across T-duality. We will see more on this in Section \ref{sec:beyond}.} the last two quantities are morally different. In fact, during the circle compactification, the six-dimensional multiplets break in a prescribed way and this determines the equality of the 5d Coulomb branch, whereas the presence of Wilson lines modifies the flavor algebra from the 6d flavor, and hence affects the Higgs branch structure, and preserves only the total rank of the flavor algebra.

Even though the quantities in equation \eqref{eqn:T-dual-invariants} seem to provide promising indicators to identify dual theories, they do not uniquely identify a theory. There are examples of LSTs where the quantities in equation \eqref{eqn:T-dual-invariants} are identical, but a geometric approach demonstrates that the LSTs are not T-dual. Consider for instance the family of theories
\begin{equation}
    \mathcal{K}_{12-N_R, N_R, 0}(1^8, 1^8)=[E_8]\,\underbrace{1\ 2 \ \cdots \ 2 \ 1}_{12 \text{ curves}}\,[E_8] \,,
\end{equation}
for which the T-dual-invariant quantities are independent of $N_R$, and value:
\begin{equation}
     \quad \kappa_P =2 \,,\quad \kappa_R = 12 \,,\quad \kappa_F=1 \,,\quad \operatorname{dim}(\mathcal{C}) = 12 \,,\quad \operatorname{rank}(\mathfrak{f}) =18 \,.
\end{equation}
One could engineer a T-dual $\mathrm{Spin}(32)/\mathbb{Z}_2$ model by the lattice argument, but then would this theory be T-dual to all the elements in this family, i.e., for all values of $N_R$? Unsurprisingly, the answer is no. As shown in \cite{Morrison:1996na,Morrison:1996pp}, only the $\mathcal{K}_{4, 8, 0}(1^8,1^8)$ model has a $\mathrm{Spin}(32)/\mathbb{Z}_2$  heterotic dual. This result has been achieved by considering compact geometries and explicitly matching the lattices, hinting at the reality that only by explicitly comparing geometries one can extract T-dual models.

\begin{figure}[H]
    \centering
    \begin{tikzpicture}
    \filldraw (0,0) circle (3pt);
    \filldraw (0,-1) circle (3pt);
    \filldraw (1,-2) circle (3pt);
    \filldraw (-1,-2) circle (3pt);
    \filldraw (0,-3) circle (3pt);
    \filldraw (0,-4) circle (3pt);
    \draw (0,0)--(0,-1)--(1,-2)--(0,-3)--(0,-4);
    \draw (0,-1)--(-1,-2)--(0,-3);

        \filldraw (3,0) circle (3pt);
    \filldraw (3,-1) circle (3pt);
    \filldraw (3,-2) circle (3pt);
    \filldraw (3,-3) circle (3pt);
    \filldraw (3,-4) circle (3pt);
    \draw (3,0)--(3,-1)--(3,-2)--(3,-3)--(3,-4);
    \end{tikzpicture}
    \caption{A schematic example of a subdiagram of the Hasse diagrams of $\mathcal{K}_{N_L, N_R, K}(\rho_R, \rho_L)$ for different choices of $N_L$ and $N_R$ that show how the different assignment of the contraction maps affect the foliation of the Higgs branch moduli space.}
    \label{fig:different-hasse-same-curves}
\end{figure}
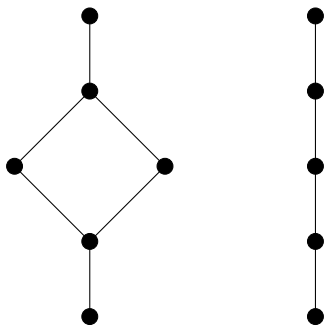

Nevertheless, the invariant quantities in equation \eqref{eqn:T-dual-invariants} already perform a good job in cutting down the number of models for which a geometric comparison is needed. One can also improve this list by introducing a further structure that is sensitive to different model geometries: the Higgs branch Hasse diagram structure. In fact, even if the Higgs branch itself is modified under compactification, RG-flows on it automatically know about the base geometry: see Figure \ref{fig:different-hasse-same-curves} where anticipating some of the results of this paper, different Hasse diagrams are associated with theories with the same curve configurations and same T-dual invariants.

Hence even if this quantity by itself does not explicitly determine which of the $E_8\times E_8$ heterotic models has a T-dual in the $\mathrm{Spin}(32)/\mathbb{Z}_2$ counterpart, it can be gainfully employed to shed light on how different geometries realising the same tensor branch curve configuration are actually different even when the invariants in equation \eqref{eqn:T-dual-invariants} are all equal.

\section{The Higgs Branch of A-type Orbi-instantons}\label{sec:Higgs_Branch_OI}

Before turning to the central topic of this paper, the Higgs branch of heterotic LSTs, it is useful to revisit the study of the Higgs branch of the 6d $(1,0)$ SCFTs that we discussed in Section \ref{sec:AtypeOI}: the rank $N$ $(\mathfrak{e}_8, \mathfrak{su}_K)$ orbi-instanton theories. In particular, we use this section to introduce and review the relevant concepts that also arise in the SCFT context that are necessary to tackle the case of little string theories.

In addition to the Higgs branch of the SCFT, there are a variety of different Higgs branches which are associated with $\mathcal{O}_{N,K}$. We have addressed the SCFT that lives at the origin of the tensor branch and the field theory that lives at the generic point of the tensor branch.  At non-generic points of the tensor branch, there are also effective field theory descriptions, perhaps coupled to conformal sectors unlike at the generic point. Emanating from each point of the tensor branch is a Higgs branch associated with the theory that lives at that point. These Higgs branches are often distinct hyperk\"ahler spaces, and they often have different dimensions.

From the tensor branch description, we can use the 't Hooft anomaly matching techniques of \cite{Ohmori:2014pca,Ohmori:2014kda,Intriligator:2014eaa,Baume:2021qho} to determine the anomaly polynomial of the associated SCFT, and thus to determine the dimension of the Higgs branch. We can also use the tensor branch description to determine the anomaly polynomial of the effective field theory on the generic point of the tensor branch, and thus the dimension of the Higgs branch of that effective field theory; for orbi-instanton theories, the dimension of the Higgs branch typically increases as one moves from the generic point of the tensor branch to the SCFT at the origin; this is related to the famed small instanton transition \cite{Witten:1995gx}. In particular, for the SCFT $\mathcal{O}_{N,K}$, we have
\begin{equation}
    d_H = \frac{K(K+1)}{2} + 30(N+K) -1  \,,
\end{equation}
whereas, at the generic point of the tensor branch, the Higgs branch dimension of the effective field theory is
\begin{equation}
    d_H^\text{TB} = \frac{K(K+1)}{2} + (N+K) -1 \,. 
\end{equation}

Going beyond just the dimension, a natural question to now ask is: what is the structure of the Higgs branch of the orbi-instanton theory $\mathcal{O}_{N,K}$? For sufficiently large $N$, it is clear that some of the interacting SCFT fixed points on the Higgs branch should correspond to all possible pairs of $E_8$-homomorphism $\rho$ and nilpotent orbit $O$, though generically these are only a subset of the fixed points. Furthermore, we would like to understand whether, if we have fixed points labelled by $(\rho, O)$ and $(\rho', O')$, there exists a renormalization group flow between these two fixed points. For fixed $\rho$, the dominance ordering between the partitions associated with $O$ and $O'$ is equivalent to the ordering of the Higgs branch RG flows,\footnote{In \cite{Cremonesi:2014uva}, an explicit correspondence between nilpotent orbits, $\rho$, of classical algebras $G$, and Gaiotto--Witten $T_\rho(G)$ theories has been given. This has been extended in  \cite{Cabrera:2016vvv,Cabrera:2018ldc} by connecting the RG flows with Kraft--Procesi transitions in the context of Type IIB D3-D5-NS5-brane systems engineering 3d $\mathcal{N}=4$ quiver gauge theories. As we will see, the 3d mirror for the $(\mathfrak{e}_8, \mathfrak{su}(K))$ orbi-instanton involves a copy of the $T\left(SU(K)\right)$ there, where the $\mathfrak{su}(K)$ flavor symmetry comes from the Coulomb symmetry of the 3d mirror. Therefore the equivalence between the dominance ordering and the Higgs branch RG flows can be established from the brane system.} however, mathematically no such similar ordering is known for the homomorphisms from $\mathbb{Z}_K \rightarrow E_8$.\footnote{However, the 6d $(1,0)$ SCFT perspective has been gainfully used to write a putative partial ordering on the set of $E_8$-homomorphisms \cite{Frey:2018vpw}.}

The SCFT $\mathcal{O}_{N,K}$ compactified on a torus gives rise to a 4d $\mathcal{N}=2$ SCFT. Interestingly, the resulting SCFT has an alternative interpretation as a class $\mathcal{S}$ theory, which we can exploit to learn about the Hasse diagram of interacting fixed points on the Higgs branch. Class $\mathcal{S}$ \cite{Gaiotto:2009we,Gaiotto:2009hg} is a method for generating 4d $\mathcal{N}=2$ SCFTs starting from the 6d $(2,0)$ SCFT of ADE-type $\mathfrak{g}$ and performing a twisted compactification on a $n$-punctured genus $g$ Riemann surface, $C_{g,n}$; we refer to these 4d SCFTs as
\begin{equation}
    \mathcal{S}_\mathfrak{g} \langle C_{g,n} \rangle \{ \cdots \} \,.
\end{equation}
The $\cdots$ encodes the information describing the nature of each of the $n$ punctures. In this section, we consider only $\mathfrak{g} = \mathfrak{su}(K)$, and punctures which are both regular and untwisted; the puncture data is then a choice of nilpotent orbit of $\mathfrak{su}(K)$, which are in one-to-one correspondence to integer partitions of $K$. The $T^2$ compactification of the $\mathcal{O}_{N,K}$ is \cite{Mekareeya:2017jgc}
\begin{equation}\label{eqn:OIclassS}
    \mathcal{O}_{N,K} \quad \xlongrightarrow{T^2} \quad \mathcal{S}_{\mathfrak{su}(6(N+K))}\langle C_{0,3} \rangle \{ [1^K, N, (N+K)^5], [(3N + 3K)^2], [(2N + 2K)^3] \} \,.
\end{equation}
We can see that the class $\mathcal{S}$ description generally makes manifest a flavor algebra which is
\begin{equation}
    \mathfrak{s}[\mathfrak{u}(K) \oplus \mathfrak{u}(1) \oplus \mathfrak{u}(5)] \oplus 
    \mathfrak{su}(2) \oplus \mathfrak{su}(3) \,.
\end{equation}
Using the standard class $\mathcal{S}$ machinery, we can compute the Hall--Littlewood index and count the number of moment-maps of the theory, which reproduces the dimension of the expected flavor algebra that is given in equation \eqref{eqn:OIflavor}.

Given a 4d $\mathcal{N}=2$ SCFT, we can consider the $S^1$-reduction to obtain a 3d $\mathcal{N}=4$ theory; and the Higgs branch of the 3d theory is identical to the Higgs branch of the original 4d SCFT. Under 3d mirror symmetry, we obtain another 3d $\mathcal{N}=4$ theory, which has Higgs and Coulomb branches swapped from the $S^1$-reduction. For class $\mathcal{S}$ theories coming from the A-type $(2,0)$ theory with only regular and untwisted punctures, the 3d mirrors are well-known \cite{Benini:2010uu}. For class $\mathcal{S}$ of type $\mathfrak{su}(K)$ on a sphere with $n$ regular untwisted punctures associated to nilpotent orbits $\rho_1, \cdots, \rho_n$ of $\mathfrak{su}(K)$ the 3d mirror is a Lagrangian star-shaped quiver obtained by gauging the common $\mathfrak{su}(K)$ flavor symmetry of the 3d theories known as $T_{\rho_i}(SU(K))$ studied in \cite{Gaiotto:2008ak}. Therefore, the 3d mirror of the class $\mathcal{S}$ theory in equation \eqref{eqn:OIclassS} is\footnote{The 3d quivers that we write in this paper all have unitary gauge nodes, which we denote by writing the rank below the node. A quiver consisting of only unitary gauge nodes and no flavor nodes has an overall $\mathfrak{u}(1)$ gauge algebra under which none of the matter is charged and which thus decouples. This $\mathfrak{u}(1)$ can be decoupled from any of the $\mathfrak{u}(M)$ gauge nodes.}
\begin{equation}
\begin{gathered}
    \begin{tikzpicture}
      \node[node, label=below:{\footnotesize $1$}] (Z1)  {};
      \node[node, label=below:{\footnotesize $2$}] (Z2) [right=8mm of Z1] {};
      \node[tnode] (dots) [right=8mm of Z2] {\footnotesize $\cdots$};
      \node[node, label=below:{\footnotesize $K$}] (Zk) [right=8mm of dots] {};
      \node[node, label={[below right=5pt and -9pt]:\rotatebox{-30}{\footnotesize $(N+K)$}}] (A1)  [right=8mm of Zk] {};
      \node[node, label={[below right=5pt and -9pt]:\rotatebox{-30}{\footnotesize $2(N+K)$}}] (A2) [right=8mm of A1] {};
      \node[node, label={[below right=5pt and -9pt]:\rotatebox{-30}{\footnotesize $3(N+K)$}}] (A3) [right=8mm of A2] {};
      \node[node, label={[below right=5pt and -9pt]:\rotatebox{-30}{\footnotesize $4(N+K)$}}] (A4) [right=8mm of A3] {};
      \node[node, label={[below right=5pt and -9pt]:\rotatebox{-30}{\footnotesize $5(N+K)$}}] (A5) [right=8mm of A4] {};
      \node[node, label={[below right=5pt and -9pt]:\rotatebox{-30}{\footnotesize $6(N+K)$}}] (N3) [right=8mm of A5] {};
      \node[node, label={[below right=5pt and -9pt]:\rotatebox{-30}{\footnotesize $4(N+K)$}}] (B4) [right=8mm of N3] {};
      \node[node, label={[below right=5pt and -9pt]:\rotatebox{-30}{\footnotesize $2(N+K)$}}] (B2) [right=8mm of B4] {};
      \node[node, label=right:{\footnotesize $3(N+K)$}] (Nu) [above=7mm of N3] {};
      \draw (Z1.east) -- (Z2.west);
      \draw (Z2.east) -- (dots.west);
      \draw (dots.east) -- (Zk.west);
      \draw (Zk.east) -- (A1.west);
      \draw (A1.east) -- (A2.west);
      \draw (A2.east) -- (A3.west);
      \draw (A3.east) -- (A4.west);
      \draw (A4.east) -- (A5.west);
      \draw (A5.east) -- (N3.west);
      \draw (N3.east) -- (B4.west);
      \draw (B4.east) -- (B2.west);
      \draw (N3.north) -- (Nu.south); 
    \end{tikzpicture}
  \end{gathered} \,.
\end{equation}
In fact, the class $\mathcal{S}$ description, and thus the 3d mirror, for any $\mathcal{O}_{N,K}(\rho, O)$ was given in \cite{Mekareeya:2017jgc}. Let $O$ be associated with an integer partition of $K$, which we refer to as $P_O$. Then, consider the following integer partitions:
\begin{equation}
  \begin{aligned}
    \rho_1 &= &[&N_{inst} - m_6, \\ 
      & & &N_{inst} - m_6 - m_5, \\ 
      & & &N_{inst} - m_6 - m_5 - m_4, \\
      & & &N_{inst} - m_6 - m_5 - m_4 - m_3, \\
      & & &N_{inst} - m_6 - m_5 - m_4 - m_3 - m_2, \\
      & & &N_{inst} - m_6 - m_5 - m_4 -m_3 - m_2 - m_1] \,\sqcup\, P_O \,, \\
    \rho_2 &= &[&2(N_{inst}) + 2m_4' + m_3' + m_2', \\
    & & &2(N_{inst}) + m_4' + m_3' + m_2', \\
    & & &2(N_{inst}) + m_4' + m_3'] \,, \\
    \rho_3 &= &[&3(N_{inst}) + 2m_4' + 2m_3' + m_2', \\
    & & &3(N_{inst}) + 2m_4' + m_3' + m_2'
    ] \,,
  \end{aligned}
\end{equation}
where the $m_i$, $m_i'$ are the Kac labels defining the $E_8$-homomorphism $\rho$ in equation \eqref{eqn:thems}. It is easy to see that each of the $\rho_i$ is an integer partition of 
\begin{equation}
    N_{\mathcal{S}} \,\,\coloneqq\,\, 6(N_{inst}) + K - \sum_{i=1}^6 i m_i \,\,=\,\, 6(N_{inst}) + \sum_{i=2}^4 i m_i' \,,
\end{equation}
where we have used equation \eqref{eqn:sumKac} to show the equality, and we have defined this quantity as $N_{\mathcal{S}}$ in terms of the number\footnote{In fact, $N_{inst}$ counts the number of curves in the six-dimensional tensor branch configuration.} 
\begin{equation}
    N_{inst}=N+K-\sum \limits_{i=1}^{6} (i-1) m_i - \sum \limits_{i=2}^{4} i m'_{i} \,.
\end{equation}
The class $\mathcal{S}$ description of the torus-compactification of $\mathcal{O}_{N,K}(\rho, O)$ is then
\begin{equation}
    \mathcal{O}_{N,K}(\rho, O) \quad \xlongrightarrow{T^2} \quad \mathcal{S}_{\mathfrak{su}(N_{\mathcal{S}})}\langle C_{0,3} \rangle \{ \rho_1, \rho_2, \rho_3 \} \,.
\end{equation}
This proposal has been derived and verified extensively, for example by comparing the 't Hooft anomalies from both perspectives, in \cite{Mekareeya:2017jgc}. Of particular interest to us are the cases where $P_O = [1^K]$; then, there is a naive $\mathfrak{su}(K)$ flavor algebra with flavor central charge
\begin{equation}
    k = 2K + 12 \,.
\end{equation}
Once the class $\mathcal{S}$ description is known, the 3d mirror is also known, following \cite{Benini:2010uu}, and it is
\begin{equation}\label{eqn:3dmirror}
    \begin{gathered}
    \begin{tikzpicture}
    \node[tnode] (T1) {$T_{\rho_1}(SU(N_{\mathcal{S}}))$};
    \node[node, label=below:{\footnotesize $N_{\mathcal{S}}$}] (N3) [right=6mm of T1] {};
    \node[tnode] (T2) [right=6mm of N3] {$T_{\rho_2}(SU(N_{\mathcal{S}}))$};
    \node[tnode] (Nu) [above=5mm of N3] {$T_{\rho_3}(SU(N_{\mathcal{S}}))$};
    
    \draw (T1.east) -- (N3.west);
    \draw (N3.east) -- (T2.west);    
    \draw (N3.north) -- (Nu.south);
\end{tikzpicture}
    \end{gathered}
    \,.
\end{equation}
In fact, the 3d mirror constructed in this way is a Lagrangian quiver, and thus we can use well-developed and sophisticated techniques to explore the structure of the Coulomb branch. 

Alternatively, a Lagrangian quiver describing a SQFT which has the same Coulomb branch as the 3d mirror is given by the ``magnetic quiver'' \cite{Cabrera:2019izd}, which can be engineered from the brane realization of the 6d $(1,0)$ SCFT.\footnote{Whether the magnetic quiver constructed in this way actually reproduces the 3d mirror is moot for our purposes; to understand the Higgs branch of the 6d SCFT, it is sufficient to understand the Coulomb branch of the 3d mirror.} Reducing the M-theory description of $\mathcal{O}_{N,K}$ to Type IIA, we obtain a configuration consisting of $N$ NS5-branes, $K$ D6-branes, and the M9-brane becomes an 8 D8/O8$^-$ stack. The magnetic phase \cite{Cabrera:2019izd}, involves suspending the D6-branes between D8-branes, instead of the NS5s; from this description, there are straightforward rules to read off the Lagrangian quiver gauge theory which is proposed to have the same Coulomb branch as the 3d mirror of $\mathcal{O}_{N,K}$.

More generally, for $\mathcal{O}_{N,K}(\rho, [1^K])$, this procedure has been carried out in \cite{Cabrera:2019izd}; that is, the authors determine the quiver gauge theory describing the Higgs branch of the orbi-instanton theories where the $\mathfrak{e}_8$ flavor symmetry is Higgsed, but the $\mathfrak{su}(K)$ flavor symmetry is not. This was determined for each phase, i.e., for the SCFT where all of the M5-branes sit on top of the M9-brane wall, and for all tensor branch loci where the M5-branes are separated from each other, and from the M9-brane. We review the results for three cases here.\footnote{There are many more phases, corresponding to the different ways of combining/separating the brane stacks, but only the following three will be important in what follows.}
\begin{itemize}
    \item The ``finite-coupling theory'' is the gauge theory which describes the magnetic quiver of the effective 6d theory at the generic point of the tensor branch; in M-theory, this is where the M5-branes and the M9-brane are all separated. The finite-coupling magnetic quiver for $\mathcal{O}_{N,K}(\rho, [1^K])$ is
    \begin{equation}
       \begin{gathered}
    \begin{tikzpicture}
      \node[node, label=below:{\footnotesize $1$}] (Z1)  {};
      \node[node, label=below:{\footnotesize $2$}] (Z2) [right=6mm of Z1] {};
      \node[tnode] (dots) [right=6mm of Z2] {\footnotesize $\cdots$};
      \node[node, label=below:{\footnotesize $K$}] (Zk) [right=6mm of dots] {};
      \node[node, label=left:{\footnotesize $1$}] (Nu1) [above left=7mm of Zk] {};
      \node[tnode] (dots1) [above=3mm of Zk] {\footnotesize $\cdots$};
      \node[node, label=right:{\footnotesize $1$}] (Nu2) [above right=7mm of Zk] {};
      \node[node, label=below:{\footnotesize $g_1$}] (A1)  [right=6mm of Zk] {};
      \node[node, label=below:{\footnotesize $g_2$}] (A2) [right=6mm of A1] {};
      \node[node, label=below:{\footnotesize $g_3$}] (A3) [right=6mm of A2] {};
      \node[node, label=below:{\footnotesize $g_4$}] (A4) [right=6mm of A3] {};
      \node[node, label=below:{\footnotesize $g_5$}] (A5) [right=6mm of A4] {};
      \node[node, label=below:{\footnotesize $g_6$}] (N3) [right=6mm of A5] {};
      \node[node, label=below:{\footnotesize $g_7$}] (B4) [right=6mm of N3] {};
      \node[node, label=below:{\footnotesize $2r$}] (B2) [right=6mm of B4] {};
      \node[node, label=right:{\footnotesize $g_8$}] (Nu) [above=5mm of N3] {};
      \draw (Z1.east) -- (Z2.west);
      \draw (Z2.east) -- (dots.west);
      \draw (dots.east) -- (Zk.west);
      \draw (Zk) -- (Nu1);
      \draw (Zk) -- (Nu2);
      \draw (Zk.east) -- (A1.west);
      \draw (A1.east) -- (A2.west);
      \draw (A2.east) -- (A3.west);
      \draw (A3.east) -- (A4.west);
      \draw (A4.east) -- (A5.west);
      \draw (A5.east) -- (N3.west);
      \draw (N3.east) -- (B4.west);
      \draw (B4.east) -- (B2.west);
      \draw (N3.north) -- (Nu.south); 
      \draw[decorate,decoration={brace,raise=2mm}] (Nu1.west)--(Nu2.east) node[midway,above=2mm] () {\footnotesize $n+p$};
    \end{tikzpicture}
  \end{gathered} \,.
    \end{equation}
    \item The ``semi-infinite-coupling theory'' is the gauge theory describing the magnetic quiver at the subloci of the tensor branch where all but one curve in equation \eqref{eqn:AOI_TB} have been shrunk to zero-volume. In M-theory, this is equivalent to the M5-branes being together in one stack but displaced from the M9-brane. The semi-infinite-coupling magnetic quiver for $\mathcal{O}_{N,K}(\rho, [1^K])$ is
    \begin{equation}
        \begin{gathered}
    \begin{tikzpicture}
      \node[node, label=below:{\footnotesize $1$}] (Z1)  {};
      \node[node, label=below:{\footnotesize $2$}] (Z2) [right=6mm of Z1] {};
      \node[tnode] (dots) [right=6mm of Z2] {\footnotesize $\cdots$};
      \node[node, label=below:{\footnotesize $K$}] (Zk) [right=6mm of dots] {};
      \node[node, label=left:{\footnotesize $n+p$}] (Nu1) [above=5mm of Zk] {};
      \node[node, label=below:{\footnotesize $g_1$}] (A1)  [right=6mm of Zk] {};
      \node[node, label=below:{\footnotesize $g_2$}] (A2) [right=6mm of A1] {};
      \node[node, label=below:{\footnotesize $g_3$}] (A3) [right=6mm of A2] {};
      \node[node, label=below:{\footnotesize $g_4$}] (A4) [right=6mm of A3] {};
      \node[node, label=below:{\footnotesize $g_5$}] (A5) [right=6mm of A4] {};
      \node[node, label=below:{\footnotesize $g_6$}] (N3) [right=6mm of A5] {};
      \node[node, label=below:{\footnotesize $g_7$}] (B4) [right=6mm of N3] {};
      \node[node, label=below:{\footnotesize $2r$}] (B2) [right=6mm of B4] {};
      \node[node, label=right:{\footnotesize $g_8$}] (Nu) [above=5mm of N3] {};
      \draw (Z1.east) -- (Z2.west);
      \draw (Z2.east) -- (dots.west);
      \draw (dots.east) -- (Zk.west);
      \draw (Zk) -- (Nu1);
      \draw (Nu1) to[out=130, in=410, looseness=12] (Nu1);
      \draw (Zk.east) -- (A1.west);
      \draw (A1.east) -- (A2.west);
      \draw (A2.east) -- (A3.west);
      \draw (A3.east) -- (A4.west);
      \draw (A4.east) -- (A5.west);
      \draw (A5.east) -- (N3.west);
      \draw (N3.east) -- (B4.west);
      \draw (B4.east) -- (B2.west);
      \draw (N3.north) -- (Nu.south); 
    \end{tikzpicture}
  \end{gathered}  \,.
    \end{equation}
    \item Finally, the ``infinite-coupling theory'' is the gauge theory describing the magnetic quiver of the 6d SCFT itself; i.e., where the stack of $N$ M5-branes is contained inside of the M9-brane. The infinite-coupling magnetic quiver for $\mathcal{O}_{N,K}(\rho, [1^K])$ is
    \begin{equation}\label{eqn:infcplingMQ}
       \begin{gathered}
    \begin{tikzpicture}
      \node[node, label=below:{\footnotesize $1$}] (Z1)  {};
      \node[node, label=below:{\footnotesize $2$}] (Z2) [right=6mm of Z1] {};
      \node[tnode] (dots) [right=6mm of Z2] {\footnotesize $\cdots$};
      \node[node, label=below:{\footnotesize $K$}] (Zk) [right=6mm of dots] {};
      \node[node, label={[below right=5pt and -9pt]:\rotatebox{-30}{\footnotesize $g_1+(n+p)$}}] (A1)  [right=6mm of Zk] {};
      \node[node, label={[below right=5pt and -9pt]:\rotatebox{-30}{\footnotesize $g_2+2(n+p)$}}] (A2) [right=6mm of A1] {};
      \node[node, label={[below right=5pt and -9pt]:\rotatebox{-30}{\footnotesize $g_3+3(n+p)$}}] (A3) [right=6mm of A2] {};
      \node[node, label={[below right=5pt and -9pt]:\rotatebox{-30}{\footnotesize $g_4+4(n+p)$}}] (A4) [right=6mm of A3] {};
      \node[node, label={[below right=5pt and -9pt]:\rotatebox{-30}{\footnotesize $g_5+5(n+p)$}}] (A5) [right=6mm of A4] {};
      \node[node, label={[below right=5pt and -9pt]:\rotatebox{-30}{\footnotesize $g_6+6(n+p)$}}] (N3) [right=6mm of A5] {};
      \node[node, label={[below right=5pt and -9pt]:\rotatebox{-30}{\footnotesize $g_7+4(n+p)$}}] (B4) [right=6mm of N3] {};
      \node[node, label={[below right=5pt and -9pt]:\rotatebox{-30}{\footnotesize $2r+2(n+p)$}}] (B2) [right=6mm of B4] {};
      \node[node, label=right:{\footnotesize $g_8+3(n+p)$}] (Nu) [above=5mm of N3] {};
      \draw (Z1.east) -- (Z2.west);
      \draw (Z2.east) -- (dots.west);
      \draw (dots.east) -- (Zk.west);
      \draw (Zk.east) -- (A1.west);
      \draw (A1.east) -- (A2.west);
      \draw (A2.east) -- (A3.west);
      \draw (A3.east) -- (A4.west);
      \draw (A4.east) -- (A5.west);
      \draw (A5.east) -- (N3.west);
      \draw (N3.east) -- (B4.west);
      \draw (B4.east) -- (B2.west);
      \draw (N3.north) -- (Nu.south); 
    \end{tikzpicture}
  \end{gathered} \,.
    \end{equation}
\end{itemize}
In each case, the $g_i$, $r$, $p$, and $n$ are defined in terms of $N$, $K$ and the integers appearing in equation \eqref{eqn:thems} that define the $E_8$-homomorphism $\rho$; their specific forms are given in \cite{Cabrera:2019izd}, however, it is not illuminating to repeat them here. Instead, we merely note that the infinite-coupling magnetic quiver as derived from the Type IIA brane system is the same as that derived from the 3d mirror of the class $\mathcal{S}$ description of the torus-compactification as given in equation \eqref{eqn:3dmirror}.\footnote{
	For convenience, we summarize how to determine the Lagrangian quiver description of the $T_O(SU(K))$ theories. Let $P$ be a partition of $K$ written in weakly-increasing order and zero-extended to be of length $K$: $P = [n_1, n_2, \cdots, n_K]$. Let $O$ be the nilpotent orbit of $\mathfrak{su}(K)$ associated with the partition $P$. Then, $T_O(SU(K))$ is a linear quiver of $\mathfrak{u}(q_i)$ gauge nodes and a $\mathfrak{u}(K)$ flavor node:
\begin{equation}
     \begin{gathered}
    \begin{tikzpicture}
      \node[node, label=below:{\footnotesize $q_1$}] (Z1)  {};
      \node[node, label=below:{\footnotesize $q_2$}] (Z2) [right=6mm of Z1] {};
      \node[tnode] (dots) [right=6mm of Z2] {\footnotesize $\cdots$};
      \node[node, label=below:{\footnotesize $q_K$}] (Zk) [right=6mm of dots] {};
      \node[fnode, label=below:{\footnotesize $K$}] (Fk) [right=6mm of Zk] {};
      \draw (Z1.east) -- (Z2.west);
      \draw (Z2.east) -- (dots.west);
      \draw (dots.east) -- (Zk.west); 
      \draw (Zk.east) -- (Fk.west);
    \end{tikzpicture}
  \end{gathered} \,, \qquad \begin{gathered} \text{ where } \qquad q_i = \sum_{j < i} n_i \end{gathered} \,.
\end{equation}
When $q_i = 0$, the corresponding gauge node is absent.} Therefore, we can simply use equation \eqref{eqn:3dmirror} for the magnetic quiver. The brane system approach to determining the magnetic quiver will become particularly relevant in the next section, Section \ref{sec:LSTs}, where we derive the magnetic quiver for LSTs from that perspective, bypassing the lack of a class $\mathcal{S}$ description.

Once the Lagrangian quiver description of the magnetic quiver has been determined, we can study the Hasse diagram of the interacting fixed points on the Coulomb branch using the technique of quiver subtraction \cite{Cabrera:2018ann,Gledhill:2021cbe,Bourget:2019aer}. 
The terminology employed in the subtraction algorithm relies on the usual definitions formulated in the context of symplectic singularities \cite{Slodowy_1980}. Thus we provide a brief dictionary between the physical and the mathematical perspectives, given a (possibly reductive) gauge group $G$ together with some matter.
\begin{itemize}
    \item The space of VEVs of the scalars inside of the 3d $\mathcal{N}=4$ vector multiplets associated to $G$ is the \textit{symplectic singularity} we are interested in studying.
    \item A \textit{leaf}, defined with respect to a choice of gauge group $G'$, is the set of VEVs needed to be turned on to break the original gauge group to $G'\subset G$. Hence, specifying a point on a certain leaf is equivalent to referring to the Higgsed theory with gauge group $G'$ together with some matter. 
    \item Running the renormalization group flow corresponds to studying the \textit{foliation} of the singularity in \textit{symplectic leaves} $\mathcal{L}_i$. The partial ordering structure arises by considering the \textit{closure of a leaf} $\overline{\mathcal{L}}$, defined as the set of theories with gauge group $G^*\subseteq G$ such that $G^*$ can be broken to $G'$.
    \item Two \textit{leaves} $\mathcal{L}$ and $\mathcal{L}'$ are adjacent if there is no other leaf $\mathcal{L}''$ such that $\overline{\mathcal{L}}' \subset \overline{\mathcal{L}}'' \subset \overline{\mathcal{L}}$. The set of VEVs needed to be tuned to move from the theory living in a point of $\mathcal{L}'$ to the theory living on a point in $\mathcal{L}$ is called \textit{transverse slice}, and when the leaves are adjacent this slice is \textit{minimal}: its action cannot be broken in the action of two consecutive slices.   
    \item In this language, the Coulomb branch of a theory living on a leaf $\mathcal{L}$ is the transverse slice from that leaf to the top of the Hasse diagram under the partial ordering, i.e., to the fully Higgsed theory.
\end{itemize}
The Higgs branch of a $d \geq 3$ eight-supercharge quiver, which is the space of VEVs of the scalars inside of the hypermultiplets, is also a symplectic singularity and thus has the same description in terms of the foliation/leaf structure.

We summarize the quiver subtraction algorithm, as relevant for the cases that we discuss in this paper, in Algorithm \ref{alg:QS}; we gloss over technical details involving non-trivial decoration subtleties and slice subtraction that are unnecessary for the quivers studied in this paper.

\begin{alg}\label{alg:QS}
\textbf{Quiver Subtraction.} Given a 3d $\mathcal{N}=4$ unitary quiver $\mathcal{Q}$, its Coulomb branch is a symplectic singularity whose foliation can be derived as follows via quiver subtraction.
\begin{enumerate}
    \item The transverse slice between two adjacent leaves is the closure of a minimal nilpotent orbit:\begin{enumerate}
    \item Start looking for subquivers $\mathcal{S}$ of $\mathcal{Q}$ shaped as affine Dynkin diagrams of a Lie algebra $\mathfrak{g}$ and with nodes with rank greater or equal than the affine Dynkin labels of $\mathfrak{g}$.
    \item Subtract the affine Dynkin labels of $\mathfrak{g}$ from the rank of the nodes of $\mathcal{S}$, leaving each link untouched and add an extra $\mathfrak{u}(1)$ node connected in such a way that all the nodes untouched by the subtraction have their balance $b_i= - 2 \, \mathrm{rank}\left( node \right) + \sum \mathrm{rank}\left( adjacent \ nodes \right)$ unchanged.
    \item In this case, the transverse slice is $\overline{ \mathrm{min.} \mathfrak{g} }$.
    \end{enumerate}
    \item The transverse slice between two adjacent leaves is a Klenian singularity of type $A$:\begin{enumerate}
    \item Start looking for subquivers $\mathcal{S}$ of $\mathcal{Q}$ consisting of a $\mathfrak{u}(1)$ gauge node attached to another $\mathfrak{u}(1)$ node with $l\ge 2$ hypermultiplets between them.
    \item Delete $\mathcal{S}$ from $\mathcal{Q}$. Leaving each link untouched, add an extra $\mathfrak{u}(1)$ node connected in such a way that all the nodes untouched by the subtraction have their balance $b_i= - 2 \, \mathrm{rank}\left( node \right) + \sum \mathrm{rank}\left( adjacent \ nodes \right)$ unchanged.
    \item In this case, the transverse slice is $A_{l-1}$.
    \end{enumerate}

    \item The resulting quiver $\mathcal{Q}'$ is associated with the closure of the largest non-trivial leaf contained in the closure of $\mathcal{Q}$,\footnote{The quiver subtraction algorithm engineers a quiver for the symplectic singularity associated with the closure of a certain leaf, this singularity foliates from the leaf all the way down to the unHiggsed theory, i.e., the bottom of the Hasse diagram. The shortcoming of this approach consists in the fact that the Higgs branch of the theory living on the leaf foliates from that leaf to the top of the Hasse diagram, i.e., the fully Higgsed theory. A new subtraction algorithm \cite{DecayAndFissionAlg} partially solves this problem, engineering a quiver that foliates and is thus ordered, in the physical direction of the Higgs branch \cite{DecayAndFissionHiggs}. Hence we will draw Hasse diagrams either using the foliation ordering introduced here or, when stated, in the reverse ordering given by Higgs branch dimension depending on the analysis we want to perform.} and the transverse slice between the two highest leaves is $\overline{ \mathrm{min.} \mathfrak{g} }$ or $A_{l-1}$. \begin{equation*}
        \begin{tikzpicture}
        \node[node, fill=black, draw] at (0,0) (1) {};
        \node[node, fill=black, draw] at (0,-1) (2) {};
        \node[tnode] at (0,-2) (3) {\footnotesize $\vdots$};
        \node[node, fill=black, draw] at (0,-3) (4) {};

        \draw (1)--(2) node[midway,right] () {\footnotesize $\overline{ \mathrm{min.} \mathfrak{g} }$ or $A_{l-1}$};
        \draw (2)--(0,-1.8) (3)--(4);

        \draw (-0.4,0)--(-0.6,0)--(-0.6,-3)--(-0.4,-3);
        \draw (0.4,-1)--(0.6,-1)--(0.6,-3)--(0.4,-3);

        \node[] at (0.9,-2) () {\footnotesize $\mathcal{Q}'$};
        \node[] at (-0.9,-1.5) () {\footnotesize $\mathcal{Q}$};
        
    \end{tikzpicture}
    \end{equation*}
\end{enumerate}
\end{alg}

Notice however that the quiver $\mathcal{Q}'$ obtained by quiver subtraction cannot be associated with a physical electric theory when the quiver $\mathcal{Q}$ is a magnetic quiver. In fact, it is fundamental that the highest point in the theory is included in the foliation in order to have an electric counterpart as it includes the instanton transition that needs to be resummed in the higher dimensional Higgs branch. Another caveat involves the case where multiple subtractions are done on the same subquiver of the quiver \cite{Bourget:2022ehw}.
\begin{alg}\label{alg:Decoration}
	\textbf{Decoration.} The rebalancing $\mathfrak{u}(1)$ node introduced when subtracting a slice has to be considered identical to the subtracted section of the quiver. Hence when multiple subtractions of the same slice are performed on the same set of nodes the new rebalancing nodes are \textit{decorated} all in the same way. Decorated $\mathfrak{u}(1)$ nodes with a link of lacety $l$ can then be merged, allowing for new Higgsing directions according to the rule:
\begin{equation}\begin{matrix}
   \begin{gathered}
       \begin{tikzpicture}
           \node[] at (0,0) (q) {\footnotesize $\mathcal{Q}$};
           \node[node,label=above:{\footnotesize $\mathfrak{u}(1)$}] [above right=5mm of q] (n1) {};
           \node[node,label=above:{\footnotesize $\mathfrak{u}(1)$}] [above left=5mm of q] (n2) {};
           \draw (q)--(n1) node[midway, below right] () {\footnotesize $l$};
           \draw (q)--(n2) node[midway, below left] () {\footnotesize $k$};
            \draw[->] (1,0.5)--(2.5,0.5) node[midway,above] () {\footnotesize$s$};
           \node[] at (3,0) (q1) {\footnotesize $\mathcal{Q}$};
           \node[node,label=above:{\footnotesize $\mathfrak{u}(1)$}] [above =5mm of q1] (n11) {};
           \draw (q1)--(n11) node[midway, right] () {\footnotesize $l+k$};

   \def\radius{0.35cm}
   \begin{pgfonlayer}{background}
    \foreach \coord in {n1,n2,n11} {
      \fill[blue!20] (\coord) circle (\radius);
    }
  \end{pgfonlayer}
           
       \end{tikzpicture}
   \end{gathered} &   \text{ with }  s=\begin{cases}
        A_1 & k=l \\
        m & k\neq l \,.
    \end{cases}
\end{matrix}
\end{equation}
\end{alg}
We need to point out that the identical nature of the decorated nodes automatically identifies every choice of different $\mathfrak{u}(1)$ node chosen when subtracting a slice. Moreover, when multiple subtractions of a slice $s$ are repeated $n$ times, the slice is denoted as $ns$ to identify the discrete $S_n$ action quotienting the introduced $\mathfrak{u}(1)$ rebalancing nodes.

Despite the lack of a general proof, on a case-by-case basis, it is easy to see that the Hasse diagram of the Coulomb branch of the infinite-coupling magnetic quiver, obtained via quiver subtraction, and the Hasse diagram produced by complex-structure deformations of the singular non-compact Calabi--Yau threefold that engineers the SCFT in F-theory, are identical.\footnote{Similarly, the finite-coupling and semi-infinite-coupling magnetic quiver produce Hasse diagrams which are related to the complex-structure deformations of the relevant desingularizations of the Calabi--Yau threefold engineering the SCFT.} In general, the quiver subtraction algorithm produces a Hasse diagram where the edges are labelled by a symplectic singularity, however this labelling is (naively) absent from the Hasse diagram obtained geometrically from the complex-structure deformations.\footnote{See \cite{DKL} for an algorithm, in certain cases, to extract the symplectic singularity labelling the edges under Higgs branch RG flow from the tensor branch configuration.}

\subsection{Example: \texorpdfstring{$\mathcal{O}_{1, 2}(\rho, \sigma)$}{O\_\{1,2\}(rho, sigma)}}

To emphasize the procedure for constructing the Higgs branch Hasse diagram, we
consider the rank one $(\mathfrak{e}_8, \mathfrak{su}(2))$ orbi-instanton as an
example. In this case, we can study the structure of the Higgs branch of the 6d
$(1,0)$ SCFT $\mathcal{O}_{1,2}$ directly from the geometric description via
F-theory. The tensor branch geometry corresponding to the rank one
$(\mathfrak{e}_8, \mathfrak{su}(2))$ orbi-instanton is:
\begin{equation}
    1\,2\,\overset{\mathfrak{su}_2}{2} \,.
\end{equation}
We can then consider the singular Calabi--Yau threefolds obtained by shrinking
the three curves to zero volume, performing a complex-structure deformation of
that geometry, and then desingularize to obtain the tensor branch descriptions
of the resulting SCFTs after Higgsing. The Hasse diagram of the Higgs branch of
$\mathcal{O}_{1,2}$ obtained in such a way is depicted in Figure
\ref{fig:Hasse_OI}; in the figure, we denote each vertex, corresponding to an
interacting SCFT fixed point, by the tensor branch description of the SCFT, but
we emphasize that this is the Hasse diagram of the Higgs branch of the 6d SCFT,
not of the Higgs branch of the effective field theory on the tensor
branch.\footnote{Similar methods can be used to determine the Hasse diagram of
the Higgs branch of the effective field theory on the tensor branch.} 

\begin{figure}[t]
    \centering
    \begin{tikzpicture}
    \node[tnode] (null) [] {\footnotesize $\varnothing$};
    \node[tnode] (1) [below=0.8cm of null] {\footnotesize $1$};
    \node[tnode,draw=red,fill=red!30] (11) [below right=2cm and 4cm of 1] {\footnotesize $1 \sqcup 1$};
    \node[tnode] (1su) [below=0.8cm of 1] {\footnotesize $\stackon{$1$}{$\mathfrak{su}_2$}$};
    \node[tnode] (12) [below left=0.2cm and 1.6cm of 11] {\footnotesize $1 \ 2$};
    \node[tnode,draw=red,fill=red!30] (111) [below right=2cm and 4cm of 11] {\footnotesize $1 \sqcup 1 \sqcup 1$};
    \node[tnode] (12su) [below=1.6cm of 1su] {\footnotesize $1\stackon{$2$}{$\mathfrak{su}_2$}$};
    \node[tnode,draw=red,fill=red!30] (121) [below right=1.1cm and 6cm of 12su] {\footnotesize $1 \ 2 \sqcup 1$};
    \node[tnode] (122) [below right=1.9cm and 2.8cm of 12su] {\footnotesize $1 \ 2 \ 2$};
    \node[tnode] (122su) [below=2.5cm of 12su] {\footnotesize $1 \ 2\stackon{$2$}{$\mathfrak{su}_2$}$};

    \draw (null)--(1) node[midway,left] () {\footnotesize $\mathfrak{e}_8$};
    \draw (1)--(1su) node[midway,left] () {\footnotesize $d_{10}$};
    \draw (1su)--(12su) node[midway,left] () {\footnotesize $\mathfrak{e}_7$};
    \draw (12su)--(122su) node[midway,left] () {\footnotesize $\mathfrak{e}_8$};
    \draw (1)--(11) node[midway,above right] () {\footnotesize $\mathfrak{e}_8$};
    \draw (11)--(111) node[midway,above right] () {\footnotesize $\mathfrak{e}_8$};
    \draw (12)--(11) node[midway,above left] () {\footnotesize $A_1$};
    \draw (12su)--(12) node[midway,above left] () {\footnotesize $b_3$};
    \draw (121)--(111) node[midway,above left] () {\footnotesize $A_1$};
    \draw (122)--(121) node[midway,above left] () {\footnotesize $m$};
    \draw (122su)--(122) node[midway,above left] () {\footnotesize $g_2$};
    \draw (121)--(12) node[midway,above right] () {\footnotesize $\mathfrak{e}_8$};
    \end{tikzpicture}
	\caption{The Higgs branch Hasse diagram of $\mathcal{O}_{1,2}$. Vertices highlighted in red indicate that the theory at that point of the Higgs branch is a product of multiple interacting SCFTs.}
    \label{fig:Hasse_OI}
\end{figure}

\begin{figure}[p]
    \centering
    \includegraphics[width=\textwidth,page=1]{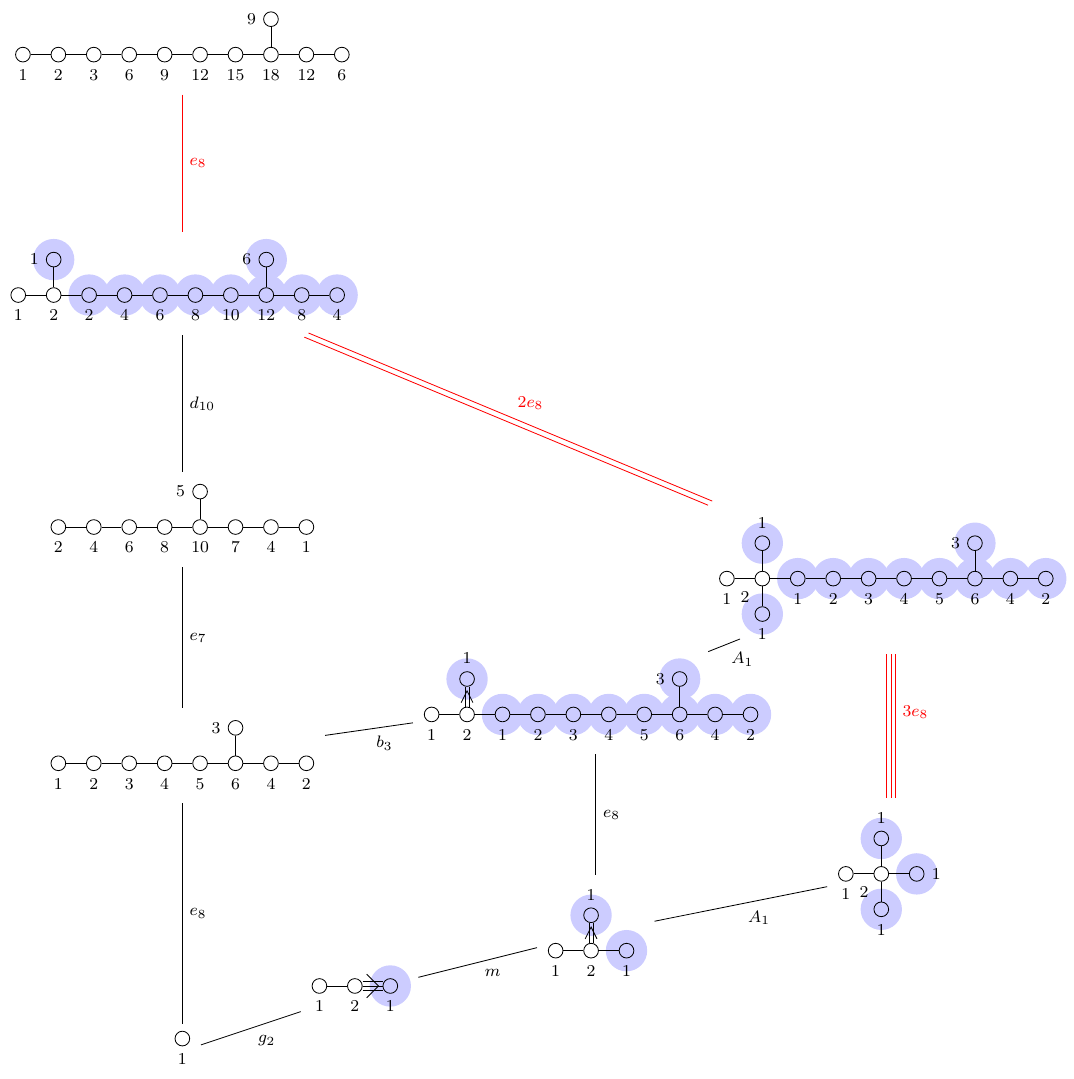}
	\caption{The Hasse diagram for the Coulomb branch of the magnetic quiver in equation \eqref{eqn:O12MQ} obtained via quiver subtraction. This is the same structure as the Hasse diagram of the Higgs branch of $\mathcal{O}_{1,2}$ shown in Figure \ref{fig:Hasse_OI}. The blue shading is used to indicate the decorated section of the quiver.}
    \label{fig:Hasse_O.IviaQuiverSubtraction}
\end{figure}

\begin{figure}[p]
    \centering
    \includegraphics[width=0.99\textwidth,page=2]{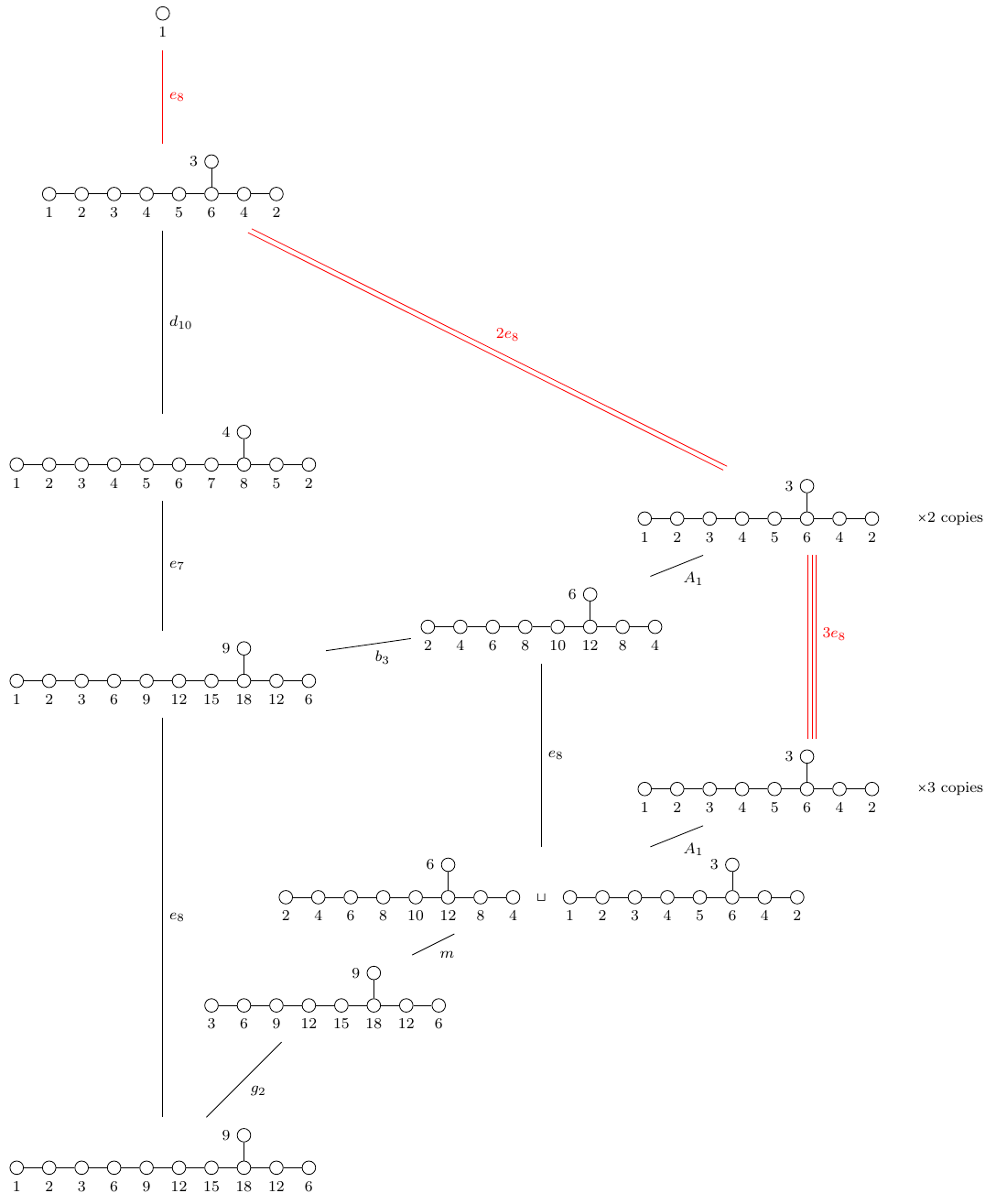}
	\caption{The Hasse diagram of $\mathcal{O}_{1,2}$. At each vertex, we have reverse-engineered the magnetic quiver associated with the symplectic leaf.}
    \label{fig:Hasse_O.IviaReverseQuiverSubtraction}
\end{figure} 

The magnetic quiver describing the Higgs branch of this 6d SCFT (which we have called the infinite-coupling magnetic quiver) is easily obtained from equation \eqref{eqn:3dmirror} using that the only non-trivial Kac label describing the $E_8$-homomorphism is $m_1 = 3$. The 3d $\mathcal{N}=4$ magnetic quiver is then  
\begin{equation}\label{eqn:O12MQ}
\begin{gathered}
    \begin{tikzpicture}
 	\tikzset{node distance = 0.6cm};
    \node (t1) [node, label=below:{\footnotesize{$1$}}] {};
    \node (t2) [node, right of=t1,label=below:{\footnotesize{$2$}}] {};  
	\node (g1) [node,right of=t2,label=below:{\footnotesize{$3$}}] {};
	\node (g2) [node, right of=g1,label=below:{\footnotesize{$6$}}] {};
	\node (g3) [node, right of=g2,label=below:{\footnotesize{$9$}}] {};
	\node (g4) [node, right of=g3,label=below:{\footnotesize{$12$}}] {};
	\node (g5) [node, right of=g4,label=below:{\footnotesize{$15$}}] {};
	\node (g6) [node, right of=g5,label=below:{\footnotesize{$18$}}] {};
	\node (g4') [node, right of=g6,label=below:{\footnotesize{$12$}}] {};
	\node (g2') [node, right of=g4',label=below:{\footnotesize{$6$}}] {};
    \node (g3') [node, above of=g6,label=left:{\footnotesize{$9$}}] {};

	\draw (t1)--(t2) (t2)--(g1) (g1)--(g2) (g2)--(g3) (g3)--(g4) (g4)--(g5) (g5)--(g6) 
(g6)--(g4') (g4')--(g2') (g6)--(g3');
	\end{tikzpicture}
\end{gathered}
      \,.
\end{equation}
Quiver subtraction yields the Hasse diagram of the interacting fixed point of the Coulomb branch of this quiver. We depict the result in Figure \ref{fig:Hasse_O.IviaQuiverSubtraction}. 

First, we note that, as a pair of directed acyclic graphs, the Hasse diagrams in Figures \ref{fig:Hasse_OI} and \ref{fig:Hasse_O.IviaQuiverSubtraction} have the same structure. Furthermore, it is straightforward to read off the dimension of the Coulomb branch and the (Coulomb) flavor symmetry algebra of each interacting fixed point of the Hasse diagram in Figure \ref{fig:Hasse_O.IviaQuiverSubtraction}. By comparing the Coulomb branch dimension and Coulomb symmetries to the 6d Higgs branch dimension and flavor algebras, we can relate the 3d $\mathcal{N}=4$ quivers in Figure \ref{fig:Hasse_O.IviaQuiverSubtraction} vertex-by-vertex to the 6d $(1,0)$ SCFT in Figure \ref{fig:Hasse_OI}. In this way, we can write the known magnetic quivers of the 6d SCFTs at each vertex themselves in the Hasse diagram, which we depict in Figure \ref{fig:Hasse_O.IviaReverseQuiverSubtraction}; this process thus reverse engineers the 3d $\mathcal{N}=4$ Coulomb branch Hasse diagram in Figure \ref{fig:Hasse_O.IviaReverseQuiverSubtraction} utilizing the 6d perspective. 

In fact, an alternative, modified version of the quiver
subtraction algorithm has been proposed \cite{DecayAndFissionAlg} which directly engineers the transverse slices from each leaf to the top, i.e., the physical Higgs branches, starting from the magnetic quiver of the smallest leaf. This method recovers the magnetic quivers for the Higgs branches \cite{DecayAndFissionHiggs} of the interacting fixed points in the Hasse diagram in Figure
\ref{fig:Hasse_O.IviaReverseQuiverSubtraction} directly, without reverse engineering via
the 6d SCFT perspective. 

\subsection{Example: \texorpdfstring{$N$}{N} \texorpdfstring{$E_8$}{E8} Instantons}\label{sec:sym}

A particularly interesting example is the worldvolume theory that lives on a stack of $N$ M5-branes contained inside of an M9-brane. This can be taken as a limiting case of the orbi-instanton theories, where the orbifold singularity is trivial. These theories are also known as the rank $N$ E-string theories, and have the following curve configuration:
\begin{equation}
    1 \underbrace{\ 2 \ 2 \ \cdots \ 2}_{N-1 \text{ curves}} \,.
\end{equation}
From their 't Hooft anomalies \cite{Ohmori:2014pca}, we can see that they have the following Higgs branch dimensions:
\begin{equation}
    d_H = 30N - 1 \,.
\end{equation}
The infinite coupling magnetic quiver for these theories can be derived from the Type IIA brane system using the aforementioned rules and the result is an affine $E_8$ quiver with its entries multiplied by $N$:
\begin{equation} \label{eqn:Magnetic_undeco_C^2}
    \left(E_{8,N} \right) := \begin{aligned}
        \begin{tikzpicture}
      \node[node, label=below :{\footnotesize $N$}] (A1)  [] {};
      \node[node, label=below :{\footnotesize $2N$}] (A2) [right=6mm of A1] {};
      \node[node, label=below :{\footnotesize $3N$}] (A3) [right=6mm of A2] {};
      \node[node, label=below :{\footnotesize $4N$}] (A4) [right=6mm of A3] {};
      \node[node, label=below :{\footnotesize $5N$}] (A5) [right=6mm of A4] {};
      \node[node, label=below :{\footnotesize $6N$}] (N3) [right=6mm of A5] {};
      \node[node, label=below :{\footnotesize $4N$}] (B4) [right=6mm of N3] {};
      \node[node, label=below :{\footnotesize $2N$}] (B2) [right=6mm of B4] {};
      \node[node, label=right:{\footnotesize $3N$}] (Nu) [above=5mm of N3] {};

      \draw (A1.east) -- (A2.west);
      \draw (A2.east) -- (A3.west);
      \draw (A3.east) -- (A4.west);
      \draw (A4.east) -- (A5.west);
      \draw (A5.east) -- (N3.west);
      \draw (N3.east) -- (B4.west);
      \draw (B4.east) -- (B2.west);
      \draw (N3.north) -- (Nu.south); 
    \end{tikzpicture}
    \end{aligned} \,.
\end{equation}
Here, we have introduced the notation $\left(E_{8,N} \right)$ to refer to such magnetic quivers compactly.

\begin{figure}[t]
    \centering
    \begin{tikzpicture}
    \node[tnode] (null) [] {\footnotesize $\varnothing$ };
    \node[tnode] (1) [below=1cm of null] {\footnotesize $1$ };
    \node[tnode] (11) [below=1cm of 1] {\footnotesize $1 \sqcup 1 $ };
    \node[tnode] (12) [below right=0.5cm and 5cm of 11] {\footnotesize $1 \ 2$ };
    \node[tnode] (111) [below=1cm of 11] {\footnotesize $1 \sqcup 1 \sqcup 1 $};
    \node[tnode] (121) [below right=0.5cm and 1.25cm of 111] {\footnotesize $1 \ 2 \sqcup 1$ };
    \node[tnode] (122) [below right=1.7cm and 2.2cm of 121] {\footnotesize $1 \ 2 \ 2$ };
    \node[tnode] (vdots) [below=1cm of 111] {\footnotesize $\vdots$};
    \node[tnode] (vdots1) [below right=0.7cm and 2.5cm of vdots] {\footnotesize $\vdots$};
    \node[tnode] (vdots2) [below right=0.3cm and 3cm of vdots1] {\footnotesize $\vdots$};
    \node[tnode] (1N1) [below=1cm of vdots] {\footnotesize $\underbrace{1 \sqcup \cdots \sqcup 1 }_{N-1 \text{ curves}}$ };
    \node[tnode] (SymN1) [below right=0.3cm and 2cm of 1N1] {\footnotesize $\ddots$};
    \node[tnode] (1N2) [below right=0.3cm and 2cm of SymN1] {\footnotesize $\underbrace{1 \ 2 \cdots 2}_{N-1 \text{ curves}}$};
    \node[tnode] (1N11) [below=1cm of 1N1] {\footnotesize $\underbrace{1 \sqcup \cdots \sqcup 1 }_{N \text{ curves}}$ };
    \node[tnode] (SymN2) [below right=0.3cm and 2.5cm of 1N11] {\footnotesize $\ddots$};
    \node[tnode] (1N22) [below right=0.3cm and 2.5cm of SymN2] {\footnotesize $\underbrace{1 \ 2 \cdots 2}_{N \text{ curves}}$ };

    \draw (null)--(1) node[midway,left] () {};
    \draw (1)--(11) node[midway,left] () {};
    \draw (11)--(111) node[midway,left] () {};
    \draw (111)--(vdots)--(1N1);
    \draw (1N1)--(1N11) node[midway,left] () {};
    \draw (11)--(12) node[midway,above right] () {};
    \draw (12)--(121) node[midway,left] () {};
    \draw (111)--(121) node[midway,above right] () {};
    \draw (121)--(122) node[midway,above right] () {};
    \draw (121)--(vdots1)--(SymN1);
    \draw (vdots1)--(122);
    \draw (1N1)--(SymN1)--(1N2);
    \draw (SymN1)--(vdots2);
    \draw (1N11)--(SymN2)--(1N22);
    \draw (SymN2)--(SymN1) node[midway,left] () {};
    \draw (SymN2)--(1N2) node[midway,above left] () {};
    \end{tikzpicture}
    \caption{The Higgs branch Hasse diagram of the rank $N$ E-string SCFT derived from the M-theory picture. On each vertex in the Hasse diagram, we have written the tensor branch curve configuration for the SCFT at that vertex.}
    \label{fig:Hasse_OI_undercorated}
\end{figure}
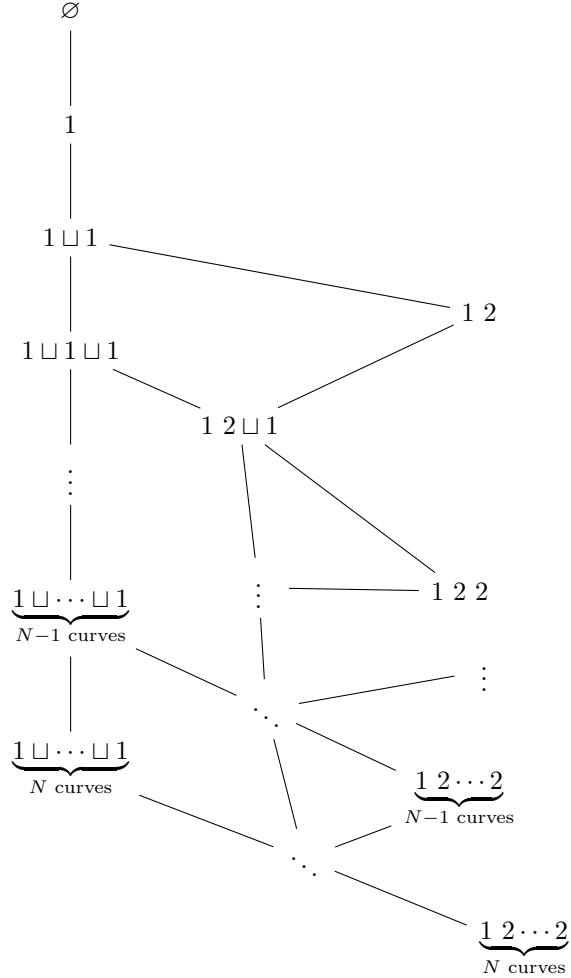

We would now like to study the Higgs branch Hasse diagram for the rank $N$ E-string theory, taking advantage of the quiver subtraction algorithm as applied to the magnetic quiver. In this case, the Hasse diagram of the Higgs branch can also be determined directly from the M5-brane perspective. There are two operations of interest: one can perform the small instanton transition and dissolve one M5-brane, or else one can separate the stack of $N$ M5-brane into two separate stacks of $N = N' + N''$, where both stacks are still contained inside of the M9-brane. The former typically leads to interacting SCFTs with a single energy-momentum tensor, whereas the latter transitions create product theories. In summary, the transitions are:
\begin{itemize}
    \item pick one stack of $N_t$ M5-branes and replace it with a stack of $N_t - 1$ M5-branes.
    \item pick one stack of $N_t$ M5-branes and replace with two separate stacks of $N'$ and $N''$ M5-branes, such that $N_t = N' + N''$.
\end{itemize}
The transitive reduction of combinations of such transitions provides the Higgs branch of the rank $N$ E-string; we have schematically depicted this structure in Figure \ref{fig:Hasse_OI_undercorated}.

A naive application of quiver subtraction applied to the magnetic quiver in equation \eqref{eqn:Magnetic_undeco_C^2}, following the rules that we have delineated in Algorithm \ref{alg:QS}, does not reproduce the Higgsing pattern in Figure \ref{fig:Hasse_OI_undercorated}. In particular, we can simply subtract the affine $E_8$ Dynkin diagram from equation \eqref{eqn:Magnetic_undeco_C^2} repeatedly, leading to a linear Hasse diagram. This is a subtlety specific to magnetic quivers which are (multiples of) affine Dynkin diagrams, as has been noted in \cite{Cremonesi:2014xha}.

Conveniently, we can formally consider the rank $N$ E-string as the worldvolume theory on a stack of $N$ M5-branes, inside an M9-wall, and probing $\mathbb{C}^2 / \mathbb{Z}_1$. Geometrically $\mathbb{C}^2/\mathbb{Z}_1$ is is isomorphic to $\mathbb{C}^2$, but it leads to a slightly different Type IIA brane description. From this brane description, we can extract the following magnetic quiver for the rank $N$ E-string:
\begin{equation}\label{eqn:Magnetic_undeco_C^2/Z_1}
    \begin{gathered}
         \begin{tikzpicture}
      \node[node, label=below :{\footnotesize $1$}] (A0)  [] {};   
      \node[node, label=below :{\footnotesize $N$}] (A1)  [right=6mm of A0] {};
      \node[node, label=below :{\footnotesize $2N$}] (A2) [right=6mm of A1] {};
      \node[node, label=below :{\footnotesize $3N$}] (A3) [right=6mm of A2] {};
      \node[node, label=below :{\footnotesize $4N$}] (A4) [right=6mm of A3] {};
      \node[node, label=below :{\footnotesize $5N$}] (A5) [right=6mm of A4] {};
      \node[node, label=below :{\footnotesize $6N$}] (N3) [right=6mm of A5] {};
      \node[node, label=below :{\footnotesize $4N$}] (B4) [right=6mm of N3] {};
      \node[node, label=below :{\footnotesize $2N$}] (B2) [right=6mm of B4] {};
      \node[node, label=right:{\footnotesize $3N$}] (Nu) [above=5mm of N3] {};

      \draw (A0.east) -- (A1.west);
      \draw (A1.east) -- (A2.west);
      \draw (A2.east) -- (A3.west);
      \draw (A3.east) -- (A4.west);
      \draw (A4.east) -- (A5.west);
      \draw (A5.east) -- (N3.west);
      \draw (N3.east) -- (B4.west);
      \draw (B4.east) -- (B2.west);
      \draw (N3.north) -- (Nu.south); 
    \end{tikzpicture}
    \end{gathered} \,.
\end{equation}
Extra care needs to be taken with respect to the additional, leftmost $\mathfrak{u}(1)$ gauge node. When $N = 1$, this $\mathfrak{u}(1)$ is ugly in the sense of \cite{Gaiotto:2008ak}; in fact, in that case the quiver in equation \eqref{eqn:Magnetic_undeco_C^2/Z_1} is the magnetic quiver for the rank one E-string coupled to a free hypermultiplet. For general $N$, the dimension of the magnetic quiver in equation \eqref{eqn:Magnetic_undeco_C^2/Z_1} is larger by one than the dimension of the Higgs branch of the rank $N$ E-string; this fictitious extra $\mathfrak{u}(1)$ direction should be discarded when using this magnetic quiver to study the structure of the Coulomb branch of the magnetic quiver in equation \eqref{eqn:Magnetic_undeco_C^2}.

The physical meaning of the extra $\mathfrak{u}(1)$ node in equation \eqref{eqn:Magnetic_undeco_C^2/Z_1} has been explained in \cite{Cremonesi:2014xha}. When constructing a magnetic quiver whose Coulomb branch engineers the moduli space $M_{n,G}$ of $n$ instantons for the gauge group $G$ one encounters overextended Dynkin diagram quivers. The extra $\mathfrak{u}(1)$ node is associated with the $\mathfrak{su}(2)$ global symmetry acting on $\mathbb{C}^2$, which naturally mixes with the parameterization of the instanton solution in $G$, and therefore it must be factored out in any Hilbert series computation.

\begin{figure}[p]
    \centering
    \includegraphics[page=3, width=0.9\textwidth]{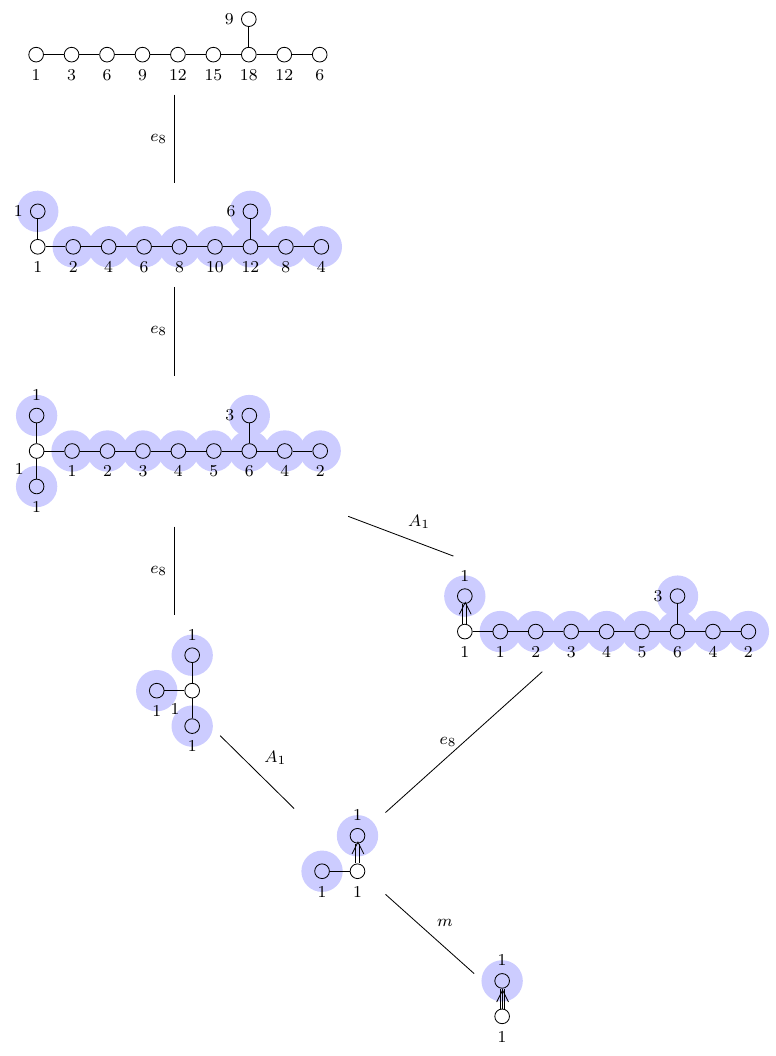}
	\caption{The Hasse diagram for the rank three E-string obtained from applying the quiver subtraction algorithm to the magnetic quiver given in equation \eqref{eqn:Magnetic_undeco_C^2/Z_1}.}
    \label{fig:rank3E-stringHasse}
\end{figure} 

Applying the quiver subtraction algorithm to the quiver in equation
\eqref{eqn:Magnetic_undeco_C^2/Z_1}, we obtain a Hasse diagram which recovers the Hasse diagram drawn from the 6d perspective in Figure
\ref{fig:Hasse_OI_undercorated}. For $N=3$, we have depicted the quiver subtraction process in Figure
\ref{fig:rank3E-stringHasse}. Thus, we observe, from both the quiver subtraction and 6d SCFT perspectives, that each interacting fixed point on the Higgs branch of the rank $N$ E-string is associated with a partition of $1 \leq n \leq N$. We write partitions in multiplicative form
\begin{equation}
    [1^{m_1}, 2^{m_2}, \cdots, N^{m_N}] \qquad \text{ where } \qquad \sum_{i=1}^N i m_i = n \,.
\end{equation}
The 6d SCFT at each point is a union of lower-rank E-string theories. Let $P$ be an arbitrary partition, then the associated 6d SCFT is
\begin{equation}
    P \qquad \longleftrightarrow \qquad \bigsqcup_{i=1}^N \, \bigsqcup_{j = 1}^{m_i} \, 1 \underbrace{\,2 \cdots 2\,}_{i - 1} \,.
\end{equation}
The partial ordering on such partitions is defined as follows. Let $P$ be a partition of $1 \leq n \leq N$ which we write in weakly-decreasing order, zero-extended to be of length $N$:
\begin{equation}
    P = [p_1, p_2, \cdots, p_N] \,.
\end{equation}
Then $P \geq P'$ if there exist partitions $P_{i}$ of some $p \leq p_i$, for each element $p_i$ in $P$, such that 
\begin{equation}
    \bigsqcup_{i=1}^N P_i = P' \,.
\end{equation}

When constructing the Hasse diagram of such partitions under the partial ordering that we have just defined, there are two kinds of elementary transitions. These correspond to either splitting a single stack of M5-branes into two stacks or performing a small instanton transition and dissolving a single M5-brane inside of the M9-brane.
In the former case, the elementary transition involves picking an element $i$ inside of the partition $P$ and connecting it to a partition $P'$ obtained from $P$ by replacing $i$ with two elements, $j$ and $i-j$. The transverse slice is
\begin{equation}
    \begin{cases}
        A_1 &\quad \text{ if } j = \tfrac{i}{2} \,, \\
        m & \quad \text{ otherwise. }
    \end{cases}
\end{equation}
For the small instanton elementary transition, the edge in the Hasse diagram is between partitions
\begin{equation}
    [1^{m_1 \geq 1}, \cdots] \quad \rightarrow \quad [1^{m_1-1}, \cdots] \,,
\end{equation}
and the associated transverse slice is
\begin{equation}
    m_1 \, \mathfrak{e}_8 \,.
\end{equation}

\begin{figure}[p]
    \centering
    \resizebox{\textwidth}{!}{
    \begin{tikzpicture}
    \draw[dashed,rotate=-12,cyan,fill=cyan!5](8,-13.9) ellipse (8cm and 1.3cm);
    \draw[dashed,rotate=-11.5,orange,fill=orange!5](6.7,-11.2) ellipse (8cm and 1.2cm);
    \draw[dashed,rotate=-11.5,magenta,fill=magenta!5](6.7,-6.6) ellipse (7.5cm and 1cm);
    \draw[dashed,rotate=-11.5,red,fill=red!5](3.2,-4.3) ellipse (4.5cm and 0.7cm);
    \node[tnode] (null) [] {\footnotesize $\varnothing$ , $\mathrm{MQ}:\left(E_{8,0} \right)$};
    \node[tnode] (1) [below=1.5cm of null] {\footnotesize $1$ , $\mathrm{MQ}:\left(E_{8,1} \right)$};
    \node[tnode] (11) [below=1.5cm of 1] {\footnotesize $1 \sqcup 1 $ , $\mathrm{MQ}:\left(E_{8,1} \right)^2$};
    \node[tnode] (label11) [left=1cm of 11] {\color{red}\footnotesize $\mathrm{Sym}^2\left(\mathbb{C}^2\right)$};
    \node[tnode] (12) [below right=0.5cm and 1.5cm of 11] {\footnotesize $1 \ 2$ , $\mathrm{MQ}:\left(E_{8,2} \right)$};
    \node[tnode] (111) [below=1.5cm of 11] {\footnotesize $1 \sqcup 1 \sqcup 1 $ , $\mathrm{MQ}:\left(E_{8,1} \right)^3$};
    \node[tnode] (label111) [left=1cm of 111] {\color{magenta}\footnotesize $\mathrm{Sym}^3\left(\mathbb{C}^2\right)$};
    \node[tnode] (121) [below right=0.5cm and 1.25cm of 111] {\footnotesize $1 \ 2 \sqcup 1$ , $\mathrm{MQ}:\left(E_{8,2} \right) \cdot \left(E_{8,1} \right)$};
    \node[tnode] (122) [below right=0.5cm and 1.5cm of 121] {\footnotesize $1 \ 2 \ 2$ , $\mathrm{MQ}:\left(E_{8,3} \right)$};
    \node[tnode] (vdots) [below=1.5cm of 111] {\footnotesize $\vdots$};
    \node[tnode] (vdots1) [below right=0.3cm and 5cm of vdots] {\footnotesize $\vdots$};
    \node[tnode] (vdots2) [below right=0.3cm and 5cm of vdots1] {\footnotesize $\vdots$};
    \node[tnode] (1N1) [below=1.5cm of vdots] {\footnotesize $\underbrace{1 \sqcup \cdots \sqcup 1 }_{N-1 \text{ curves}}$ , $\mathrm{MQ}:\left(E_{8,1} \right)^{N-1}$};
        \node[tnode] (label1N1) [left=1cm of 1N1] {\color{orange}\footnotesize $\mathrm{Sym}^{N-1}\left(\mathbb{C}^2\right)$};
    \node[tnode] (SymN1) [below right=0.3cm and 2cm of 1N1] {\footnotesize $\ddots$};
    \node[tnode] (1N2) [below right=0.3cm and 2cm of SymN1] {\footnotesize $\underbrace{1 \ 2 \cdots 2}_{N-1 \text{ curves}}$ , $\mathrm{MQ}:\left(E_{8,N-1} \right)$};
    \node[tnode] (1N11) [below=1.5cm of 1N1] {\footnotesize $\underbrace{1 \sqcup \cdots \sqcup 1 }_{N \text{ curves}}$ , $\mathrm{MQ}:\left(E_{8,1} \right)^N$};
            \node[tnode] (label1N11) [left=1cm of 1N11] {\color{cyan}\footnotesize $\mathrm{Sym}^{N}\left(\mathbb{C}^2\right)$};
    \node[tnode] (SymN2) [below right=0.3cm and 2.5cm of 1N11] {\footnotesize $\ddots$};
    \node[tnode] (1N22) [below right=0.3cm and 2.5cm of SymN2] {\footnotesize $\underbrace{1 \ 2 \cdots 2}_{N \text{ curves}}$ , $\mathrm{MQ}:\left(E_{8,N} \right)$};

    \draw (null)--(1) node[midway,left] () {\footnotesize $\mathfrak{e}_8$};
    \draw (1)--(11) node[midway,left] () {\footnotesize $2\mathfrak{e}_8$};
    \draw (11)--(111) node[midway,left] () {\footnotesize $3\mathfrak{e}_8$};
    \draw (111)--(vdots)--(1N1);
    \draw (1N1)--(1N11) node[midway,left] () {\footnotesize $N\mathfrak{e}_8$};
    \draw (11)--(12) node[midway,above right] () {\footnotesize $A_1$};
    \draw (12)--(121) node[midway,left] () {\footnotesize $\mathfrak{e}_8$};
    \draw (111)--(121) node[midway,above right] () {\footnotesize $A_1$};
    \draw (121)--(122) node[midway,above right] () {\footnotesize $m$};
    \draw (121)--(vdots1)--(SymN1);
    \draw (vdots1)--(122);
    \draw (1N1)--(SymN1)--(1N2);
    \draw (SymN1)--(vdots2);
    \draw (1N11)--(SymN2)--(1N22);
    \draw (SymN2)--(SymN1) node[midway,left] () {};
    \draw (SymN2)--(1N2) node[midway,above left] () {\footnotesize $\mathfrak{e}_8$};
    \end{tikzpicture}
    }
    \caption{The Higgs branch Hasse diagram for the rank $N$ E-string theory. The blocks labelled by $\mathrm{Sym}^{n}\left(\mathbb{C}^2 \right)$ denote that there is a subdiagram of the Hasse diagram which is the Hasse diagram of the foliation of $\mathrm{Sym}^{n}\left(\mathbb{C}^2 \right)$. In addition to the curve configuration for the 6d $(1,0)$ SCFTs, we also write the magnetic quiver for the theory at that vertex. The transverse slices are obtained from the quiver subtraction algorithm.}
    \label{fig:Hasse_OI_undercorated_MQ}
\end{figure}
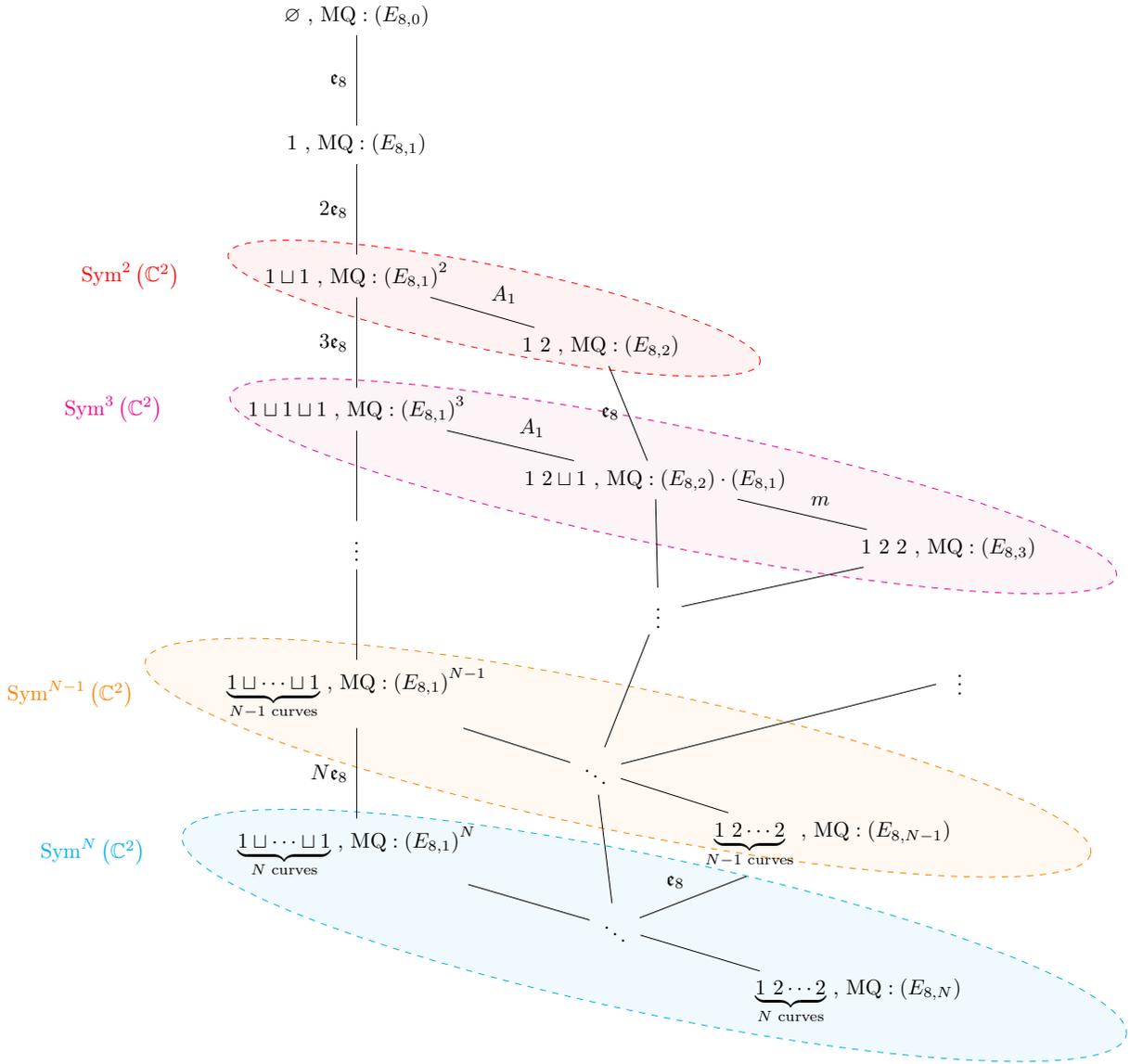

For fixed $n$, the subset of the Hasse diagram is identical to that obtained from the foliation of $\operatorname{Sym}^n(\mathbb{C}^2)$, as studied in \cite{Bourget:2022ehw}. This is unsurprising, as $\operatorname{Sym}^n(\mathbb{C}^2)$ describes the Higgs branch of a stack of $n$ M5-branes and captures how a single stack can be split into multiple stacks. In our construction of the rank $N$ E-string, we begin with a single stack of $N$ M5-branes, and these can be split into distinct stacks while remaining inside of the M9-brane; such splitting is blind to the presence of the M9-brane. The M9-brane allows us to also dissolve M5-branes; this creates transitions between $\operatorname{Sym}^n(\mathbb{C}^2)$ and $\operatorname{Sym}^{m < n}(\mathbb{C}^2)$. In Figure \ref{fig:Hasse_OI_undercorated_MQ}, we draw an enhanced version of the Hasse diagram of the rank $N$ E-string which appeared in Figure \ref{fig:Hasse_OI_undercorated}, where we have included the magnetic quivers, the transverse slices associated to the elementary transitions, and we have indicated which 6d SCFTs belong to each $\operatorname{Sym}^n(\mathbb{C}^2)$ subdiagram of the Hasse diagram of the Higgs branch.

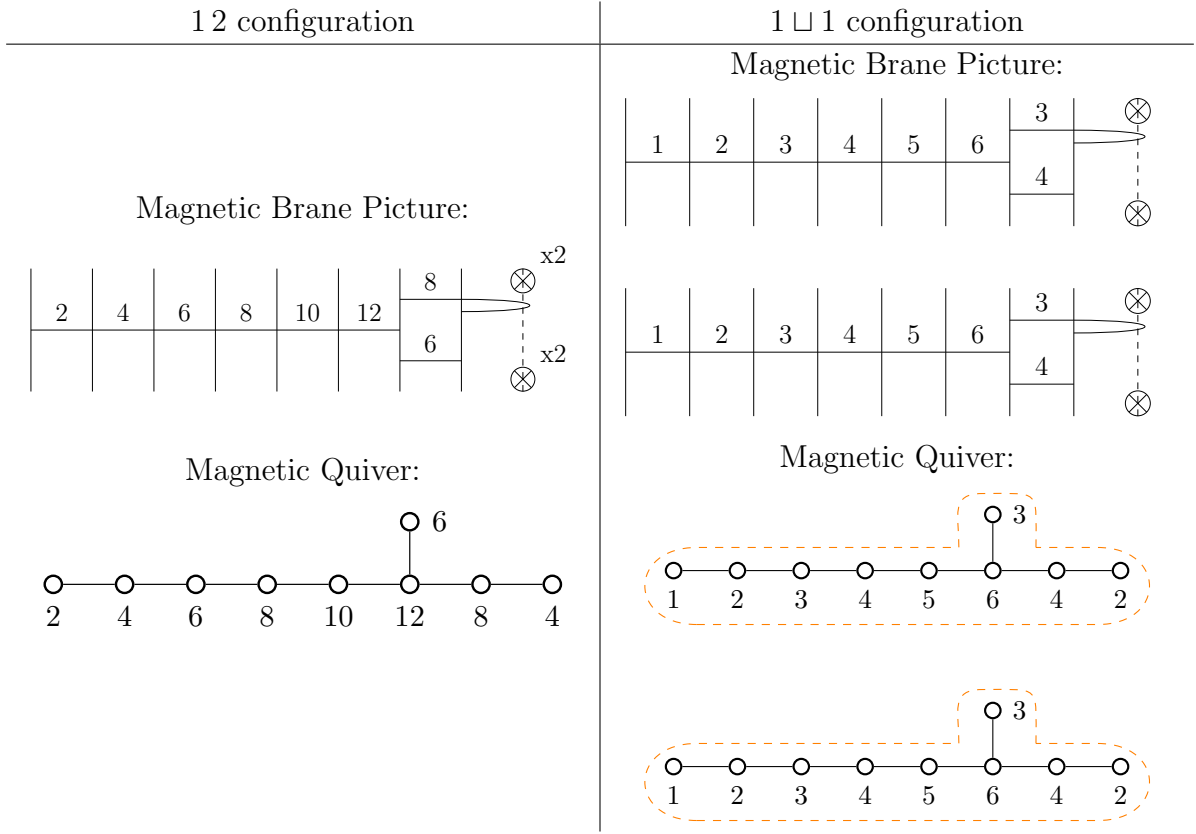
\begin{figure}[t]

    \centering
    \begin{tabular}{c|c}
    $1\,2 $ configuration     & $1 \sqcup 1 $ configuration \\ \hline
      \centering
      $
      \begin{gathered}
    \text{Magnetic Brane Picture:} \\
    \vspace{0.5cm}
    \resizebox{0.45\textwidth}{!}{
    \begin{tikzpicture}[cross/.style={path picture={ 
  \draw[black]
(path picture bounding box.south east) -- (path picture bounding box.north west) (path picture bounding box.south west) -- (path picture bounding box.north east);
}}]
        \draw (0,-1)--(0,1);
        \draw (1,-1)--(1,1);
        \draw (2,-1)--(2,1);
        \draw (3,-1)--(3,1);
        \draw (4,-1)--(4,1);
        \draw (5,-1)--(5,1);
        \draw (6,-1)--(6,1);
        \draw (7,-1)--(7,1);
        \draw[dashed] (8,-1)--(8,1);

        \draw (0,0)--(1,0) node[midway,above] () {$2$};
        \draw (1,0)--(2,0) node[midway,above] () {$4$};
        \draw (2,0)--(3,0) node[midway,above] () {$6$};
        \draw (3,0)--(4,0) node[midway,above] () {$8$};
        \draw (4,0)--(5,0) node[midway,above] () {$10$};
        \draw (5,0)--(6,0) node[midway,above] () {$12$};
        \draw (6,0.5)--(7,0.5) node[midway,above] () {$8$};
        \draw (6,-0.5)--(7,-0.5) node[midway,above] () {$6$};
        \draw (7,0.5) to[out=0, in=0, looseness=19] (7,0.3) node[] () {};

        \node[draw,circle,cross,label=above right:{$\mathrm{x}2$}] () at (8,0.8) {};
        \node[draw,circle,cross,label=above right:{$\mathrm{x}2$}] () at (8,-0.8) {};
    \end{tikzpicture}
    }\\
        \text{Magnetic Quiver:} \\
    \vspace{0.5cm}
    \resizebox{0.45\textwidth}{!}{
       \begin{tikzpicture}
      \node[node, label=below :{\footnotesize $2$}] (A1)  [] {};
      \node[node, label=below :{\footnotesize $4$}] (A2) [right=6mm of A1] {};
      \node[node, label=below :{\footnotesize $6$}] (A3) [right=6mm of A2] {};
      \node[node, label=below :{\footnotesize $8$}] (A4) [right=6mm of A3] {};
      \node[node, label=below :{\footnotesize $10$}] (A5) [right=6mm of A4] {};
      \node[node, label=below :{\footnotesize $12$}] (N3) [right=6mm of A5] {};
      \node[node, label=below :{\footnotesize $8$}] (B4) [right=6mm of N3] {};
      \node[node, label=below :{\footnotesize $4$}] (B2) [right=6mm of B4] {};
      \node[node, label=right:{\footnotesize $6$}] (Nu) [above=5mm of N3] {};

      \draw (A1.east) -- (A2.west);
      \draw (A2.east) -- (A3.west);
      \draw (A3.east) -- (A4.west);
      \draw (A4.east) -- (A5.west);
      \draw (A5.east) -- (N3.west);
      \draw (N3.east) -- (B4.west);
      \draw (B4.east) -- (B2.west);
      \draw (N3.north) -- (Nu.south); 
    \end{tikzpicture}
    }  
         \end{gathered} 
    $
    
    &

    \centering
    $
    \begin{gathered}
    \text{Magnetic Brane Picture:} \\
    \vspace{0.5cm}
    \resizebox{0.45\textwidth}{!}{
    \begin{tikzpicture}[cross/.style={path picture={ 
  \draw[black]
(path picture bounding box.south east) -- (path picture bounding box.north west) (path picture bounding box.south west) -- (path picture bounding box.north east);
}}]
        \draw (0,-1)--(0,1);
        \draw (1,-1)--(1,1);
        \draw (2,-1)--(2,1);
        \draw (3,-1)--(3,1);
        \draw (4,-1)--(4,1);
        \draw (5,-1)--(5,1);
        \draw (6,-1)--(6,1);
        \draw (7,-1)--(7,1);
        \draw[dashed] (8,-1)--(8,1);

        \draw (0,0)--(1,0) node[midway,above] () {$1$};
        \draw (1,0)--(2,0) node[midway,above] () {$2$};
        \draw (2,0)--(3,0) node[midway,above] () {$3$};
        \draw (3,0)--(4,0) node[midway,above] () {$4$};
        \draw (4,0)--(5,0) node[midway,above] () {$5$};
        \draw (5,0)--(6,0) node[midway,above] () {$6$};
        \draw (6,0.5)--(7,0.5) node[midway,above] () {$3$};
        \draw (6,-0.5)--(7,-0.5) node[midway,above] () {$4$};
        \draw (7,0.5) to[out=0, in=0, looseness=19] (7,0.3) node[] () {};

        \node[draw,circle,cross] () at (8,0.8) {};
        \node[draw,circle,cross] () at (8,-0.8) {};
    \end{tikzpicture}
    } \\
        \resizebox{0.45\textwidth}{!}{
    \begin{tikzpicture}[cross/.style={path picture={ 
  \draw[black]
(path picture bounding box.south east) -- (path picture bounding box.north west) (path picture bounding box.south west) -- (path picture bounding box.north east);
}}]
        \draw (0,-1)--(0,1);
        \draw (1,-1)--(1,1);
        \draw (2,-1)--(2,1);
        \draw (3,-1)--(3,1);
        \draw (4,-1)--(4,1);
        \draw (5,-1)--(5,1);
        \draw (6,-1)--(6,1);
        \draw (7,-1)--(7,1);
        \draw[dashed] (8,-1)--(8,1);

        \draw (0,0)--(1,0) node[midway,above] () {$1$};
        \draw (1,0)--(2,0) node[midway,above] () {$2$};
        \draw (2,0)--(3,0) node[midway,above] () {$3$};
        \draw (3,0)--(4,0) node[midway,above] () {$4$};
        \draw (4,0)--(5,0) node[midway,above] () {$5$};
        \draw (5,0)--(6,0) node[midway,above] () {$6$};
        \draw (6,0.5)--(7,0.5) node[midway,above] () {$3$};
        \draw (6,-0.5)--(7,-0.5) node[midway,above] () {$4$};
        \draw (7,0.5) to[out=0, in=0, looseness=19] (7,0.3) node[] () {};

        \node[draw,circle,cross] () at (8,0.8) {};
        \node[draw,circle,cross] () at (8,-0.8) {};
    \end{tikzpicture}
    }\\
        \text{Magnetic Quiver:} \\
    \vspace{0.5cm}
    \resizebox{0.45\textwidth}{!}{
       \begin{tikzpicture}
      \node[node, label=below :{\footnotesize $1$}] (A1)  [] {};
      \node[node, label=below :{\footnotesize $2$}] (A2) [right=6mm of A1] {};
      \node[node, label=below :{\footnotesize $3$}] (A3) [right=6mm of A2] {};
      \node[node, label=below :{\footnotesize $4$}] (A4) [right=6mm of A3] {};
      \node[node, label=below :{\footnotesize $5$}] (A5) [right=6mm of A4] {};
      \node[node, label=below :{\footnotesize $6$}] (N3) [right=6mm of A5] {};
      \node[node, label=below :{\footnotesize $4$}] (B4) [right=6mm of N3] {};
      \node[node, label=below :{\footnotesize $2$}] (B2) [right=6mm of B4] {};
      \node[node, label=right:{\footnotesize $3$}] (Nu) [above=5mm of N3] {};

      \draw (A1.east) -- (A2.west);
      \draw (A2.east) -- (A3.west);
      \draw (A3.east) -- (A4.west);
      \draw (A4.east) -- (A5.west);
      \draw (A5.east) -- (N3.west);
      \draw (N3.east) -- (B4.west);
      \draw (B4.east) -- (B2.west);
      \draw (N3.north) -- (Nu.south); 

\draw[dashed,orange] (0.3,-0.7)--(5.5,-0.7) to[out=0, in=0, looseness=2.3] (5.5,0.3)--(4.7,0.3);
\draw[dashed,orange] (4.7,0.3) to[out=90, in=0, looseness=2] (4,1) to[out=180, in=90, looseness=1.4] (3.7,0.3) -- (0.3,0.3) to[out=180, in=180, looseness=2.3] (0.3,-0.7);
      
    \end{tikzpicture}
    } \\
        \resizebox{0.45\textwidth}{!}{
       \begin{tikzpicture}
      \node[node, label=below :{\footnotesize $1$}] (A1)  [] {};
      \node[node, label=below :{\footnotesize $2$}] (A2) [right=6mm of A1] {};
      \node[node, label=below :{\footnotesize $3$}] (A3) [right=6mm of A2] {};
      \node[node, label=below :{\footnotesize $4$}] (A4) [right=6mm of A3] {};
      \node[node, label=below :{\footnotesize $5$}] (A5) [right=6mm of A4] {};
      \node[node, label=below :{\footnotesize $6$}] (N3) [right=6mm of A5] {};
      \node[node, label=below :{\footnotesize $4$}] (B4) [right=6mm of N3] {};
      \node[node, label=below :{\footnotesize $2$}] (B2) [right=6mm of B4] {};
      \node[node, label=right:{\footnotesize $3$}] (Nu) [above=5mm of N3] {};

      \draw (A1.east) -- (A2.west);
      \draw (A2.east) -- (A3.west);
      \draw (A3.east) -- (A4.west);
      \draw (A4.east) -- (A5.west);
      \draw (A5.east) -- (N3.west);
      \draw (N3.east) -- (B4.west);
      \draw (B4.east) -- (B2.west);
      \draw (N3.north) -- (Nu.south); 

\draw[dashed,orange] (0.3,-0.7)--(5.5,-0.7) to[out=0, in=0, looseness=2.3] (5.5,0.3)--(4.7,0.3);
\draw[dashed,orange] (4.7,0.3) to[out=90, in=0, looseness=2] (4,1) to[out=180, in=90, looseness=1.4] (3.7,0.3) -- (0.3,0.3) to[out=180, in=180, looseness=2.3] (0.3,-0.7);
      
    \end{tikzpicture}
    }
        \end{gathered}
        $
    \end{tabular}
    \caption{Type IIA brane system and magnetic quiver before the $A_1$ transition that takes from the $1\,2$ theory to the product theory $1 \sqcup 1$. The vertical solid line segments are D8-branes whereas the dashed vertical line is an $\mathrm{O8}^-$-plane, the horizontal segments are D6-branes, and the crossed circles are NS5-branes. Note: here we have depicted the true magnetic quivers, not the overextended magnetic quivers that we need to use the determine the Hasse diagram.}
    \label{fig:11to1_Branes}
\end{figure}

Throughout this section, we have used the artifice of over-extending the magnetic quiver for the rank $N$ E-string from that in equation \eqref{eqn:Magnetic_undeco_C^2} to that in equation \eqref{eqn:Magnetic_undeco_C^2/Z_1}. This was necessary to derive the Hasse diagram using the quiver subtraction algorithm that we have presented, due to the emergent $\mathfrak{su}(2)$ Coulomb symmetry that is hidden in equation \eqref{eqn:Magnetic_undeco_C^2} for $N > 1$. We can recover the same results from the true magnetic quiver by studying its Type IIA brane realization, instead of directly applying the quiver subtraction algorithm. We depict this in Figure \ref{fig:11to1_Branes} for the $A_1$ transition between the SCFT associated to the curve configuration $1\,2$ and that associated to $1 \sqcup 1$; we simply mimic the separation of M5-branes inside of the M9-brane via the separation NS5-branes inside the D8/O8 stack.
An NS5-brane brought on top of the $\operatorname{O8}^-$-plane obtained from the M9-brane is pulled away together with its set of D6-branes created when passing through the $8$ D8-branes associated with the $\operatorname{O8}^-$-plane. Notice that the fact that the slice is one-dimensional corresponds to the single distance modulus between the stacks.

\section{The Higgs Branch of Heterotic Orbifold LSTs}\label{sec:LSTs}

In Section \ref{sec:AtypeOI}, we introduced the rank $N$ $(\mathfrak{e}_8, \mathfrak{su}(K))$ orbi-instanton SCFTs, and then, in Section \ref{sec:Higgs_Branch_OI}, we explored their Higgs branches. To understand the Higgs branch, the principle methodology was to determine the 3d $\mathcal{N}=4$ Lagrangian quiver describing the Coulomb branch of the 3d mirror of the orbi-instanton. We derived this quiver via two methods, using the class $\mathcal{S}$ description of the torus compactification of the orbi-instantons, and reducing the M-theory description to Type IIA then passing to the (infinite-coupling) magnetic phase of the resulting brane system. 

In Section \ref{sec:HeteroticOrbitfoldLST}, we introduced a class of 6d $(1,0)$ little string theories that were constructed via the gluing of two orbi-instanton SCFTs. In this section, we determine the Higgs branches of these LSTs. We start with the M-theory description and reduce it to Type IIA (or Type I'), pass to the magnetic phase, and take the infinite-coupling limit. We see that this produces the same magnetic quiver as that obtained from a naive Coulomb gauging of the 3d mirrors of the orbi-instanton theories that were glued together to form the LST under discussion.

A second class of LSTs are those arising on the worldvolume of NS5-branes, probing a $\mathbb{Z}_K$ orbifold singularity, in the heterotic $\mathrm{Spin}(32)/\mathbb{Z}_2$ string theory. We introduce these theories in Section \ref{sec:spin32}, and similarly derive the magnetic quivers that capture their Higgs branches from a dual brane description. 

\subsection{A Brane System for Heterotic \texorpdfstring{$E_8 \times E_8$}{E8 x E8} LSTs}\label{sec:LSTs_E_8xE_8}

\begin{figure}[t]
\begin{multicols}{2}
\centering
    \begin{tikzpicture}
    \draw[dashed] (1,0)--(5,0) node[left=4] () {$\mathbb{C}^2/\mathbb{Z}_K$};
    \draw (5,-2.5)--(5,2.5) node[above,right] () {M9};
    \draw (1,-2.5)--(1,2.5) node[above,right] () {M9};
    \node at (3,0) [red,above=0.3] () {M5};
    \draw[red] (2.8,0.2)--(3.2,-0.2);
    \draw[red] (2.8,-0.2)--(3.2,0.2);

    \draw[->] (-0.5,1.5)--(0,1.5) node [right] () {$x^6$};
    \draw[->] (-0.5,1.5)--(-0.5,2) node [above] () {$x^{789}$};
    \end{tikzpicture}
    \vfill
    \resizebox{0.4\textwidth}{!}{
    \begin{tabular}{|c|c|}\hline
             Brane & Spatial Orientation  \\ \hline
             M5 & $x^0,x^1,x^2,x^3,x^4,x^5$ \\ \hline 
             M9 & $x^0,x^1,x^2,x^3,x^4,x^5,x^7,x^8,x^9,x^{10}$ \\ \hline 
             $\mathbb{C}^2/\mathbb{Z}_K$ & $x^0,x^1,x^2,x^3,x^4,x^5,x^6$ \\ \hline
    \end{tabular}
    }
\end{multicols}
    \caption{General point on the tensor branch of a $\mathcal{K}_{N_L,N_R,K}(\rho_L,\rho_R)$ theory in the M-theory picture. At ``infinite'' coupling only one of the two M5-M9 free moduli is taken to zero size, hence each M5-brane can be put either on the left or the right M9-wall. In the table, the spatial extension of the various elements of the M-theory engineering is indicated.}
    \label{fig:M-theory_LST}
\end{figure}
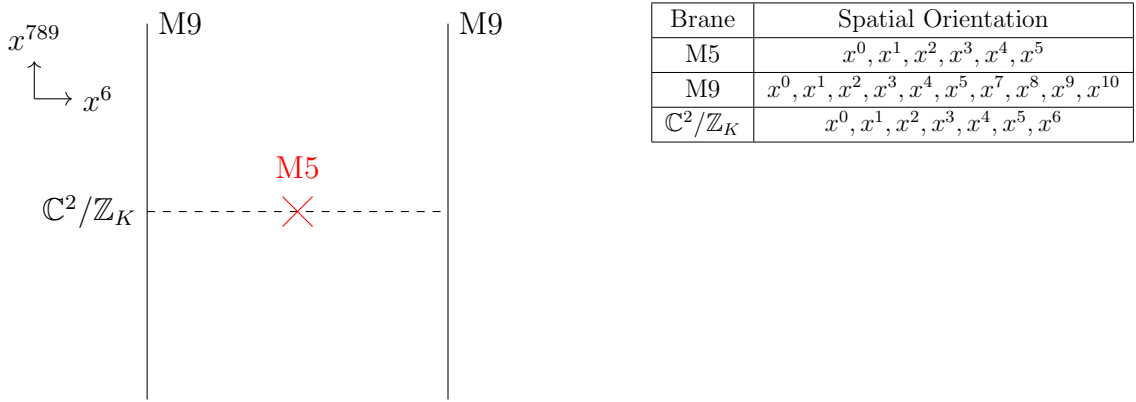

The heterotic $E_8 \times E_8$ $\mathbb{Z}_K$ orbifold LSTs, $\mathcal{K}_{N_L, N_R, K}(\rho_L, \rho_R)$, can be realized in heterotic M-theory where M5-branes are suspended between a pair of end-of-the-world Ho\v{r}ava--Witten branes, and probing a $\mathbb{C}^2/\mathbb{Z}_K$ orbifold singularity. We depict this setup in Figure \ref{fig:M-theory_LST}. When the $N$ M5-branes are separated along the $x^6$-direction between the two M9-branes, there are a number of distance moduli and we must move to the infinite-coupling phase where there is only one remaining distance modulus to realize the LST itself, as opposed to the effective field theory on the tensor branch of the LST.

Descending to a Type IIA description can be achieved by substituting the singular space with a Taub-NUT space of charge $K$ and using the Taub-NUT circle as the M-theory circle \cite{Polchinski:1995df}. The brane system obtained in this way is exactly an NS5-D6-D8-brane one: the M5-branes correspond to NS5-branes, whereas the Taub-NUT space becomes a stack of $K$ coincident D6-branes with an $S^1/\mathbb{Z}_2$ direction, for details see \cite{DelZotto:2022ohj}. The M9-walls become two $O8^-$-planes $+8$ D8-branes each, put at the antipodes of the $S^1/\mathbb{Z}_2$ interval. The embedding choices $\rho_L$ and $\rho_R$ are boundary conditions of $K$ D6-branes terminating on the $8$ D8-branes coming from the left or right M9-brane, respectively.   

Moving to the magnetic phase of the theory can be achieved by the usual procedure of suspending the D6-branes between D8-branes, the only caveat relies on the moduli tuning for the NS5-branes. In principle, there are at most $N_L+N_R+2K-1$ curves in the tensor branch configuration of the $\mathcal{K}_{N_L,N_R,K}(\rho_L,\rho_R)$ theory, hence we have at most $N_L+N_R+2K-1$ different ways to reach infinite coupling by sending all the scales in the theory apart from one to zero.\footnote{Here, we have counted the cases where zero M5-branes are put inside of one of the M9-branes. We explain this special case in more detail in Section \ref{sec:beyond}.} This reflects choosing how to split the NS5-branes between the left and the right $\mathrm{O8}^-$-planes. However to realise the embedding $\rho_L$ and $\rho_R$ a certain number of NS5-branes is obliged to be assigned to one of the planes, actually reducing the possible infinite coupling limits one can explore. Each of these limits corresponds to a different base space geometry in the F-theory model, as analysed in Section \ref{sec:HeteroticOrbitfoldLST}. Therefore for the general $\mathcal{K}_{N_L,N_R,K}(\rho_L,\rho_R)$ LST, the magnetic quiver at infinite coupling can be determined using the same analysis as appeared in \cite{Cabrera:2019dob}, using the notation therein, and it takes the form:
\begin{equation} \label{eqn:heterotic_Infinite_MQ}
    \begin{gathered}
    \resizebox{\textwidth}{!}{
    \begin{tikzpicture}
      \node[node, label=below:{\footnotesize $K$}] (Zk) [] {};
      \node[node, label=below right:\rotatebox{-30}{\footnotesize $g_{1,R}+(n_R+p_R)$}] (A1)  [right=6mm of Zk] {};
      \node[node, label=below right:\rotatebox{-30}{\footnotesize $g_{2,R}+2(n_R+p_R)$}] (A2) [right=6mm of A1] {};
      \node[node, label=below right:\rotatebox{-30}{\footnotesize $g_{3,R}+3(n_R+p_R)$}] (A3) [right=6mm of A2] {};
      \node[node, label=below right:\rotatebox{-30}{\footnotesize $g_{4,R}+4(n_R+p_R)$}] (A4) [right=6mm of A3] {};
      \node[node, label=below right:\rotatebox{-30}{\footnotesize $g_{5,R}+5(n_R+p_R)$}] (A5) [right=6mm of A4] {};
      \node[node, label=below right:\rotatebox{-30}{\footnotesize $g_{6,R}+6(n_R+p_R)$}] (N3) [right=6mm of A5] {};
      \node[node, label=below right:\rotatebox{-30}{\footnotesize $g_{7,R}+4(n_R+p_R)$}] (B4) [right=6mm of N3] {};
      \node[node, label=below right:\rotatebox{-30}{\footnotesize $2r_R+2(n_R+p_R)$}] (B2) [right=6mm of B4] {};
      \node[node, label=right:{\footnotesize $g_{8,R}+3(n_R+p_R)$}] (Nu) [above=5mm of N3] {};
      \node[node, label=below left:\rotatebox{30}{\footnotesize $g_{1,L}+(n_L+p_L)$}] (A1s)  [left=6mm of Zk] {};
      \node[node, label=below left:\rotatebox{30}{\footnotesize $g_{2,L}+2(n_L+p_L)$}] (A2s) [left=6mm of A1s] {};
      \node[node, label=below left:\rotatebox{30}{\footnotesize $g_{3,L}+3(n_L+p_L)$}] (A3s) [left=6mm of A2s] {};
      \node[node, label=below left:\rotatebox{30}{\footnotesize $g_{4,L}+4(n_L+p_L)$}] (A4s) [left=6mm of A3s] {};
      \node[node, label=below left:\rotatebox{30}{\footnotesize $g_{5,L}+5(n_L+p_L)$}] (A5s) [left=6mm of A4s] {};
      \node[node, label=below left:\rotatebox{30}{\footnotesize $g_{6,L}+6(n_L+p_L)$}] (N3s) [left=6mm of A5s] {};
      \node[node, label=below left:\rotatebox{30}{\footnotesize $g_{7,L}+4(n_L+p_L)$}] (B4s) [left=6mm of N3s] {};
      \node[node, label=below left:\rotatebox{30}{\footnotesize $2r_L+2(n_L+p_L)$}] (B2s) [left=6mm of B4s] {};
      \node[node, label=left:{\footnotesize $g_{8,L}+3(n_L+p_L)$}] (Nus) [above=5mm of N3s] {};
      \node[tnode] (node) [right=6mm of B2] {$\,.$};

      \draw (Zk.east) -- (A1.west);
      \draw (A1.east) -- (A2.west);
      \draw (A2.east) -- (A3.west);
      \draw (A3.east) -- (A4.west);
      \draw (A4.east) -- (A5.west);
      \draw (A5.east) -- (N3.west);
      \draw (N3.east) -- (B4.west);
      \draw (B4.east) -- (B2.west);
      \draw (N3.north) -- (Nu.south); 

    \draw (Zk.west) -- (A1s.east);
      \draw (A1s.west) -- (A2s.east);
      \draw (A2s.west) -- (A3s.east);
      \draw (A3s.west) -- (A4s.east);
      \draw (A4s.west) -- (A5s.east);
      \draw (A5s.west) -- (N3s.east);
      \draw (N3s.west) -- (B4s.east);
      \draw (B4s.west) -- (B2s.east);
      \draw (N3s.north) -- (Nus.south); 
    \end{tikzpicture} }
  \end{gathered} 
\end{equation}
The parameters here are the same as those appearing around equation \eqref{eqn:infcplingMQ}, Here, $n_L+n_R+p_L+p_R=N_L+N_R+N_{\rho_L}+N_{\rho_R}$ is the total number of NS5-branes in the theory including the $N_{\rho_L}$ and $N_{\rho_R}$ instantonic NS5-branes bounded to one of the orientifold planes. Notice that $n_{L/R}+p_{L/R}\ge N_{\rho_{L/R}}$ so that the instanton boundary condition can always be realized, and the free subdivision between left and right planes applies only to non-instantonic NS5-branes. This magnetic quiver has been derived also in \cite{DelZotto:2023myd,DelZotto:2023nrb}. Importantly, exactly as in the orbi-instanton case of Section \ref{sec:Higgs_Branch_OI}, when sending an NS5-brane on top of the orientifold plane, one performs the $29$-dimensional $\mathfrak{e}_8$ transition.

\subsection{Coulomb Gauging, Fusion, and Magnetic Quivers}

In the framework of six-dimensional SCFTs it is well-known that given two theories with a common non-Abelian flavor factor, one can often gauge this symmetry at the price of introducing an additional tensor multiplet. In the F-theory realization, this implies that we can fuse two theories by rendering the non-compact flavor curve compact, and use it to connect the two theories. This is precisely how we have introduced the $\mathcal{K}_{N_L, N_R, L}(\rho_L, \rho_R)$ orbifold LSTs, except that we have fused the common flavor symmetry of two orbi-instanton theories in such a way that the resulting intersection matrix is negative semi-definite, instead of negative definite. This fusion is depicted in equation \eqref{eqn:swiper}.

It is then natural to ask: how is this fusion procedure reflected in an action on the magnetic quivers of the fused theories?\footnote{Gauging a subset of the flavor symmetry is translated at the level of Hilbert series as a hyperk\"ahler quotient. Differently from \cite{Hanany:2023tvn}, where this operation is studied as an extended subtraction algorithm on a single magnetic quiver, the operation of fusion we are proposing in this subsection regards the identification of two symmetry factors in two distinct quiver theories. Therefore, it can not be worked out directly from the subtraction rules proposed in the aforementioned paper.} First, we determine how the fusion process acts on the electric brane system. Then, we move to the magnetic phase and compare the resulting magnetic quiver with that of the two SCFTs before fusion. 

As an example, we first consider the fusion of rank $N$ $(\mathfrak{su}(K), \mathfrak{su}(K))$ conformal matter with rank $M$ $(\mathfrak{su}(K), \mathfrak{su}(K))$ conformal matter. The result will be a new SCFT: rank $N+M$ $(\mathfrak{su}(K), \mathfrak{su}(K))$ conformal matter. Rank $N$  $(\mathfrak{g}, \mathfrak{g})$ conformal matter theory is the 6d $(1,0)$ SCFT living on the worldvolume of a stack of $N$ coincident M5-branes probing a $\mathbb{C}^2/\Gamma$ singularity, where $\Gamma$ is the finite subgroup of $SU(2)$ of the same ADE-type as $\mathfrak{g}$ \cite{DelZotto:2014hpa}.  For $\mathfrak{g}=\mathfrak{su}(K)$, the Type IIA NS5-D6-D8 brane system engineering this theory can be obtained via the circle compactification of M-theory, and the associated magnetic quiver is \cite{Hanany:2018vph}:
\begin{equation}
    \begin{gathered}
    \begin{tikzpicture}
      \node[node, label=below:{\footnotesize $1$}] (Z1)  {};
      \node[node, label=below:{\footnotesize $2$}] (Z2) [right=6mm of Z1] {};
      \node[tnode] (dots) [right=6mm of Z2] {\footnotesize $\cdots$};
      \node[node, label=below:{\footnotesize $K-1$}] (Zk-1) [right=6mm of dots] {};
      \node[node, label=below:{\footnotesize $K$}] (Zk) [right=6mm of Zk-1] {};
      \node[node, label=right:{\footnotesize $N$}] (Fk) [above=6mm of Zk] {};
      \node[node, label=below:{\footnotesize $K-1$}] (Zk1) [right=6mm of Zk] {};
      \node[tnode] (dots1) [right=6mm of Zk1] {\footnotesize $\cdots$};
      \node[node, label=below:{\footnotesize $2$}] (Z21) [right=6mm of dots1] {};
      \node[node, label=below:{\footnotesize $1$}] (Z11) [right=6mm of Z21] {};
      \draw (Z1.east) -- (Z2.west);
      \draw (Z2.east) -- (dots.west);
      \draw (dots.east) -- (Zk-1.west);
      \draw (Zk-1.east) -- (Zk.west);
      \draw (Zk.north) -- (Fk.south);
      \draw (Zk.east) -- (Zk1.west);
      \draw (Zk1.east) -- (dots1.west);
      \draw (dots1.east) -- (Z21.west);
      \draw (Z21.east) -- (Z11.west);
      \draw (Fk) to[out=130, in=410, looseness=12] (Fk);
    \end{tikzpicture}
  \end{gathered} \quad \,.
\end{equation}

Let us focus on the case $K = 3$. The electric Type IIA brane systems for the rank $N$ and rank $M$ conformal matter theories are depicted in Figure \ref{fig:electricbranes}. We can fuse the two theories by recombining the D6-branes on the right of the rank $N$ theory with the D6-branes on the left of the rank $M$ theory and eliminating the D8-branes. This induces a new length scale in the theory between the rightmost NS5-brane of the former theory and the leftmost of the latter, this is exactly the size of the curve associated with the extra tensor multiplet introduced. Recall that at the SCFT point, all the curve volumes must be taken to zero, so after the fusion process one obtains a 6d theory which must be taken to the SCFT point. 

\begin{figure}[p]
    \centering
    \begin{multicols}{2}
        \resizebox{0.45\textwidth}{!}{\begin{tikzpicture}[cross/.style={path picture={ 
  \draw[black]
(path picture bounding box.south east) -- (path picture bounding box.north west) (path picture bounding box.south west) -- (path picture bounding box.north east);
}}]
        \draw (0,-1)--(0,1);
        \draw (1,-1)--(1,1);
        \draw (2,-1)--(2,1);
        \draw (8,-1)--(8,1);
        \draw (9,-1)--(9,1);
        \draw (10,-1)--(10,1);

        \node[draw,circle,cross] (NS1) at (3,0) {};
        \node[draw,circle,cross] (NS2) at (4,0) {};
        \node[draw,circle,cross] (NS3) at (6,0) {};
        \node[draw,circle,cross] (NS4) at (7,0) {};
        \node[tnode] (dots) at (5,0) {$\cdots$};

        \draw[decorate,decoration={brace,raise=4mm}] (NS1.west)--(NS4.east) node[midway,above=6mm] () {$N$ NS5};
        \draw (NS1)--(NS2) node[midway,below] () {$3$};
        \draw (NS2)--(dots)--(NS3);
        \draw (NS3)--(NS4) node[midway,below] () {$3$};
        
        \draw (NS1)--(2,0);
        \draw (NS1)--(2.5,0.5)--(1,0.5);
        \draw (NS1)--(2.5,-0.5)--(0,-0.5);
        \draw (NS4)--(8,0);
        \draw (NS4)--(7.5,0.5)--(9,0.5);
        \draw (NS4)--(7.5,-0.5)--(10,-0.5);
		\node[] (Title) at (5,1.8) {rank $N$ $\left(\mathfrak{su}(3),\mathfrak{su}(3)\right)$ conformal matter};
    \end{tikzpicture}
    } \vfill
            \resizebox{0.45\textwidth}{!}{\begin{tikzpicture}[cross/.style={path picture={ 
  \draw[black]
(path picture bounding box.south east) -- (path picture bounding box.north west) (path picture bounding box.south west) -- (path picture bounding box.north east);
}}]
        \draw (0,-1)--(0,1);
        \draw (1,-1)--(1,1);
        \draw (2,-1)--(2,1);
        \draw (8,-1)--(8,1);
        \draw (9,-1)--(9,1);
        \draw (10,-1)--(10,1);

        \node[draw,circle,cross] (NS1) at (3,0) {};
        \node[draw,circle,cross] (NS2) at (4,0) {};
        \node[draw,circle,cross] (NS3) at (6,0) {};
        \node[draw,circle,cross] (NS4) at (7,0) {};
        \node[tnode] (dots) at (5,0) {$\cdots$};

        \draw[decorate,decoration={brace,raise=4mm}] (NS1.west)--(NS4.east) node[midway,above=6mm] () {$M$ NS5};
        \draw (NS1)--(NS2) node[midway,below] () {$3$};
        \draw (NS2)--(dots)--(NS3);
        \draw (NS3)--(NS4) node[midway,below] () {$3$};
        
        \draw (NS1)--(2,0);
        \draw (NS1)--(2.5,0.5)--(1,0.5);
        \draw (NS1)--(2.5,-0.5)--(0,-0.5);
        \draw (NS4)--(8,0);
        \draw (NS4)--(7.5,0.5)--(9,0.5);
        \draw (NS4)--(7.5,-0.5)--(10,-0.5);
		    \node[] (Title) at (5,1.8) {rank $M$ $\left(\mathfrak{su}(3),\mathfrak{su}(3)\right)$ conformal matter};
    \end{tikzpicture}
    } \,.
    \end{multicols}
    \caption{The Type IIA electric brane systems engineering rank $N$ and rank $M$ $(\mathfrak{su}(3), \mathfrak{su}(3))$ conformal matter.}\label{fig:electricbranes}

\vspace{1cm}
     \begin{tikzpicture}[cross/.style={path picture={ 
  \draw[black]
(path picture bounding box.south east) -- (path picture bounding box.north west) (path picture bounding box.south west) -- (path picture bounding box.north east);
}},scale=0.92, every node/.style={scale=0.92}]
        \draw (0,-1)--(0,1);
        \draw (1,-1)--(1,1);
        \draw (2,-1)--(2,1);
        \draw (4,-1)--(4,1);
        \draw (5,-1)--(5,1);
        \draw (6,-1)--(6,1);

        \node[draw,circle,cross] (NS1) at (3,0) {};
        \node[draw,circle,cross] (NS2) at (3,1) {};
        \node[tnode] (dots) at (3,0.6) {\footnotesize  $\vdots$};

        \draw[decorate,decoration={brace,raise=4mm}] (NS2.west)--(NS2.east) node[midway,above=6mm] () {\footnotesize $N$  NS5-branes};

        \draw (0,-0.5)--(1,-0.5) node[midway,below] () {\footnotesize $1$};
        \draw (1,-0.5)--(2,-0.5) node[midway,below] () {\footnotesize $2$};
        \draw (2,-0.5)--(4,-0.5) node[midway,below] () {\footnotesize $3$};
        \draw (4,-0.5)--(5,-0.5) node[midway,below] () {\footnotesize $2$};
        \draw (5,-0.5)--(6,-0.5) node[midway,below] () {\footnotesize $1$};

	     \node[] (Title) at (3,2.5) {\footnotesize rank $N$ $\left(\mathfrak{su}(3),\mathfrak{su}(3)\right)$ conformal matter};
        \draw (8,-1)--(8,1);
        \draw (9,-1)--(9,1);
        \draw (10,-1)--(10,1);
        \draw (12,-1)--(12,1);
        \draw (13,-1)--(13,1);
        \draw (14,-1)--(14,1);

        \node[draw,circle,cross] (NS1) at (11,0) {};
        \node[draw,circle,cross] (NS2) at (11,1) {};
        \node[tnode] (dots) at (11,0.6) {\footnotesize $\vdots$};

        \draw[decorate,decoration={brace,raise=4mm}] (NS2.west)--(NS2.east) node[midway,above=6mm] () {\footnotesize $M$ NS5-branes};

        \draw (8,-0.5)--(9,-0.5) node[midway,below] () {\footnotesize $1$};
        \draw (9,-0.5)--(10,-0.5) node[midway,below] () {\footnotesize $2$};
        \draw (10,-0.5)--(12,-0.5) node[midway,below] () {\footnotesize $3$};
        \draw (12,-0.5)--(13,-0.5) node[midway,below] () {\footnotesize $2$};
        \draw (13,-0.5)--(14,-0.5) node[midway,below] () {\footnotesize $1$};

	     \node[] (Title) at (11,2.5) {\footnotesize rank $M$ $\left(\mathfrak{su}(3),\mathfrak{su}(3)\right)$ conformal matter.};
        \draw[->] (3,-1.5)--(6,-2.5);
        \draw[->] (11,-1.5)--(8,-2.5);

        \draw (4,-7)--(4,-5);
        \draw (5,-7)--(5,-5);
        \draw (6,-7)--(6,-5);
        \draw (8,-7)--(8,-5);
        \draw (9,-7)--(9,-5);
        \draw (10,-7)--(10,-5);

        \node[draw,circle,cross] (NS1) at (6.5,-6) {};
        \node[draw,circle,cross] (NS2) at (6.5,-5) {};
        \node[tnode] (dots) at (6.5,-5.4) {\footnotesize $\vdots$};

        \draw[decorate,decoration={brace,raise=4mm}] (NS2.west)--(NS2.east) node[midway,above=6mm] () {\footnotesize $N$};
        \node[draw,circle,cross] (NS1) at (7.5,-6) {};
        \node[draw,circle,cross] (NS2) at (7.5,-5) {};
        \node[tnode] (dots) at (7.5,-5.4) {\footnotesize $\vdots$};

        \draw[decorate,decoration={brace,raise=4mm}] (NS2.west)--(NS2.east) node[midway,above=6mm] () {\footnotesize $M$};
        
        \draw (4,-6.5)--(5,-6.5) node[midway,below] () {\footnotesize $1$};
        \draw (5,-6.5)--(6,-6.5) node[midway,below] () {\footnotesize $2$};
        \draw (6,-6.5)--(8,-6.5) node[midway,below] () {\footnotesize $3$};
        \draw (8,-6.5)--(9,-6.5) node[midway,below] () {\footnotesize $2$};
        \draw (9,-6.5)--(10,-6.5) node[midway,below] () {\footnotesize $1$};

        \node[] (Title) at (7,-3) {\footnotesize Magnetic Phase After Fusion};
        \draw[->] (7,-8)--(7,-9);

        \draw (4,-11)--(4,-13);
        \draw (5,-11)--(5,-13);
        \draw (6,-11)--(6,-13);
        \draw (8,-11)--(8,-13);
        \draw (9,-11)--(9,-13);
        \draw (10,-11)--(10,-13);

        \node[draw,circle,cross] (NS1) at (7,-12) {};
        \node[draw,circle,cross] (NS2) at (7,-11) {};
        \node[tnode] (dots) at (7,-11.4) {\footnotesize $\vdots$};

        \draw[decorate,decoration={brace,raise=4mm}] (NS2.west)--(NS2.east) node[midway,above=6mm] () {\footnotesize $N+M$ NS5-branes};

        \draw (4,-12.5)--(5,-12.5) node[midway,below] () {\footnotesize $1$};
        \draw (5,-12.5)--(6,-12.5) node[midway,below] () {\footnotesize $2$};
        \draw (6,-12.5)--(8,-12.5) node[midway,below] () {\footnotesize $3$};
        \draw (8,-12.5)--(9,-12.5) node[midway,below] () {\footnotesize $2$};
        \draw (9,-12.5)--(10,-12.5) node[midway,below] () {\footnotesize $1$};

        \node[] (Title) at (7,-9.5) {\footnotesize rank $N+M$ $\left(\mathfrak{su}(3),\mathfrak{su}(3)\right)$ conformal matter};
    \end{tikzpicture}
    \caption{Fusion process between rank $N$ and rank $M$ $\left(\mathfrak{su}(3),\mathfrak{su}(3)\right)$ conformal matter from the magnetic phase of the brane system. In the last step, we go to the infinite-coupling phase by recombining the two NS5-brane stacks.}
    \label{fig:Magnetic_Fusion_Branes}
\end{figure} 

It is clear looking at the magnetic brane phase pictured in Figure \ref{fig:Magnetic_Fusion_Branes}, that the fusion process deletes the common subset of balanced nodes giving rise to the gauged flavor and merges the two $\mathfrak{u}(3)$ nodes. The $\mathfrak{u}(N)$ and $\mathfrak{u}(M)$ gauge nodes, coming from the NS5-brane stacks, and attached to the $\mathfrak{u}(3)$ node are now separated along the direction where the D6-branes are extended objects and the NS5-branes are pointlike. Thus to shrink to zero volume the curve associated with the introduced tensor multiplet, and thus to move to the superconformal point, we need to make all the NS5-branes coincident. This recovers the rank $N+M$ $\left(\mathfrak{su}(3),\mathfrak{su}(3)\right)$ SCFT magnetic quiver.

With this example in mind, it is straightforward to formulate a set of rules that implements the fusion of two 6d SCFTs via Coulomb gauging directly on the magnetic quiver. This algorithm is unsurprising, as it is precisely the procedure of Coulomb gauging as described in \cite{Benini:2010uu}.

\begin{alg}\label{alg:CoulombGauging} 
Consider two 3d $\mathcal{N}=4$ unitary quivers $\mathcal{Q}$ and $\mathcal{Q}'$, such that they exhibit a tail of balanced gauge nodes giving rise to the same non-Abelian flavor symmetry factor. If the first unbalanced node $\mathrm{N}$ next to the tail in $\mathcal{Q}$ and $\mathcal{Q}'$ is the same, it is possible to perform the \textit{Coulomb gauging} procedure, creating the quiver $\mathcal{F}$, whereby the common tail of balanced nodes is deleted and the $\mathrm{N}$ nodes are identified.\footnote{We could also consider the Coulomb gauging of two tails of balanced nodes of the same quiver; in this case, we are required to include an adjoint-valued hypermultiplet on the node N, after gauging. See \cite{Benini:2010uu} for more details, though we do not study such gaugings here.} Schematically:
\begin{equation}
\begin{aligned}
    \mathcal{F} &:= \operatorname{Coulomb Gauging}\left( \mathcal{Q},\mathcal{Q}'\right) \\
   &= \operatorname{Coulomb Gauging}\left(
    \begin{gathered}
        \begin{tikzpicture}
            \node[tnode] (E)  {\footnotesize $\mathcal{E}$};
            \node[tnode] (N) [right=5mm of E] {\footnotesize $\mathrm{N}$};
            \node[tnode] (BalancedTail) [right=5mm of N] {\footnotesize \textit{Balanced Tail}};
            \draw (BalancedTail)--(N)--(E);
        \end{tikzpicture}
    \end{gathered} ,
        \begin{gathered}
        \begin{tikzpicture}
            \node[tnode] (BalancedTail) {\footnotesize \textit{Balanced Tail}};
            \node[tnode] (N) [right=5mm of BalancedTail] {\footnotesize $\mathrm{N}$};
            \node[tnode] (E') [right=5mm of N] {\footnotesize $\mathcal{E}'$};
            \draw (BalancedTail)--(N)--(E');
        \end{tikzpicture}
    \end{gathered}
    \right)\\
    &  =       \begin{gathered}
        \begin{tikzpicture}
            \node[tnode] (E) {\footnotesize $\mathcal{E}$};
            \node[tnode] (N) [right=5mm of E] {\footnotesize $\mathrm{N}$};
            \node[tnode] (E') [right=5mm of N] {\footnotesize $\mathcal{E}'$};
            \draw (E)--(N)--(E');
        \end{tikzpicture} \,.
    \end{gathered}
\end{aligned}
\end{equation}
\end{alg}

After Coulomb gauging, we find the magnetic quiver of the theory where there is a single volume modulus remaining. If we are interested in fusing two 6d $(1,0)$ SCFTs in such a way that we obtain an SCFT, we need to subsequently take the infinite-coupling limit, where this curve volume is taken to zero. For LSTs which are obtained via the fusion of two SCFTs, the single introduced volume modulus is exactly the LST string tension, and thus we obtain the magnetic quiver for the LST on the nose. In particular, we notice that we can apply the Coulomb gauging algorithm to the 3d mirrors of the orbi-instanton theories, as given in equation \eqref{eqn:3dmirror}, and we recover exactly the magnetic quiver given in equation \eqref{eqn:heterotic_Infinite_MQ} that was derived directly from the Type IIA brane engineering.

\subsection{Heterotic \texorpdfstring{$\mathrm{Spin}(32)/\mathbb{Z}_2$}{Spin(32)/Z2} Orbifold LSTs}\label{sec:spin32}

There is yet another family of LSTs that can be studied from the brane perspective, those are the so-called heterotic $\mathrm{Spin}(32)/\mathbb{Z}_2$ orbifold LSTs. Such LSTs describe the worldvolume theory of a stack of $N$ NS5-branes in the $\mathrm{Spin}(32)/\mathbb{Z}_2$ heterotic string, in the presence of an orbifold singularity $\mathbb{C}^2/\Gamma$, where $\Gamma$ is again a finite subgroup of $SU(2)$.

We first consider the case where the orbifold is trivial. The worldvolume theory governing such a stack of $N$ NS5-branes is simply an $\mathfrak{sp}(N)$ gauge algebra with sixteen hypermultiplets in the fundamental representation and one hypermultiplet in the antisymmetric representation. In the F-theory description of LSTs, we can depict this theory as
\begin{equation}
    \overset{\mathfrak{sp}_N}{0} \,,
\end{equation}
where the finite-volume $(0)$-curve is the curve providing the LST string tension. This is a Lagrangian theory, which we can see directly by noting that there are no non-minimal singularities located over any points of the $(0)$-curve. Due to this description, we can determine the worldvolume theories in the presence of $\mathbb{C}^2/\Gamma$ simply by orbifolding the Lagrangian theory; this was done extensively in \cite{Blum:1997mm,Intriligator:1997dh,Blum:1997fw}. The resulting theories depend on a choice of flat connection at infinity for the $\mathrm{Spin}(32)/\mathbb{Z}_2$ gauge bundle, and thus on a choice of homomorphism:\footnote{As $\mathrm{Spin}(32)/\mathbb{Z}_2$ is non-simply-connected, $\mathrm{Spin}(32)/\mathbb{Z}_2$-bundles admit a non-trivial generalized Stiefel--Whitney class. A trivial/non-trivial Stiefel--Whitney class corresponds to a configuration with/without ``vector structure'', as described in detail in \cite{Berkooz:1996iz}. In this paper, we focus only on cases with vector structure.}
\begin{equation}
    \sigma : \Gamma \rightarrow \mathrm{Spin}(32)/\mathbb{Z}_2 \,.
\end{equation}
We denote the resulting LSTs as
\begin{equation}
    \widetilde{K}_{N, \mathfrak{g}}(\sigma) \,,
\end{equation}
where $\mathfrak{g}$ is the simple Lie algebra of the same ADE-type as $\Gamma$.

We focus on LSTs of this class where $\Gamma = \mathbb{Z}_K$. These particular models can be obtained from the aforementioned M-theory setting used to engineer $\mathcal{K}_{N_L, N_R, K}(\rho_L, \rho_R)$ theories in Section \ref{sec:LSTs_E_8xE_8}, but performing a different duality/compactification chain \cite{DelZotto:2022ohj}. Starting from the M-theory configuration in Section \ref{sec:LSTs_E_8xE_8}, one can consider the $x^5$-direction to be an $S^1$, and use this circle to compactify M-theory to Type IIA. Then we can perform T-duality along the Taub-NUT circle obtaining a Type IIB model that can be T-dualized to a Type IIA model via the $x^6$-circle; this describes the heterotic $\mathrm{Spin}(32)/\mathbb{Z}_2$ setting. The LST model engineered in this way is labelled by
\begin{equation}\label{eqn:SO(32)model}
    \widetilde{K}_{N,K}(\sigma) \,,
\end{equation}
where $N$ is the original number of M5-branes that are now mapped to a stack of coincident D6-branes wrapped on the interval $S^1/\mathbb{Z}_2$, and $K$ is the order of the singularity $\mathbb{C}^2/\mathbb{Z}_K$ that in this picture corresponds to NS5-branes. The two M9-walls are recombined and then split again during the chain of dualities and realise $2$ $O8^-$-planes with $8$ D8-branes each on the antipodes of the $S^1/\mathbb{Z}_2$ D6-brane interval. The final element in equation \eqref{eqn:SO(32)model} is the embedding $\sigma : \mathbb{Z}_K \rightarrow \mathrm{Spin}(32)/\mathbb{Z}_2$ of the discrete group $\mathbb{Z}_K$ in $\mathrm{Spin}(32)/\mathbb{Z}_2$.  This involved chain of operations on the M-theory setting establishes a possible T-dual connection between the following models:
\begin{equation}
    \mathcal{K}_{N_L, N_R, K}(\rho_L, \rho_R) \leftrightarrow \widetilde{K}_{N,K}(\sigma) \,.
\end{equation}
The precise matching between the numbers of NS5-branes on both sides and the mapping between the $\rho_L$, $\rho_R$ and the $\sigma$ is not explicit in this sequence of T-dualities. Instead, we can compute the invariants in equation \eqref{eqn:T-dual-invariants} in order to look for putative T-dual models \cite{DelZotto:2022ohj}, a process to which we return in Section \ref{sec:tduality}.

Having engineered a Type IIA NS5-D6-D8-brane system for $\widetilde{K}_{N,K}(\sigma)$ theories, we can move to the magnetic phase with the usual procedure of \cite{Cabrera:2019izd} and extract a magnetic quiver. This has been done extensively in \cite{DelZotto:2023nrb,DelZotto:2023myd}, therefore here we just sketch the general properties and some relevant subtleties of such magnetic quivers. For instance we can consider the embedding $\widetilde{\sigma}$ of $\mathbb{Z}_{2k}$ in $\mathrm{Spin}(32)/\mathbb{Z}_2$ that engineers an $SO(16)^2$ flavor symmetry in the $\widetilde{K}_{N,2k}(\widetilde{\sigma})$ theory. In terms of the curve configuration in F-theory, this model has the description:
\begin{equation}\label{eqn:so16so16}
    \widetilde{K}_{N,2k}(\widetilde{\sigma}) =[SO(16)] \,\ \underbrace{\stackon{1}{$\mathfrak{sp}_N$} \ \stackon{2}{$\mathfrak{su}_{2N}$} \ \cdots \ \stackon{2}{$\mathfrak{su}_{2N}$} \ \stackon{1}{$\mathfrak{sp}_{N}$}}_{k +1 \, \text{ curves }} \ \, [SO(16)] \,.
\end{equation}
The magnetic quiver at finite coupling associated with this theory can be worked out from the brane system, and it is:
\begin{equation}\label{eqn:finite_quiver}
    \begin{gathered}
            \begin{tikzpicture}
      \node[node, label=below:{\footnotesize $2N$}] (Zk) [] {};
      \node[node, label=below :{\footnotesize $2N$}] (A1)  [right=6mm of Zk] {};
      \node[node, label=below :{\footnotesize $2N$}] (A2) [right=6mm of A1] {};
      \node[node, label=below :{\footnotesize $2N$}] (A3) [right=6mm of A2] {};
      \node[node, label=below :{\footnotesize $2N$}] (A4) [right=6mm of A3] {};
      \node[node, label=below :{\footnotesize $2N$}] (A5) [right=6mm of A4] {};
      \node[node, label=below :{\footnotesize $2N$}] (N3) [right=6mm of A5] {};
      \node[node, label=below :{\footnotesize $N$}] (B4) [right=6mm of N3] {};
      \node[node, label=right:{\footnotesize $N$}] (Nu) [above=5mm of N3] {};
      \node[node, label=below :{\footnotesize $2N$}] (A1s)  [left=6mm of Zk] {};
      \node[node, label=below :{\footnotesize $2N$}] (A2s) [left=6mm of A1s] {};
      \node[node, label=below :{\footnotesize $2N$}] (A3s) [left=6mm of A2s] {};
      \node[node, label=below :{\footnotesize $2N$}] (A4s) [left=6mm of A3s] {};
      \node[node, label=below :{\footnotesize $2N$}] (A5s) [left=6mm of A4s] {};
      \node[node, label=below :{\footnotesize $2N$}] (N3s) [left=6mm of A5s] {};
      \node[node, label=below :{\footnotesize $N$}] (B4s) [left=6mm of N3s] {};
      \node[node, label=left:{\footnotesize $N$}] (Nus) [above=5mm of N3s] {};
      \node[node, label=above:{\footnotesize $1$}] (NS1) [above left=5mm and 5mm of Zk] {};
      \node[node, label=above:{\footnotesize $1$}] (NSn) [above right=5mm and 5mm of Zk] {};
      \node[tnode] (node) [above=3mm of Zk] {\footnotesize $\cdots$};

    \draw[decorate,decoration={brace,raise=6mm}] (NS1.west)--(NSn.east) node[midway,above=6mm] () {\footnotesize $k$};

      \draw (Zk.east) -- (A1.west);
      \draw (A1.east) -- (A2.west);
      \draw (A2.east) -- (A3.west);
      \draw (A3.east) -- (A4.west);
      \draw (A4.east) -- (A5.west);
      \draw (A5.east) -- (N3.west);
      \draw (N3.east) -- (B4.west);
      \draw (N3.north) -- (Nu.south); 

    \draw (Zk.west) -- (A1s.east);
      \draw (A1s.west) -- (A2s.east);
      \draw (A2s.west) -- (A3s.east);
      \draw (A3s.west) -- (A4s.east);
      \draw (A4s.west) -- (A5s.east);
      \draw (A5s.west) -- (N3s.east);
      \draw (N3s.west) -- (B4s.east);
      \draw (N3s.north) -- (Nus.south);

      \draw (Zk) -- (NS1);
      \draw (Zk) -- (NSn);
        \end{tikzpicture}
    \end{gathered} \,,
\end{equation}
where we notice the general property that the shape of the quiver resembles an affine $D_{16}$ Dynkin diagram, with the caveat that when antisymmetric matter is present in the theory, i.e., when the $(-1)$-curves carry an $\mathfrak{su}$-type algebra, an extra node $\mathfrak{u}(1)$ is attached to the corresponding tail of the respective magnetic quiver, realizing a balance symmetry of unitary-type instead of special-orthogonal. The central bouquet of $\mathfrak{u}(1)$s is due to the NS5-branes; for general embedding choice these floating $\mathfrak{u}(1)$ are scattered along the quiver due to Hanany--Witten transitions in the brane model, and determine the breaking of the maximal $\mathfrak{so}_{32}$ flavor algebra in subgroups of the form $\mathfrak{so}(4q_k)\times \mathfrak{su}(2q_{k-1})\times \cdots \times \mathfrak{su}(2q_1) \times \mathfrak{so}(32-4q)\subset \mathfrak{so}(32)$ or its equivalent with $\mathfrak{su}$ algebras instead of $\mathfrak{so}$ when antisymmetric matter is present.

The novel characteristic of these $\mathrm{Spin}(32)/\mathbb{Z}_2$ magnetic quivers is shown in the infinite coupling magnetic phase: moving an NS5-brane on top of just one of the orientifold planes would realize an $\mathfrak{e}_8$ transition, which contrasts the fact that the minimal 29-dimensional instantonic transition should be a $\mathfrak{d}_{16}$ transition, as pointed out in \cite{Witten:1995gx,Blum:1997mm}. So the proposed \textit{correct} instantonic transitions are realized by adding an affine $D_{16}$ Dynkin diagram to the quiver for each NS5-brane available. Hence for the theory in equation \eqref{eqn:so16so16} the proposed infinite coupling quiver is:
\begin{equation}\label{eqn:infinite_quiver}
    \begin{gathered}
            \begin{tikzpicture}
      \node[node, label=below:{\footnotesize $2N+2k$}] (Zk) [] {};
      \node[node, label={[below right=5pt and -7pt]:\rotatebox{-30}{\footnotesize $2N+2k$}}] (A1)  [right=6mm of Zk] {};
      \node[node, label={[below right=5pt and -7pt]:\rotatebox{-30}{\footnotesize $2N+2k$}}] (A2) [right=6mm of A1] {};
      \node[node, label={[below right=5pt and -7pt]:\rotatebox{-30}{\footnotesize $2N+2k$}}] (A3) [right=6mm of A2] {};
      \node[node, label={[below right=5pt and -7pt]:\rotatebox{-30}{\footnotesize $2N+2k$}}] (A4) [right=6mm of A3] {};
      \node[node, label={[below right=5pt and -7pt]:\rotatebox{-30}{\footnotesize $2N+2k$}}] (A5) [right=6mm of A4] {};
      \node[node, label={[below right=5pt and -7pt]:\rotatebox{-30}{\footnotesize $2N+2k$}}] (N3) [right=6mm of A5] {};
      \node[node, label={[below right=5pt and -7pt]:\rotatebox{-30}{\footnotesize $N+k$}}] (B4) [right=6mm of N3] {};
      \node[node, label=right:{\footnotesize $N+k$}] (Nu) [above=5mm of N3] {};
      \node[node, label={[below left=5pt and -7pt]:\rotatebox{30}{\footnotesize $2N+2k$}}] (A1s)  [left=6mm of Zk] {};
      \node[node, label={[below left=5pt and -7pt]:\rotatebox{30}{\footnotesize $2N+2k$}}] (A2s) [left=6mm of A1s] {};
      \node[node, label={[below left=5pt and -7pt]:\rotatebox{30}{\footnotesize $2N+2k$}}] (A3s) [left=6mm of A2s] {};
      \node[node, label={[below left=5pt and -7pt]:\rotatebox{30}{\footnotesize $2N+2k$}}] (A4s) [left=6mm of A3s] {};
      \node[node, label={[below left=5pt and -7pt]:\rotatebox{30}{\footnotesize $2N+2k$}}] (A5s) [left=6mm of A4s] {};
      \node[node, label={[below left=5pt and -7pt]:\rotatebox{30}{\footnotesize $2N+2k$}}] (N3s) [left=6mm of A5s] {};
      \node[node, label={[below left=5pt and -7pt]:\rotatebox{30}{\footnotesize $N+k$}}] (B4s) [left=6mm of N3s] {};
      \node[node, label=left:{\footnotesize $N+k$}] (Nus) [above=5mm of N3s] {};

      \draw (Zk.east) -- (A1.west);
      \draw (A1.east) -- (A2.west);
      \draw (A2.east) -- (A3.west);
      \draw (A3.east) -- (A4.west);
      \draw (A4.east) -- (A5.west);
      \draw (A5.east) -- (N3.west);
      \draw (N3.east) -- (B4.west);
      \draw (N3.north) -- (Nu.south); 

    \draw (Zk.west) -- (A1s.east);
      \draw (A1s.west) -- (A2s.east);
      \draw (A2s.west) -- (A3s.east);
      \draw (A3s.west) -- (A4s.east);
      \draw (A4s.west) -- (A5s.east);
      \draw (A5s.west) -- (N3s.east);
      \draw (N3s.west) -- (B4s.east);
      \draw (N3s.north) -- (Nus.south);

        \end{tikzpicture}
    \end{gathered} \,.
\end{equation}

There are however couple of issues one can identify with this theory. As pointed out in \cite{Kapustin:1998fa}, the magnetic quiver in equation \eqref{eqn:infinite_quiver} is the 3d mirror\footnote{Notice that here we properly mean the 3d mirror as the theory is engineered by applying S-duality to a Type IIB D3-D5-NS5-brane system in presence of two $\mathrm{ON}^0$ orientifold planes.} of an $Sp(n)$ 3d $\mathcal{N}=4$ theory with $N+k$ fundamentals and one antisymmetric tensor. This theory is known to be bad in the sense of \cite{Gaiotto:2008ak} as there exist dressed monopole operators that do not respect the unitarity bound. As already noted in Section \ref{sec:sym}, balanced unitary $3d$ $\mathcal{N}=4$ quiver theories shaped as affine Dynkin diagrams have been extensively studied in \cite{Cremonesi:2014xha}, where an over-extension of the quiver was needed in order to correctly realize the reduced moduli space of instantons. Lastly, due to the badness of the theory it is expected that the flavor symmetry read from the quiver via the balance algorithm is a subset of the real flavor symmetry of the LST \cite{DelZotto:2023nrb}.

\section{A Monotonicity Theorem for \texorpdfstring{\boldmath{$\kappa_R$}}{kappaR}} \label{subsec:MonotonyThm}

We have now determined magnetic quivers describing the Higgs branches of the heterotic $E_8 \times E_8$ and heterotic $\mathrm{Spin}(32)/\mathbb{Z}_2$ $\mathbb{C}^2/\mathbb{Z}_K$ orbifold LSTs. As we have reviewed already, it is well-known that 
\begin{equation}\label{eqn:Tdual}
    \mathcal{K}_{N_L, N_R, K}(\varnothing, \varnothing) \quad \xleftrightarrow{\,\, \text{ T-duality } \,\,}  \quad \widetilde{\mathcal{K}}_{N, K}(\varnothing) \,.
\end{equation}
Here, we use $\varnothing$ to denote the trivial homomorphism of $\mathbb{Z}_K$ into either $E_8$ or $\mathrm{Spin}(32)/\mathbb{Z}_2$. To be precise, the T-duality between theories may only exist for a specific pair $(N_L, N_R)$, and for a particular choice of Wilson lines for the $E_8 \times E_8$ flavor symmetry on the left, and the $\mathrm{Spin}(32)/\mathbb{Z}_2$ on the right; this caveat is implicit whenever we discuss T-duality. Often the T-dualities between such unHiggsed theories can be shown either geometrically, or from the T-duality acting on the brane system \cite{Morrison:1996na,Witten:1996qb,Aspinwall:1996mn,Aspinwall:1996mw,Aspinwall:1996nk}. When non-trivial homomorphisms $\rho_L$, $\rho_R$, and $\sigma$ are turned on, it can be difficult to show T-dualities explicitly. Instead, we would like to follow the program of \cite{DelZotto:2020sop,DelZotto:2022ohj}, reviewed in Section \ref{sec:Tdual}, where LSTs are conjectured to be T-dual if their invariants match.

Given that we have derived the magnetic quivers for the Higgs branches of the LSTs appearing in equation \eqref{eqn:Tdual}, it is natural to ask if the invariants change in a predictable way along Higgs branch RG flows. If this is the case, then a T-duality between two parents theories, allows us to conjecture the existence of a T-duality between the relevant descendant theories. Thus, the Hasse diagram of the Higgs branch, which we have determined from the magnetic quivers in Section \ref{sec:LSTs} together with the quiver subtraction algorithm, is a powerful tool for discovering putative T-dual pairs. 

In this section, we focus on the behavior of the mixing coefficient, $\kappa_R$, between the R-symmetry and the $\mathfrak{u}(1)^{(1)}_\text{LST}$ one-form symmetry of the LST. In terms of the geometry at the generic point of the tensor branch, this quantity was defined as
\begin{equation}
    \kappa_R=\sum \limits_{i=1}^{N} \ell_{i} h^\vee_{\mathfrak{g}_i} \,,
\end{equation}
where $i$ labels the tensor multiplets, hence the curves in the LST configuration, $\ell_i$ is the null eigenvector of the intersection matrix $A^{ij}$, and $\mathfrak{g}_i$ is the gauge algebra supported over the $i$th compact curve. In this section, we \emph{do not} assume that we are considering only the $\mathcal{K}_{N_L,N_R,K}(\rho_L, \rho_R)$ LSTs.

Given a tensor branch description of an LST, we first obtain the geometry associated with the LST by shrinking all possible curves in the base.\footnote{Recall that this is not a unique procedure, but, as $\kappa_R$ is unaffected by this non-uniqueness, we do not mention it further in this section.} This gives rise to a elliptically-fibered Calabi Yau, $Y$; a Higgsing is a complex structure deformation of $Y$ to a $Y'$, which can then be blown up to $\widetilde{Y}'$, which yields the tensor branch description of the LST after Higgsing. Often, we can see that the action of the Higgsing modifies the tensor branch geometry $\widetilde{Y}$ in a straightforward manner to attain $\widetilde{Y}'$. Two such options, which appeared in \cite{Baume:2021qho}, are as follows.\footnote{An additional operation appeared in \cite{Baume:2021qho} which was referred to as ``E-string nucleation''. Even though we do not study the behavior of $\kappa_R$ under this operation, as it does not appear in the Hasse diagram of $\mathcal{K}_{N_L,N_R,K}(\rho_L, \rho_R)$ or $\widetilde{\mathcal{K}}_{N,K}(\sigma)$, it can be straightforwardly shown that the nucleation reduces the structure constant value, still leading to monotonic behavior.}
\begin{enumerate}
    \item \textit{Classical Higgsing}: the gauge algebra $\mathfrak{g}_i$ over the compact curve $C_i$ is broken to a gauge algebra $\mathfrak{h}_i$.
    \item \textit{E-string Higgsing}: an undecorated $(-1)$-curve shrinks to zero-volume, and the non-minimal singularity above the point to which the curve is shrunk is smoothed out. 
\end{enumerate}
An elementary Higgs branch transition may also correspond to a sequence of these operations.

We first consider the change of $\kappa_R$ under a Higgs branch RG flow that corresponds on the tensor branch to classical Higgsing. The intersection matrix $A^{ij}$ of curves on the tensor branch is unchanged, and thus the null vector $\ell_i$ is unchanged. Then, it is straightforward how $\kappa_R$ changes under the flow:
\begin{equation}
    \delta \kappa_R = -\ell_i (h_{\mathfrak{g}_i}^\vee - h_{\mathfrak{h}_i}^\vee) \,,
\end{equation}
where $i$ is the curve over which the gauge algebra is broken. Other invariants also change predictably, for example, the 5d Coulomb branch changes dimension as follows:
\begin{equation}
    \delta \operatorname{dim}(\mathcal{C}) = \operatorname{rank}(\mathfrak{h}_i) - \operatorname{rank}(\mathfrak{g}_i) \,.
\end{equation}

We now turn to E-string Higgsing. We first consider the case where the undecorated $(-1)$-curve that we wish to shrink intersects only one other compact curve, i.e., 
\begin{equation}
    1\,\overset{\mathfrak{g}_2}{n} \cdots \,,
\end{equation}
where we have for convenience labelled the $(-1)$-curve as $i = 1$. Let $A^{ij}$ denote the $N \times N$ pairing matrix, and then the null eigenvector satisfies
\begin{equation}\label{eqn:original}
    A^{ij}\ell_j = 0 \,.
\end{equation}
After shrinking the $(-1)$-curve, we get a configuration 
\begin{equation}
    \overset{\mathfrak{g}_2}{(n-1)} \cdots \,,
\end{equation}
and we denote the resulting negative semi-definite intersection matrix as $\widetilde{A}^{ij}$, which is an $(N-1)\times(N-1)$ matrix, as we have removed one curve. We assume that $i,j = 2 \cdots N$. We can see that $\ell_i$, where $i \geq 2$, also constitutes a null eigenvector for $\widetilde{A}^{ij}$. Write
\begin{equation}
    \widetilde{A}^{ij} = A^{ij} + \delta^{i2}\delta^{j2} \,,
\end{equation}
and then
\begin{equation}
    \widetilde{A}^{ij}\ell_j = - A^{i1}\ell_1 + \delta^{i2}\ell_2 = 0 \,,
\end{equation}
where in the last equality, we have used that $\ell_1 - \ell_2 = 0$ via equation \eqref{eqn:original}. Assuming that $\gcd(\ell_2, \cdots, \ell_N)$, we find that the change in the structure constant is
\begin{equation}
    \delta \kappa_R = -\ell_1 \,.
\end{equation}
It is straightforward to see that the same result holds when the $(-1)$-curve intersects an arbitrary number of other curves. Under E-string Higgsing, the change in the 5d Coulomb branch dimension is
\begin{equation}
    \delta \operatorname{dim}(\mathcal{C}) = -1 \,.
\end{equation}

Therefore, under the assumptions on the nature of the Higgs branch RG flows in this section, we can see that $\kappa_R$ and $\operatorname{dim}(\mathcal{C})$ decrease monotonically along the RG flow. 

\section{T-Dualities and Higgs Branch RG Flows}\label{sec:tduality}

We now use the Hasse diagram of the Higgs branch to generate vast collections of LSTs with the same T-dual-invariant properties; in particular those in equation \eqref{eqn:T-dual-invariants}. Often we find multiple LSTs which are all putatively T-dual to each other, which is sometimes referred to as $n$-ality. It is important to note that in this section, we are attempting to utilize the structure of the Higgs branch to find families of LSTs with the same T-duality-invariant properties; this is not sufficient to show that any two such LSTs are, in fact, T-dual. A useful approach to verify T-duality is that of \cite{Aspinwall:1997ye}, where one explicitly constructs the K3-fibered Calabi--Yau engineering the LST in string theory, and show that it admits an inequivalent K3-fibration corresponding to the T-dual LST. For the heterotic LSTs that we consider in this paper, when no non-trivial $E_8$-homomorphism is turned on, this geometric analysis was done in \cite{Aspinwall:1997ye}; for the $K=2$ analysis in Section \ref{sec:K2}, a similar explicit geometric approach was taken to show the T-dualities for certain non-trivial homomorphisms $\mathbb{Z}_2 \rightarrow E_8$ in \cite{DelZotto:2022xrh}.  

\subsection{A Complete \texorpdfstring{$K=2$}{K=2} Analysis}\label{sec:K2}

We give an explicit example of how to trace T-dual pairs from the Hasse diagram in the case of M5-branes probing a $\mathbb{C}^2/\mathbb{Z}_2$ orbifold in the presence of two M9-walls. This procedure reproduces the results depicted in \cite{DelZotto:2022ohj} on dual families for $K=2$, incorporating the shifts in the instanton number $N$.

Consider the heterotic $E_8 \times E_8$ LSTs $\mathcal{K}_{N_L, N_R, 2}(\rho_L, \rho_R)$ specified the embedding $\rho_{L/R}: \mathbb{Z}_2 \rightarrow E_8$. These theories can be obtained via the fusion of orbi-instanton theories, as discussed in Section \ref{sec:HeteroticOrbitfoldLST}:
\begin{equation}
    \mathcal{O}_{N_L,2}(\rho_L, [1^K]) \,\,\text{---}\,\,
    \overset{\mathfrak{su}_K}{0} \,\,\text{---}\,\, \mathcal{O}_{N_R,2}(\rho_R, [1^K]) \,.
\end{equation}
There are precisely three homomorphisms $\mathbb{Z}_2 \rightarrow E_8$, which we denote as
\begin{equation}
    1 + 1 \,, \qquad 2 \,, \qquad 2' \,,
\end{equation}
using the affine $E_8$ weighted partition discussed around equation \eqref{eqn:sumKac}. The rank one orbi-instantons Higgsed by each of these homomorphisms have the following tensor branch descriptions:
\begin{equation}
    \begin{aligned}
        \mathcal{O}_{1,2}(1+1)&= \ [\mathfrak{e}_8] \ 1 \ 2 \ \stackon{2}{$\mathfrak{su}_{2}$} [\mathfrak{g}_2] \,, \\
        \mathcal{O}_{1,2}(2)&= \ [\mathfrak{e}_7] \ 1 \ \stackon{2}{$\mathfrak{su}_{2}$} [\mathfrak{so}(7)] \,, \\
        \mathcal{O}_{1,2}(2')&= \ [\mathfrak{so}(16)] \ \stackon{1}{$\mathfrak{sp}_{1}$} [\mathfrak{so}(4)] \,,
    \end{aligned}
\end{equation}
where the extension to the rank $N$ orbi-instanton follows immediately from equation \eqref{eqn:AOIHiggsed}. The contribution to the $\kappa_R$ of the LST from each of these rank one orbi-instanton theories can also be worked out, and we find
\begin{equation}
    \delta \kappa_R(1+1) = 4 \,, \qquad \delta \kappa_R(2) = 3 \,, \qquad \delta \kappa_R(2') = 2 \,.
\end{equation}

We now construct the Higgs branch Hasse diagram for the LST $\mathcal{K}_{N_L, N_R, 2}(1+1, 1+1)$ using the quiver subtraction algorithm in Algorithm \ref{alg:QS}. We can then reverse engineer the LSTs belonging to each vertex in the Hasse diagram, as we did in Section \ref{sec:Higgs_Branch_OI}. We depict the resulting Hasse diagram in Figure \ref{fig:HasseSU2E8E8}. We note that we only draw the subset of the Hasse diagram corresponding to the vertices that are $K=2$ LSTs; the full Hasse diagram of the Higgs branch of $\mathcal{K}_{N_L, N_R, 2}(1+1, 1+1)$ also contains vertices where $K < 2$.

\begin{figure}[p]
    \centering
    \resizebox{!}{0.45\textheight}{
    \begin{tikzpicture}
        \node[draw] at (0,0) (nodo0) {$\mathcal{K}_{N_L, N_R, 2} (1+1,1+1)$};
        \node[draw] at (0,-3) (nodo1) {$\mathcal{K}_{N_L, N_R, 2} (2,1+1)$};
\draw (nodo0)--(nodo1) node[midway, right] () {$e_{8}$};
        \node[draw] at (6,-6) (nodo2r) {$\mathcal{K}_{N_L, N_R, 2} (2,2)$};
        \node[draw] at (-6,-4.5) (nodo2ll) {$\mathcal{K}_{N_L-1, N_R, 2} (1+1,1+1)$};
        \node[draw] at (0,-6) (nodo2l) {$\mathcal{K}_{N_L, N_R, 2} (2',1+1)$};
\draw (nodo1)--(nodo2r) node[midway, right=0.5cm] () {$e_{8}$};
\draw[dashed] (nodo1)--(nodo2ll) node[midway, above left] () {$A_{1}$};
\draw (nodo1)--(nodo2l) node[midway, left=0.5cm] () {$e_{7}$};
        \node[draw] at (6,-9) (nodo3r) {$\mathcal{K}_{N_L, N_R, 2} (2',2)$};
        \node[draw] at (0,-9) (nodo3l) {$\mathcal{K}_{N_L-1, N_R, 2} (2,1+1)$};
\draw (nodo2r)--(nodo3r) node[midway, right=0.5cm] () {$e_{7}$};
\draw (nodo2l)--(nodo3r) node[midway, above left=0.5cm] () {$e_{8}$};  
\draw[dashed] (nodo2l)--(nodo3l) node[midway, left=0.5cm] () {$d_{8}$};  
\draw (nodo2ll)--(nodo3l) node[midway, left=0.5cm] () {$e_{8}$};
\draw[dashed] (nodo2r)--(nodo3l) node[midway, below left=0.5cm] () {$A_{1}$};  
        \node[draw] at (6,-12) (nodo4r) {$\mathcal{K}_{N_L-1, N_R, 2} (2',1+1)$};
        \node[draw] at (0,-12) (nodo4l) {$\mathcal{K}_{N_L-1, N_R, 2} (2,2)$};
        \node[draw] at (12,-12) (nodo4rr) {$\mathcal{K}_{N_L, N_R, 2} (2',2')$};
\draw(nodo3l)--(nodo4r) node[midway, above left=0.5cm] () {$e_{7}$};
\draw(nodo3r)--(nodo4rr) node[midway, above right=0.5cm] () {$e_{7}$};
\draw (nodo3l)--(nodo4l) node[midway, left] () {$e_{8}$};
\draw[dashed] (nodo3r)--(nodo4r) node[midway, right] () {$A_{1}$};
\draw[dashed] (nodo3r)--(nodo4l) node[midway, below left=0.5cm] () {$d_{8}$};
        \node[draw] at (12,-15) (nodo5rr) {$\mathcal{K}_{N_L-1, N_R, 2} (2',2)$};
        \node[] at (0,-13) (nodoB) {$\mathrm{etc.}$};
        \node[] at (6,-13) (nodoC) {$\mathrm{etc.}$};
        \node[] at (12,-16) (nodoD) {$\mathrm{etc.}$};
        \node[] at (14,-16) (nodoD1) {$\mathrm{etc.}$};
        \node[] at (10,-16) (nodoD2) {$\mathrm{etc.}$};
        \node[] at (-2,-10) (nodoF) {$\mathrm{etc.}$};
        \node[] at (-2,-13) (nodoG) {$\mathrm{etc.}$};
\draw (nodo4r)--(nodo5rr) node[midway, above right] () {$e_{8}$};
\draw[dashed] (nodo4rr)--(nodo5rr) node[midway, right] () {$d_{8}$};

\draw (nodo4l)--(nodoB);
\draw[dashed] (nodo4l)--(nodoG);
\draw[dashed] (nodo4r)--(nodoC);
\draw (nodo5rr)--(nodoD1);
\draw[dashed] (nodo5rr)--(nodoD2);
\draw[dashed] (nodo5rr)--(nodoD);
\draw[dashed] (nodo3l)--(nodoF);
    \end{tikzpicture}
    }
    \caption{Heterotic $E_8 \times E_8$ $\mathcal{K}_{N_L, N_R, 2} (\rho_L,\rho_R)$ LST Hasse diagram. The diagram is ordered by Higgs branch dimension thus the top of this Hasse diagram corresponds to the bottom point of the Hasse diagram engineered via quiver subtraction. Dashed slices connect theories with different values of $N_L$ and $N_R$. Dashed transitions affect only $N_L$ by choice, in general, they can affect either $N_L$ or $N_R$ as long as each of them remains non-negative.}
    \label{fig:HasseSU2E8E8}
    \vspace{1cm}
    \resizebox{0.85\textwidth}{!}{
    \begin{tikzpicture}
        \node[draw] at (0,0) (nodo0) {$\widetilde{\mathcal{K}}_{N_L+N_R,2} (2p,16-2p)$};
        \node[draw] at (8,-3) (nodo1r) {$\widetilde{\mathcal{K}}_{N_L+N_R,2} \big( 2(p+1),16-2(p+1) \big)$};
        \node[draw] at (-8,-3) (nodo1l) {$\widetilde{\mathcal{K}}_{N_L+N_R-1,2} \big( 2(p-1),16-2(p-1) \big)$};
\draw[dashed] (nodo0)--(nodo1l) node[midway, left=0.5cm] () {$d_{2p}$};
\draw (nodo0)--(nodo1r) node[midway, right=0.5cm] () {$d_{16-2p}$};
\node[draw] at (0,-6) (nodo2) {$\widetilde{\mathcal{K}}_{N_L+N_R-1,2} (2p,16-2p)$};
        \node[] at (-8,-7) (nodo2l) {$\mathrm{etc.}$};
        \node[] at (-4,-7) (nodo2cl) {$\mathrm{etc.}$};
        \node[] at (4,-7) (nodo2cr) {$\mathrm{etc.}$};
        \node[] at (8,-7) (nodo2r) {$\mathrm{etc.}$};
\draw[dashed] (nodo1r)--(nodo2) node[midway, right=0.5cm] () {$d_{2(p+1)}$};
\draw (nodo1l)--(nodo2) node[midway, left=0.5cm] () {$d_{16-2(p+1)}$};      
\draw (nodo1r)--(nodo2r) node[midway, right=0.5cm] () {$d_{16-2(p-1)}$};
\draw[dashed] (nodo1l)--(nodo2l) node[midway, left=0.5cm] () {$d_{2(p-1)}$}; 
\draw[dashed] (nodo2)--(nodo2cl) node[midway, above left] () {$d_{2p}$};
\draw (nodo2)--(nodo2cr) node[midway, above right] () {$d_{16-2p}$};

    \end{tikzpicture}
    }
    \caption{A subdiagram of the heterotic $\mathrm{Spin}\left(32\right)/ \mathbb{Z}_2$ $\widetilde{\mathcal{K}}_{N_L+N_R,2} (2p,16-2p)$ LST Hasse diagram. The diagram is ordered by Higgs branch dimension thus the top of this Hasse diagram corresponds to the bottom point of the Hasse diagram engineered via quiver subtraction. Dashed slices connect theories with different values of $N = N_L + N_R$.}
    \label{fig:HasseSU2SO32}
\end{figure}

We can see that the rank of the flavor algebra is constant across this Hasse diagram, and it is straightforward using the results in Section \ref{subsec:MonotonyThm} to see that $\kappa_R$ and $\operatorname{dim}(\mathcal{C})$ both decrease by one along every elementary transition. Therefore, we can see that each LST at the same depth in the Hasse diagram has the same T-duality invariants! In particular, just from Figure \ref{fig:HasseSU2E8E8}, we could conjecture the existence of T-dualities between the following four LSTs:
\begin{equation}
        \mathcal{K}_{N_L,N_R,2}(1+1, 1+1)\,, \,\, \mathcal{K}_{N_L+1,N_R,2}(2,2)\,, \,\, \mathcal{K}_{N_L+1,N_R,2}(2',1+1)\,, \,\, \mathcal{K}_{N_L+2,N_R,2}(2',2') \,,
\end{equation}
and also between the two theories:
\begin{equation}
    \mathcal{K}_{N_L,N_R,2}(2,1+1)\,, \quad \mathcal{K}_{N_L+1,N_R,2}(2',2) \,.
\end{equation}
We summarize these T-dualities in Table \ref{tbl:K2Tduals}. In the subdiagram of the Hasse diagram that we have drawn in Figure \ref{fig:HasseSU2E8E8}, we have preferably drawn transitions decreasing $N_L$ when we have options to decrease either $N_L$ or $N_R$; thus, when we write, e.g., $\mathcal{K}_{N_L+2,N_R,2}(2',2')$, in Table \ref{tbl:K2Tduals}, this is really a shorthand for the three theories
\begin{equation}
    \mathcal{K}_{N_L+2,N_R,2}(2',2') \,, \quad \mathcal{K}_{N_L+1,N_R+1,2}(2',2') \,, \quad \mathcal{K}_{N_L,N_R+2,2}(2',2') \,.
\end{equation}
Of course, it remains to rigorously demonstrate the existence of T-dualities between the $E_8 \times E_8$ heterotic $\mathbb{C}^2/\mathbb{Z}_2$ orbifold LSTs in Table \ref{tbl:K2Tduals} by a careful analysis of the geometry engineering them. 

\begin{table}[t]
    \centering
    \begin{threeparttable}
    \begin{tabular}{ccc}
        \toprule
        Family & $E_8 \times E_8$ & $\mathrm{Spin}(32)/\mathbb{Z}_2$ \\\midrule
        $0$ & $\begin{gathered} \mathcal{K}_{N_L,N_R,2}(1+1, 1+1) \\ 
        \mathcal{K}_{N_L+1,N_R,2}(2,2) \\ \mathcal{K}_{N_L+1,N_R,2}(2',1+1) \\ \mathcal{K}_{N_L+2,N_R,2}(2',2') 
        \end{gathered}$ & 
        $\begin{gathered} \widetilde{\mathcal{K}}_{N_L+N_R,2}(0, 16) \\ 
        \widetilde{\mathcal{K}}_{N_L+N_R+1,2}(4,12) \\ 
        \widetilde{\mathcal{K}}_{N_L+N_R+2,2}(8,8) 
        \end{gathered}$ \\\midrule
        $1$ & $\begin{gathered} \mathcal{K}_{N_L,N_R,2}(2,1+1) \\ \mathcal{K}_{N_L+1,N_2,K}(2',2)
        \end{gathered}$ & 
        $\begin{gathered} \widetilde{\mathcal{K}}_{N_L+N_R,2}(2,14) \\ 
        \widetilde{\mathcal{K}}_{N_L+N_R+1,2}(6,10)
        \end{gathered}$ \\\bottomrule
    \end{tabular}
    \end{threeparttable}
    \caption{Putative T-dual pairs among the $\mathbb{C}^2/\mathbb{Z}_2$ $E_8 \times E_8$ and $\mathrm{Spin}(32)/\mathbb{Z}_2$ heterotic orbifold LSTs. The family is fixed by the value of $\kappa_R \operatorname{ mod } 2$.}
    \label{tbl:K2Tduals}
\end{table}

We now turn to the $\mathrm{Spin}\left(32\right)/ \mathbb{Z}_2$ heterotic $\mathbb{C}^2/\mathbb{Z}_2$ orbifold LSTs. In terms of the F-theory curve configuration, these theories have the following form
\begin{equation}
    \widetilde{\mathcal{K}}_{N,2} (2p,16-2p) = \  [SO(4p)] \ \stackon{1}{$\mathfrak{sp}_{N}$} \ \stackon{1}{$\mathfrak{sp}_{N+4-p}$} \ [SO(32-4p)] \,,
\end{equation}
where $p$ is the integer, in the range $[0, \cdots, 4]$, specifying the homomorphism $\mathbb{Z}_2 \rightarrow \mathrm{Spin}\left(32\right)/ \mathbb{Z}_2$, as explained in \cite{DelZotto:2022ohj}. We can now perform a similar analysis of the Hasse diagram of the Higgs branch for the $\widetilde{\mathcal{K}}_{N,K}(2p, 16-2p)$ LST, using the quiver subtraction algorithm and the magnetic quiver worked out in Section \ref{sec:spin32}. Reverse engineering the particular LST at each vertex, we find the Hasse diagram depicted in Figure \ref{fig:HasseSU2SO32}. Note that $\kappa_R$ and $\operatorname{dim}(\mathcal{C})$ both decrease by one when moving along an elementary slice in the Hasse diagram in Figure \ref{fig:HasseSU2SO32}.

Now, it is a fact well-established from geometry that the heterotic orbifold LSTs where the boundary conditions for the heterotic gauge bundles are trivial, 
\begin{equation}
    \mathcal{K}_{N_L,N_R,2}(1+1, 1+1) \qquad \text{ and } \qquad \widetilde{\mathcal{K}}_{N_L+N_R,2}(0, 16) \,,
\end{equation}
have identical T-dual-invariant properties.\footnote{Though, we emphasize once again that we expect a T-duality only to exist for certain pairs $(N_L, N_R)$.} Therefore, we can use the fact that we know both Hasse diagrams and that we know how the T-duality-invariant properties are modified along the slices in the Hasse diagram, to determine potential T-dual theories associated with non-trivial homomorphisms $\rho_L$, $\rho_R$, and $\sigma$. In the end, we find that all the heterotic $\mathbb{C}^2/\mathbb{Z}_2$ LSTs in the same family in Table \ref{tbl:K2Tduals} have the same T-duality-invariant properties, and thus are prospective T-duals.

\subsection{A Complete \texorpdfstring{$K=3$}{K=3} Analysis}\label{sec:keq3}

The number of potential T-dual families grows with $K$, the order of the orbifold. In this section, we study the $\mathbb{C}^2/\mathbb{Z}_3$ orbifold LSTs: $\mathcal{K}_{N_L, N_R, 3}(\rho_L, \rho_R)$ and $\widetilde{\mathcal{K}}_{N, 3}(\sigma)$. We find three families, each containing a large number of potential T-dual theories, summarized in Table \ref{tbl:K3}. This example illustrates the strength of the Hasse diagram for the determination of LSTs with the same T-duality-invariant properties.

\begin{table}[t]
    \centering
    \begin{threeparttable}
    \begin{tabular}{ccc}
        \toprule
        Family & $E_8 \times E_8$ & $\mathrm{Spin}(32)/\mathbb{Z}_2$ \\\midrule
        $0$ & $\begin{gathered} \mathcal{K}_{N_L,N_R,3}(2'+1, 1+1+1) \\ 
        \mathcal{K}_{N_L,N_R,3}(2+1, 2+1) \\ 
        \mathcal{K}_{N_L+1,N_R,3}(3', 2+1) \\ 
        \mathcal{K}_{N_L+1,N_R,3}(3, 2'+1) \\ 
        \mathcal{K}_{N_L+2,N_R,3}(3', 3')
        \end{gathered}$ & 
        $\begin{gathered} 
        \widetilde{\mathcal{K}}_{N-6,3}(0, 16) \\ 
        \widetilde{\mathcal{K}}_{N-5,3}(3, 13) \\ 
        \widetilde{\mathcal{K}}_{N-4,3}(6, 10) \\
        \widetilde{\mathcal{K}}_{N-3,3}(9,7) \\ 
        \widetilde{\mathcal{K}}_{N-2,3}(12, 4) \\ 
        \widetilde{\mathcal{K}}_{N-1,3}(15, 1) 
        \end{gathered}$ \\\midrule
        $1$ & $\begin{gathered} \mathcal{K}_{N_L,N_R,3}(2+1, 1+1+1) \\ 
        \mathcal{K}_{N_L+1,N_R,3}(3', 1+1+1) \\ 
        \mathcal{K}_{N_L+1,N_R,3}(3, 2+1) \\ 
        \mathcal{K}_{N_L+1,N_R,3}(2'+1, 2'+1) \\ 
        \mathcal{K}_{N_L+2,N_R,3}(3', 3)
        \end{gathered}$ & 
        $\begin{gathered}
        \widetilde{\mathcal{K}}_{N-5,3}(2, 14) \\ 
        \widetilde{\mathcal{K}}_{N-4,3}(5, 11) \\
        \widetilde{\mathcal{K}}_{N-3,3}(8,8) \\ 
        \widetilde{\mathcal{K}}_{N-2,3}(11, 5) \\ 
        \widetilde{\mathcal{K}}_{N-1,3}(14, 2) 
        \end{gathered}$ \\\midrule
        $2$ & $\begin{gathered} 
        \mathcal{K}_{N_L,N_R,3}(1+1+1, 1+1+1) \\ 
        \mathcal{K}_{N_L+1,N_R,3}(3, 1+1+1) \\ 
        \mathcal{K}_{N_L+1,N_R,3}(2'+1, 2+1) \\ 
        \mathcal{K}_{N_L+2,N_R,3}(3', 2'+1) \\ 
        \mathcal{K}_{N_L+2,N_R,3}(3, 3)
        \end{gathered}$ & 
        $\begin{gathered}
        \widetilde{\mathcal{K}}_{N-5,3}(1, 15) \\
        \widetilde{\mathcal{K}}_{N-4,3}(4,12) \\ 
        \widetilde{\mathcal{K}}_{N-3,3}(7,9) \\
        \widetilde{\mathcal{K}}_{N-2,3}(10,6) \\ 
        \widetilde{\mathcal{K}}_{N-1,3}(13, 3) \\ 
        \widetilde{\mathcal{K}}_{N,3}(16,0) 
        \end{gathered}$ \\\bottomrule
    \end{tabular}
    \end{threeparttable}
    \caption{Putative T-dual pairs among the $\mathbb{C}^2/\mathbb{Z}_3$ $E_8 \times E_8$ and $\mathrm{Spin}(32)/\mathbb{Z}_2$ heterotic orbifold LSTs. The family is fixed by the value of $\kappa_R \operatorname{ mod } 3$. Note that $N = N_L + N_R + 6$.}
    \label{tbl:K3}
\end{table}

We begin with the heterotic $E_8 \times E_8$ LSTs. First, we note that there are five different homomorphisms $\mathbb{Z}_3 \rightarrow E_8$; we list these homomorphisms, together with the associated tensor branch description of the Higgsed rank one orbi-instanton in Table \ref{tab:OIk=3}. We also include in the table the contribution of that rank one orbi-instanton to $\kappa_R$ of the LST, and the magnetic quiver. Note that the contribution from each rank one orbi-instanton theory to the rank of the flavor algebra of the LST is $8$ in all five cases.\footnote{One must carefully count the ABJ-anomaly free $\mathfrak{u}(1)$ factors; see \cite{Apruzzi:2020eqi} for more details.} Given the magnetic quivers, we are ready to perform the Coulomb gauging of the $\mathfrak{su}(3)$ Coulomb symmetry, to obtain the magnetic quivers for the LSTs associated with the fifteen different combinations of $\rho_L$ and $\rho_R$. To be explicit, we have written each of these magnetic quivers in Tables \ref{tab:1E8LSTk=3}, \ref{tab:2E8LSTk=3}, and \ref{tab:3E8LSTk=3}.

\begin{table}[t]
    \centering
    \begin{tabular}{c c c }\toprule
       \footnotesize$ \mathcal{O}_{1,3}\left( \rho \right)$  & $\delta \kappa_R$  & Magnetic Quiver\\ \midrule
        \footnotesize$ \mathcal{O}_{1,3}\left( 1+1+1 \right):= [\mathfrak{e}_8]\stackon{$1$}{}\stackon{$2$}{}\stackon{$2$}{$\mathfrak{su}_2$}\stackon{$2$}{$\mathfrak{su}_3$}[\mathfrak{su}(4)]$ & $7$ & \resizebox{0.45\textwidth}{!}{$\begin{gathered}
    \begin{tikzpicture}
      \node[node, label=below:{\footnotesize $1$}] (Z1)  {};
      \node[node, label=below:{\footnotesize $2$}] (Z2) [right=6mm of Z1] {};
      \node[node, label=below:{\footnotesize $3$}] (Zk) [right=6mm of Z2] {};
      \node[node, label=below:{\footnotesize $4$}] (A1)  [right=6mm of Zk] {};
      \node[node, label=below:{\footnotesize $8$}] (A2) [right=6mm of A1] {};
      \node[node, label=below:{\footnotesize $12$}] (A3) [right=6mm of A2] {};
      \node[node, label=below:{\footnotesize $16$}] (A4) [right=6mm of A3] {};
      \node[node, label=below:{\footnotesize $20$}] (A5) [right=6mm of A4] {};
      \node[node, label=below:{\footnotesize $24$}] (N3) [right=6mm of A5] {};
      \node[node, label=below:{\footnotesize $16$}] (B4) [right=6mm of N3] {};
      \node[node, label=below:{\footnotesize $8$}] (B2) [right=6mm of B4] {};
      \node[node, label=right:{\footnotesize $12$}] (Nu) [above=5mm of N3] {};
      \draw (Z1.east) -- (Z2.west);
      \draw (Z2.east) -- (Zk.west);
      \draw (Zk.east) -- (A1.west);
      \draw (A1.east) -- (A2.west);
      \draw (A2.east) -- (A3.west);
      \draw (A3.east) -- (A4.west);
      \draw (A4.east) -- (A5.west);
      \draw (A5.east) -- (N3.west);
      \draw (N3.east) -- (B4.west);
      \draw (B4.east) -- (B2.west);
      \draw (N3.north) -- (Nu.south); 

    \end{tikzpicture}
  \end{gathered}$ } \\ 
        \footnotesize$ \mathcal{O}_{1,3}\left( 2+1 \right):= [\mathfrak{e}_7]\stackon{$1$}{}\stackon{$2$}{$\mathfrak{su}_2$}\stackon{$2$}{$\mathfrak{su}_3$}[\mathfrak{su}(4)]$ & $6$ & \resizebox{0.45\textwidth}{!}{$\begin{gathered}
    \begin{tikzpicture}
      \node[node, label=below:{\footnotesize $1$}] (Z1)  {};
      \node[node, label=below:{\footnotesize $2$}] (Z2) [right=6mm of Z1] {};
      \node[node, label=below:{\footnotesize $3$}] (Zk) [right=6mm of Z2] {};
      \node[node, label=below:{\footnotesize $4$}] (A1)  [right=6mm of Zk] {};
      \node[node, label=below:{\footnotesize $6$}] (A2) [right=6mm of A1] {};
      \node[node, label=below:{\footnotesize $9$}] (A3) [right=6mm of A2] {};
      \node[node, label=below:{\footnotesize $12$}] (A4) [right=6mm of A3] {};
      \node[node, label=below:{\footnotesize $15$}] (A5) [right=6mm of A4] {};
      \node[node, label=below:{\footnotesize $18$}] (N3) [right=6mm of A5] {};
      \node[node, label=below:{\footnotesize $12$}] (B4) [right=6mm of N3] {};
      \node[node, label=below:{\footnotesize $6$}] (B2) [right=6mm of B4] {};
      \node[node, label=right:{\footnotesize $9$}] (Nu) [above=5mm of N3] {};
      \draw (Z1.east) -- (Z2.west);
      \draw (Z2.east) -- (Zk.west);
      \draw (Zk.east) -- (A1.west);
      \draw (A1.east) -- (A2.west);
      \draw (A2.east) -- (A3.west);
      \draw (A3.east) -- (A4.west);
      \draw (A4.east) -- (A5.west);
      \draw (A5.east) -- (N3.west);
      \draw (N3.east) -- (B4.west);
      \draw (B4.east) -- (B2.west);
      \draw (N3.north) -- (Nu.south); 

    \end{tikzpicture}
  \end{gathered}$} \\ 
        \footnotesize$ \mathcal{O}_{1,3}\left( 2'+1 \right):= [\mathfrak{so}(14)]\stackon{$1$}{$\mathfrak{sp}_1$}\stackon{$2$}{$\mathfrak{su}_3$}[\mathfrak{su}(4)]$ & $5$ & \resizebox{0.45\textwidth}{!}{$\begin{gathered}
    \begin{tikzpicture}
      \node[node, label=below:{\footnotesize $1$}] (Z1)  {};
      \node[node, label=below:{\footnotesize $2$}] (Z2) [right=6mm of Z1] {};
      \node[node, label=below:{\footnotesize $3$}] (Zk) [right=6mm of Z2] {};
      \node[node, label=below:{\footnotesize $4$}] (A1)  [right=6mm of Zk] {};
      \node[node, label=below:{\footnotesize $6$}] (A2) [right=6mm of A1] {};
      \node[node, label=below:{\footnotesize $8$}] (A3) [right=6mm of A2] {};
      \node[node, label=below:{\footnotesize $10$}] (A4) [right=6mm of A3] {};
      \node[node, label=below:{\footnotesize $12$}] (A5) [right=6mm of A4] {};
      \node[node, label=below:{\footnotesize $14$}] (N3) [right=6mm of A5] {};
      \node[node, label=below:{\footnotesize $9$}] (B4) [right=6mm of N3] {};
      \node[node, label=below:{\footnotesize $4$}] (B2) [right=6mm of B4] {};
      \node[node, label=right:{\footnotesize $7$}] (Nu) [above=5mm of N3] {};
      \draw (Z1.east) -- (Z2.west);
      \draw (Z2.east) -- (Zk.west);
      \draw (Zk.east) -- (A1.west);
      \draw (A1.east) -- (A2.west);
      \draw (A2.east) -- (A3.west);
      \draw (A3.east) -- (A4.west);
      \draw (A4.east) -- (A5.west);
      \draw (A5.east) -- (N3.west);
      \draw (N3.east) -- (B4.west);
      \draw (B4.east) -- (B2.west);
      \draw (N3.north) -- (Nu.south); 

    \end{tikzpicture}
  \end{gathered}$} \\ 
        \footnotesize$ \mathcal{O}_{1,3}\left( 3 \right):= [\mathfrak{e}_6]\stackon{$1$}{}\stackon{$2$}{$\mathfrak{su}_3$}[\mathfrak{su}(6)]$ & $4$ & \resizebox{0.45\textwidth}{!}{$\begin{gathered}
    \begin{tikzpicture}
      \node[node, label=below:{\footnotesize $1$}] (Z1)  {};
      \node[node, label=below:{\footnotesize $2$}] (Z2) [right=6mm of Z1] {};
      \node[node, label=below:{\footnotesize $3$}] (Zk) [right=6mm of Z2] {};
      \node[node, label=below:{\footnotesize $4$}] (A1)  [right=6mm of Zk] {};
      \node[node, label=below:{\footnotesize $5$}] (A2) [right=6mm of A1] {};
      \node[node, label=below:{\footnotesize $6$}] (A3) [right=6mm of A2] {};
      \node[node, label=below:{\footnotesize $8$}] (A4) [right=6mm of A3] {};
      \node[node, label=below:{\footnotesize $10$}] (A5) [right=6mm of A4] {};
      \node[node, label=below:{\footnotesize $12$}] (N3) [right=6mm of A5] {};
      \node[node, label=below:{\footnotesize $8$}] (B4) [right=6mm of N3] {};
      \node[node, label=below:{\footnotesize $4$}] (B2) [right=6mm of B4] {};
      \node[node, label=right:{\footnotesize $6$}] (Nu) [above=5mm of N3] {};
      \draw (Z1.east) -- (Z2.west);
      \draw (Z2.east) -- (Zk.west);
      \draw (Zk.east) -- (A1.west);
      \draw (A1.east) -- (A2.west);
      \draw (A2.east) -- (A3.west);
      \draw (A3.east) -- (A4.west);
      \draw (A4.east) -- (A5.west);
      \draw (A5.east) -- (N3.west);
      \draw (N3.east) -- (B4.west);
      \draw (B4.east) -- (B2.west);
      \draw (N3.north) -- (Nu.south); 

    \end{tikzpicture}
  \end{gathered}$} \\ 
        \footnotesize$ \mathcal{O}_{1,3}\left( 3' \right):=[\mathfrak{su}(9)]\stackon{$1$}{$\mathfrak{su}_3$}[\mathfrak{su}(3)]$ & $3$ &\resizebox{0.45\textwidth}{!}{ $\begin{gathered}
    \begin{tikzpicture}
      \node[node, label=below:{\footnotesize $1$}] (Z1)  {};
      \node[node, label=below:{\footnotesize $2$}] (Z2) [right=6mm of Z1] {};
      \node[node, label=below:{\footnotesize $3$}] (Zk) [right=6mm of Z2] {};
      \node[node, label=below:{\footnotesize $4$}] (A1)  [right=6mm of Zk] {};
      \node[node, label=below:{\footnotesize $5$}] (A2) [right=6mm of A1] {};
      \node[node, label=below:{\footnotesize $6$}] (A3) [right=6mm of A2] {};
      \node[node, label=below:{\footnotesize $7$}] (A4) [right=6mm of A3] {};
      \node[node, label=below:{\footnotesize $8$}] (A5) [right=6mm of A4] {};
      \node[node, label=below:{\footnotesize $9$}] (N3) [right=6mm of A5] {};
      \node[node, label=below:{\footnotesize $6$}] (B4) [right=6mm of N3] {};
      \node[node, label=below:{\footnotesize $3$}] (B2) [right=6mm of B4] {};
      \node[node, label=right:{\footnotesize $4$}] (Nu) [above=5mm of N3] {};
      \draw (Z1.east) -- (Z2.west);
      \draw (Z2.east) -- (Zk.west);
      \draw (Zk.east) -- (A1.west);
      \draw (A1.east) -- (A2.west);
      \draw (A2.east) -- (A3.west);
      \draw (A3.east) -- (A4.west);
      \draw (A4.east) -- (A5.west);
      \draw (A5.east) -- (N3.west);
      \draw (N3.east) -- (B4.west);
      \draw (B4.east) -- (B2.west);
      \draw (N3.north) -- (Nu.south); 

    \end{tikzpicture}
  \end{gathered}$ }\\ \bottomrule
    \end{tabular}
    \caption{We write the tensor branch descriptions, the contribution to $\kappa_R$, and the magnetic quiver for the rank one orbi-instanton theories $\mathcal{O}_{1,3}(\rho, [1^K])$. We drop the $[1^K]$ for ease of notation, and we label the $E_8$-homomorphism by the affine $E_8$ Dynkin labels.}
    \label{tab:OIk=3}
\end{table}

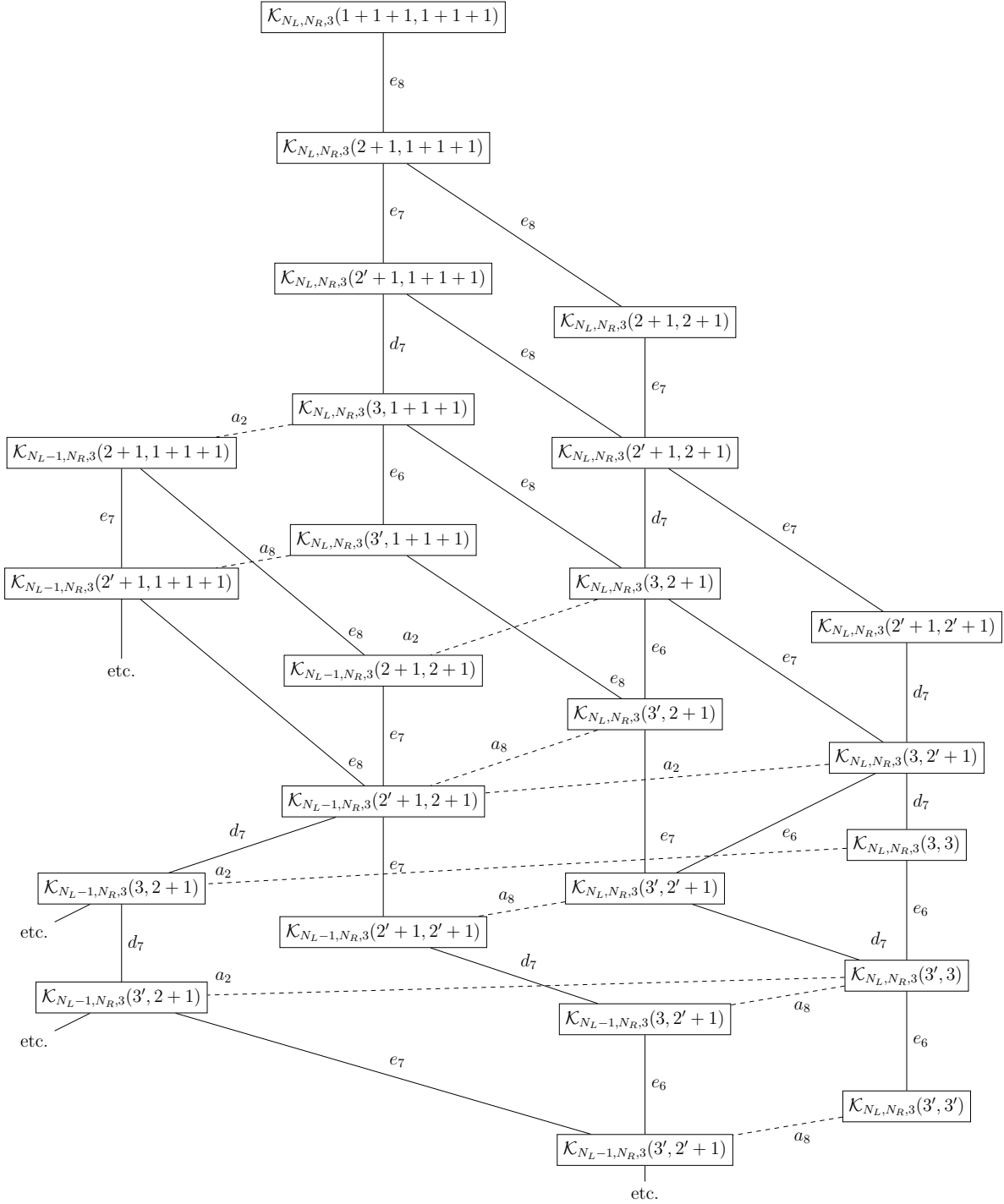
\begin{figure}[p]
    \centering
    \resizebox{!}{\textheight-3cm}{
    \begin{tikzpicture}
        \node[draw] at (0,0) (nodo0) {$\mathcal{K}_{N_L,N_R,3} (1+1+1,1+1+1)$};
        \node[draw] at (0,-3) (nodo1) {$\mathcal{K}_{N_L,N_R,3} (2+1,1+1+1)$};
\draw (nodo0)--(nodo1) node[midway, right] () {$e_{8}$};
        \node[draw] at (0,-6) (nodo2) {$\mathcal{K}_{N_L,N_R,3} (2'+1,1+1+1)$};
        \node[draw] at (6,-7) (nodo2r) {$\mathcal{K}_{N_L,N_R,3} (2+1,2+1)$};
\draw (nodo1)--(nodo2) node[midway, right] () {$e_{7}$};
\draw (nodo1)--(nodo2r) node[midway, above right] () {$e_{8}$};
        \node[draw] at (0,-9) (nodo3) {$\mathcal{K}_{N_L,N_R,3} (3,1+1+1)$};
        \node[draw] at (6,-10) (nodo3r) {$\mathcal{K}_{N_L,N_R,3} (2'+1,2+1)$};
\draw (nodo2)--(nodo3) node[midway, right] () {$d_{7}$};
\draw (nodo2)--(nodo3r) node[midway, above right] () {$e_{8}$};
\draw (nodo2r)--(nodo3r) node[midway, right] () {$e_{7}$};
        \node[draw] at (-6,-10) (nodo4l) {$\mathcal{K}_{N_L-1,N_R,3} (2+1,1+1+1)$};
        \node[draw] at (0,-12) (nodo4) {$\mathcal{K}_{N_L,N_R,3} (3',1+1+1)$};
        \node[draw] at (6,-13) (nodo4r) {$\mathcal{K}_{N_L,N_R,3} (3,2+1)$};
        \node[draw] at (12,-14) (nodo4rr) {$\mathcal{K}_{N_L,N_R,3} (2'+1,2'+1)$};
\draw (nodo3)--(nodo4) node[midway, right] () {$e_{6}$};
\draw (nodo3)--(nodo4r) node[midway, above right] () {$e_{8}$};
\draw[dashed] (nodo3)--(nodo4l) node[midway, above left] () {$a_{2}$};
\draw (nodo3r)--(nodo4r) node[midway, right] () {$d_{7}$};
\draw (nodo3r)--(nodo4rr) node[midway,above right] () {$e_{7}$};
        \node[draw] at (-6,-13) (nodo5l) {$\mathcal{K}_{N_L-1,N_R,3} (2'+1,1+1+1)$};
        \node[draw] at (0,-15) (nodo5) {$\mathcal{K}_{N_L-1,N_R,3} (2+1,2+1)$};
        \node[draw] at (6,-16) (nodo5r) {$\mathcal{K}_{N_L,N_R,3} (3',2+1)$};
        \node[draw] at (12,-17) (nodo5rr) {$\mathcal{K}_{N_L,N_R,3} (3,2'+1)$};
\draw[dashed] (nodo4)--(nodo5l) node[midway, above right] () {$a_{8}$};
\draw[dashed] (nodo4r)--(nodo5) node[above right=5mm and 3mm] () {$a_{2}$};
\draw (nodo4)--(nodo5r) node[above left=5mm and 3mm] () {$e_{8}$};
\draw (nodo4r)--(nodo5r) node[midway, right] () {$e_{6}$};
\draw (nodo4r)--(nodo5rr) node[midway,above right] () {$e_{7}$};
\draw (nodo4l)--(nodo5l) node[midway, left] () {$e_{7}$};
\draw (nodo4l)--(nodo5) node[above left=6mm and 3mm] () {$e_{8}$};
\draw (nodo4rr)--(nodo5rr) node[midway, right] () {$d_{7}$};
        \node at (-6,-15) (nodo6l) {$\mathrm{etc.}$};
        \node[draw] at (0,-18) (nodo6) {$\mathcal{K}_{N_L-1,N_R,3} (2'+1,2+1)$};
        \node[draw] at (6,-20) (nodo6r) {$\mathcal{K}_{N_L,N_R,3} (3',2'+1)$};
        \node[draw] at (12,-19) (nodo6rr) {$\mathcal{K}_{N_L,N_R,3} (3,3)$};
\draw[dashed] (nodo5r)--(nodo6) node[midway, above left] () {$a_{8}$};
\draw (nodo5)--(nodo6) node[midway, right] () {$e_{7}$};
\draw (nodo5r)--(nodo6r) node[above right=9mm and 2mm] () {$e_{7}$};
\draw (nodo5l)--(nodo6l);
\draw (nodo5l)--(nodo6) node[above left=6mm and 3mm] () {$e_{8}$};
\draw (nodo5rr)--(nodo6r) node[midway, below right] () {$e_{6}$};
\draw (nodo5rr)--(nodo6rr) node[midway, right] () {$d_{7}$};
\draw[dashed] (nodo5rr)--(nodo6) node[midway, above right] () {$a_{2}$};
        \node[draw] at (-6,-20) (nodo7l) {$\mathcal{K}_{N_L-1,N_R,3} (3,2+1)$};
        \node[draw] at (0,-21) (nodo7) {$\mathcal{K}_{N_L-1,N_R,3} (2'+1,2'+1)$};
                \node[draw] at (-6,-22.5) (nodo7z) {$\mathcal{K}_{N_L-1,N_R,3} (3',2+1)$};
        \node[draw] at (6,-23) (nodo7r) {$\mathcal{K}_{N_L-1,N_R,3} (3,2'+1)$};
        \node[draw] at (12,-22) (nodo7rr) {$\mathcal{K}_{N_L,N_R,3} (3',3)$};
\draw (nodo6)--(nodo7l) node[midway, above left] () {$d_{7}$};
\draw (nodo6)--(nodo7) node[midway, right] () {$e_{7}$};
\draw[dashed] (nodo6r)--(nodo7) node[midway, above left] () {$a_{8}$};
\draw (nodo7)--(nodo7r) node[midway, above right] () {$d_{7}$};
\draw[dashed] (nodo6rr)--(nodo7l) node[above right=1mm and 20mm] () {$a_{2}$};
\draw (nodo6rr)--(nodo7rr) node[midway, right] () {$e_{6}$};
\draw (nodo6r)--(nodo7rr) node[above left=5mm and 3mm] () {$d_{7}$};
\draw[dashed] (nodo7rr)--(nodo7r) node[midway,below right] () {$a_{8}$};
\draw[dashed] (nodo7rr)--(nodo7z) node[above right=2mm and 20mm] () {$a_{2}$};
\draw (nodo7l)--(nodo7z) node[midway,right] () {$d_{7}$};
        \node[draw] at (6,-26) (nodo8r) {$\mathcal{K}_{N_L-1,N_R,3} (3',2'+1)$};
        \node[draw] at (12,-25) (nodo8rr) {$\mathcal{K}_{N_L,N_R,3} (3',3')$};
        \node at (-8,-21) (nodo8l) {$\mathrm{etc.}$};
        \node at (-8,-23.5) (nodo8z) {$\mathrm{etc.}$};
\draw (nodo7rr)--(nodo8rr) node[midway, right] () {$e_{6}$};
\draw[dashed] (nodo8rr)--(nodo8r) node[midway, below right] () {$a_{8}$};
\draw (nodo7r)--(nodo8r) node[midway, right] () {$e_{6}$};
\draw (nodo7z)--(nodo8r) node[midway, above right] () {$e_{7}$};
\draw (nodo7l)--(nodo8l);
\draw (nodo7z)--(nodo8z);
\node at (6,-27) (nodo9r) {$\mathrm{etc.}$};
\draw (nodo8r)--(nodo9r);
    \end{tikzpicture}
    }
    \caption{Heterotic $E_8 \times E_8$ $\mathcal{K}_{N_L,N_R,3} (\rho_L,\rho_R)$ LST Hasse diagram, the diagram is ordered by Higgs branch dimension thus the top of this Hasse diagram corresponds to the bottom point of the Hasse diagram engineered via quiver subtraction. Dashed slices connect theories with different values of $N_L$ and $N_R$.}
    \label{fig:HasseLST3E8}
\end{figure}

At this stage, we can already find T-dual families within the $E_8 \times E_8$ LSTs just by looking at how $\kappa_R$, and the other invariants, vary with $N_L$ and $N_R$, thus we see that we have families given by the value of $\kappa_R \ \mathrm{mod} \ 3$. For higher $K$ this generalizes to $\kappa_R \ \mathrm{mod} \ K$. Instead, we apply the more general technique of studying the Hasse diagram. We depict the subdiagram of the Hasse diagram of $\mathcal{K}_{N_L,N_R,3} (1+1+1,1+1+1)$, consisting of the vertices that are associated to $K=3$ LSTs, in Figure \ref{fig:HasseLST3E8}. Recalling that the null eigenvector of the adjacency matrix has all entries equal to one, we can determine that all elementary slices appearing in Figure \ref{fig:HasseLST3E8} correspond to a shift of the invariants:
\begin{equation}\label{eqn:theexplorer}
    (\kappa_R, \operatorname{dim}(\mathcal{C})) \quad \rightarrow \quad (\kappa_R - 1, \operatorname{dim}(\mathcal{C}) - 1) \,.
\end{equation}
Therefore, just as in the $K=2$ case, we can observe theories with the same T-dual-invariant quantities just by measuring the depth in the Hasse diagram in Figure \ref{fig:HasseLST3E8}.

\begin{landscape}
\begin{table}[p]
    \centering
    \begin{tabular}{c c c }\toprule
       \footnotesize$ \mathcal{K}_{N_L,N_R,3} \left( \rho_L,\rho_R\right)$  & $\kappa_R$  & Magnetic quiver\\ \hline
        \footnotesize$ \mathcal{K}_{N_L,N_R,3} \left( 1+1+1, 1+1+1\right)$ & $14+3(N_L+N_R-1)$ & $    \begin{gathered}
    \resizebox{0.75\textwidth}{!}{
    \begin{tikzpicture}
      \node[node, label=below:{\footnotesize $3$}] (Zk) [] {};
      \node[node, label={[below right=5pt and -9pt]:\rotatebox{-30}{\footnotesize $4+N_R$}}] (A1)  [right=6mm of Zk] {};
      \node[node, label={[below right=5pt and -9pt]:\rotatebox{-30}{\footnotesize $8+2N_R$}}] (A2) [right=6mm of A1] {};
      \node[node, label={[below right=5pt and -9pt]:\rotatebox{-30}{\footnotesize $12+3N_R$}}] (A3) [right=6mm of A2] {};
      \node[node, label={[below right=5pt and -9pt]:\rotatebox{-30}{\footnotesize $16+4N_R$}}] (A4) [right=6mm of A3] {};
      \node[node, label={[below right=5pt and -9pt]:\rotatebox{-30}{\footnotesize $20+5N_R$}}] (A5) [right=6mm of A4] {};
      \node[node, label={[below right=5pt and -9pt]:\rotatebox{-30}{\footnotesize $24+6N_R$}}] (N3) [right=6mm of A5] {};
      \node[node, label={[below right=5pt and -9pt]:\rotatebox{-30}{\footnotesize $16+4N_R$}}] (B4) [right=6mm of N3] {};
      \node[node, label={[below right=5pt and -9pt]:\rotatebox{-30}{\footnotesize $8+2N_R$}}] (B2) [right=6mm of B4] {};
      \node[node, label=right:{\footnotesize $12+3N_R$}] (Nu) [above=5mm of N3] {};
      \node[node, label=below left:\rotatebox{30}{\footnotesize $4+N_L$}] (A1s)  [left=6mm of Zk] {};
      \node[node, label=below left:\rotatebox{30}{\footnotesize $8+2N_L$}] (A2s) [left=6mm of A1s] {};
      \node[node, label=below left:\rotatebox{30}{\footnotesize $12+3N_L$}] (A3s) [left=6mm of A2s] {};
      \node[node, label=below left:\rotatebox{30}{\footnotesize $16+4N_L$}] (A4s) [left=6mm of A3s] {};
      \node[node, label=below left:\rotatebox{30}{\footnotesize $20+5N_L$}] (A5s) [left=6mm of A4s] {};
      \node[node, label=below left:\rotatebox{30}{\footnotesize $24+6N_L$}] (N3s) [left=6mm of A5s] {};
      \node[node, label=below left:\rotatebox{30}{\footnotesize $16+4N_L$}] (B4s) [left=6mm of N3s] {};
      \node[node, label=below left:\rotatebox{30}{\footnotesize $8+2N_L$}] (B2s) [left=6mm of B4s] {};
      \node[node, label=left:{\footnotesize $12+3N_L$}] (Nus) [above=5mm of N3s] {};

      \draw (Zk.east) -- (A1.west);
      \draw (A1.east) -- (A2.west);
      \draw (A2.east) -- (A3.west);
      \draw (A3.east) -- (A4.west);
      \draw (A4.east) -- (A5.west);
      \draw (A5.east) -- (N3.west);
      \draw (N3.east) -- (B4.west);
      \draw (B4.east) -- (B2.west);
      \draw (N3.north) -- (Nu.south); 

    \draw (Zk.west) -- (A1s.east);
      \draw (A1s.west) -- (A2s.east);
      \draw (A2s.west) -- (A3s.east);
      \draw (A3s.west) -- (A4s.east);
      \draw (A4s.west) -- (A5s.east);
      \draw (A5s.west) -- (N3s.east);
      \draw (N3s.west) -- (B4s.east);
      \draw (B4s.west) -- (B2s.east);
      \draw (N3s.north) -- (Nus.south); 
    \end{tikzpicture} }
  \end{gathered} $ \\ 
        \footnotesize$ \mathcal{K}_{N_L,N_R,3} \left( 2+1, 1+1+1\right)$ & $13+3(N_L+N_R-1)$ & $    \begin{gathered}
    \resizebox{0.75\textwidth}{!}{
    \begin{tikzpicture}
      \node[node, label=below:{\footnotesize $3$}] (Zk) [] {};
      \node[node, label={[below right=5pt and -9pt]:\rotatebox{-30}{\footnotesize $4+N_R$}}] (A1)  [right=6mm of Zk] {};
      \node[node, label={[below right=5pt and -9pt]:\rotatebox{-30}{\footnotesize $8+2N_R$}}] (A2) [right=6mm of A1] {};
      \node[node, label={[below right=5pt and -9pt]:\rotatebox{-30}{\footnotesize $12+3N_R$}}] (A3) [right=6mm of A2] {};
      \node[node, label={[below right=5pt and -9pt]:\rotatebox{-30}{\footnotesize $16+4N_R$}}] (A4) [right=6mm of A3] {};
      \node[node, label={[below right=5pt and -9pt]:\rotatebox{-30}{\footnotesize $20+5N_R$}}] (A5) [right=6mm of A4] {};
      \node[node, label={[below right=5pt and -9pt]:\rotatebox{-30}{\footnotesize $24+6N_R$}}] (N3) [right=6mm of A5] {};
      \node[node, label={[below right=5pt and -9pt]:\rotatebox{-30}{\footnotesize $16+4N_R$}}] (B4) [right=6mm of N3] {};
      \node[node, label={[below right=5pt and -9pt]:\rotatebox{-30}{\footnotesize $8+2N_R$}}] (B2) [right=6mm of B4] {};
      \node[node, label=right:{\footnotesize $12+3N_R$}] (Nu) [above=5mm of N3] {};
      \node[node, label=below left:\rotatebox{30}{\footnotesize $4+N_L$}] (A1s)  [left=6mm of Zk] {};
      \node[node, label=below left:\rotatebox{30}{\footnotesize $6+2N_L$}] (A2s) [left=6mm of A1s] {};
      \node[node, label=below left:\rotatebox{30}{\footnotesize $9+3N_L$}] (A3s) [left=6mm of A2s] {};
      \node[node, label=below left:\rotatebox{30}{\footnotesize $12+4N_L$}] (A4s) [left=6mm of A3s] {};
      \node[node, label=below left:\rotatebox{30}{\footnotesize $15+5N_L$}] (A5s) [left=6mm of A4s] {};
      \node[node, label=below left:\rotatebox{30}{\footnotesize $18+6N_L$}] (N3s) [left=6mm of A5s] {};
      \node[node, label=below left:\rotatebox{30}{\footnotesize $12+4N_L$}] (B4s) [left=6mm of N3s] {};
      \node[node, label=below left:\rotatebox{30}{\footnotesize $6+2N_L$}] (B2s) [left=6mm of B4s] {};
      \node[node, label=left:{\footnotesize $9+3N_L$}] (Nus) [above=5mm of N3s] {};

      \draw (Zk.east) -- (A1.west);
      \draw (A1.east) -- (A2.west);
      \draw (A2.east) -- (A3.west);
      \draw (A3.east) -- (A4.west);
      \draw (A4.east) -- (A5.west);
      \draw (A5.east) -- (N3.west);
      \draw (N3.east) -- (B4.west);
      \draw (B4.east) -- (B2.west);
      \draw (N3.north) -- (Nu.south); 

    \draw (Zk.west) -- (A1s.east);
      \draw (A1s.west) -- (A2s.east);
      \draw (A2s.west) -- (A3s.east);
      \draw (A3s.west) -- (A4s.east);
      \draw (A4s.west) -- (A5s.east);
      \draw (A5s.west) -- (N3s.east);
      \draw (N3s.west) -- (B4s.east);
      \draw (B4s.west) -- (B2s.east);
      \draw (N3s.north) -- (Nus.south); 
    \end{tikzpicture} }
  \end{gathered} $ \\ 
        \footnotesize$ \mathcal{K}_{N_L,N_R,3} \left( 2'+1, 1+1+1\right)$ & $12+3(N_L+N_R-1)$&$    \begin{gathered}
    \resizebox{0.75\textwidth}{!}{
    \begin{tikzpicture}
      \node[node, label=below:{\footnotesize $3$}] (Zk) [] {};
      \node[node, label={[below right=5pt and -9pt]:\rotatebox{-30}{\footnotesize $4+N_R$}}] (A1)  [right=6mm of Zk] {};
      \node[node, label={[below right=5pt and -9pt]:\rotatebox{-30}{\footnotesize $8+2N_R$}}] (A2) [right=6mm of A1] {};
      \node[node, label={[below right=5pt and -9pt]:\rotatebox{-30}{\footnotesize $12+3N_R$}}] (A3) [right=6mm of A2] {};
      \node[node, label={[below right=5pt and -9pt]:\rotatebox{-30}{\footnotesize $16+4N_R$}}] (A4) [right=6mm of A3] {};
      \node[node, label={[below right=5pt and -9pt]:\rotatebox{-30}{\footnotesize $20+5N_R$}}] (A5) [right=6mm of A4] {};
      \node[node, label={[below right=5pt and -9pt]:\rotatebox{-30}{\footnotesize $24+6N_R$}}] (N3) [right=6mm of A5] {};
      \node[node, label={[below right=5pt and -9pt]:\rotatebox{-30}{\footnotesize $16+4N_R$}}] (B4) [right=6mm of N3] {};
      \node[node, label={[below right=5pt and -9pt]:\rotatebox{-30}{\footnotesize $8+2N_R$}}] (B2) [right=6mm of B4] {};
      \node[node, label=right:{\footnotesize $12+3N_R$}] (Nu) [above=5mm of N3] {};
      \node[node, label=below left:\rotatebox{30}{\footnotesize $4+N_L$}] (A1s)  [left=6mm of Zk] {};
      \node[node, label=below left:\rotatebox{30}{\footnotesize $6+2N_L$}] (A2s) [left=6mm of A1s] {};
      \node[node, label=below left:\rotatebox{30}{\footnotesize $8+3N_L$}] (A3s) [left=6mm of A2s] {};
      \node[node, label=below left:\rotatebox{30}{\footnotesize $10+4N_L$}] (A4s) [left=6mm of A3s] {};
      \node[node, label=below left:\rotatebox{30}{\footnotesize $12+5N_L$}] (A5s) [left=6mm of A4s] {};
      \node[node, label=below left:\rotatebox{30}{\footnotesize $14+6N_L$}] (N3s) [left=6mm of A5s] {};
      \node[node, label=below left:\rotatebox{30}{\footnotesize $9+4N_L$}] (B4s) [left=6mm of N3s] {};
      \node[node, label=below left:\rotatebox{30}{\footnotesize $4+2N_L$}] (B2s) [left=6mm of B4s] {};
      \node[node, label=left:{\footnotesize $7+3N_L$}] (Nus) [above=5mm of N3s] {};

      \draw (Zk.east) -- (A1.west);
      \draw (A1.east) -- (A2.west);
      \draw (A2.east) -- (A3.west);
      \draw (A3.east) -- (A4.west);
      \draw (A4.east) -- (A5.west);
      \draw (A5.east) -- (N3.west);
      \draw (N3.east) -- (B4.west);
      \draw (B4.east) -- (B2.west);
      \draw (N3.north) -- (Nu.south); 

    \draw (Zk.west) -- (A1s.east);
      \draw (A1s.west) -- (A2s.east);
      \draw (A2s.west) -- (A3s.east);
      \draw (A3s.west) -- (A4s.east);
      \draw (A4s.west) -- (A5s.east);
      \draw (A5s.west) -- (N3s.east);
      \draw (N3s.west) -- (B4s.east);
      \draw (B4s.west) -- (B2s.east);
      \draw (N3s.north) -- (Nus.south); 
    \end{tikzpicture} }
  \end{gathered} $ \\ 
        \footnotesize$ \mathcal{K}_{N_L,N_R,3} \left( 3, 1+1+1\right)$ & $11+3(N_L+N_R-1)$ &$    \begin{gathered}
    \resizebox{0.75\textwidth}{!}{
    \begin{tikzpicture}
      \node[node, label=below:{\footnotesize $3$}] (Zk) [] {};
      \node[node, label={[below right=5pt and -9pt]:\rotatebox{-30}{\footnotesize $4+N_R$}}] (A1)  [right=6mm of Zk] {};
      \node[node, label={[below right=5pt and -9pt]:\rotatebox{-30}{\footnotesize $8+2N_R$}}] (A2) [right=6mm of A1] {};
      \node[node, label={[below right=5pt and -9pt]:\rotatebox{-30}{\footnotesize $12+3N_R$}}] (A3) [right=6mm of A2] {};
      \node[node, label={[below right=5pt and -9pt]:\rotatebox{-30}{\footnotesize $16+4N_R$}}] (A4) [right=6mm of A3] {};
      \node[node, label={[below right=5pt and -9pt]:\rotatebox{-30}{\footnotesize $20+5N_R$}}] (A5) [right=6mm of A4] {};
      \node[node, label={[below right=5pt and -9pt]:\rotatebox{-30}{\footnotesize $24+6N_R$}}] (N3) [right=6mm of A5] {};
      \node[node, label={[below right=5pt and -9pt]:\rotatebox{-30}{\footnotesize $16+4N_R$}}] (B4) [right=6mm of N3] {};
      \node[node, label={[below right=5pt and -9pt]:\rotatebox{-30}{\footnotesize $8+2N_R$}}] (B2) [right=6mm of B4] {};
      \node[node, label=right:{\footnotesize $12+3N_R$}] (Nu) [above=5mm of N3] {};
      \node[node, label=below left:\rotatebox{30}{\footnotesize $4+N_L$}] (A1s)  [left=6mm of Zk] {};
      \node[node, label=below left:\rotatebox{30}{\footnotesize $5+2N_L$}] (A2s) [left=6mm of A1s] {};
      \node[node, label=below left:\rotatebox{30}{\footnotesize $6+3N_L$}] (A3s) [left=6mm of A2s] {};
      \node[node, label=below left:\rotatebox{30}{\footnotesize $8+4N_L$}] (A4s) [left=6mm of A3s] {};
      \node[node, label=below left:\rotatebox{30}{\footnotesize $10+5N_L$}] (A5s) [left=6mm of A4s] {};
      \node[node, label=below left:\rotatebox{30}{\footnotesize $12+6N_L$}] (N3s) [left=6mm of A5s] {};
      \node[node, label=below left:\rotatebox{30}{\footnotesize $8+4N_L$}] (B4s) [left=6mm of N3s] {};
      \node[node, label=below left:\rotatebox{30}{\footnotesize $4+2N_L$}] (B2s) [left=6mm of B4s] {};
      \node[node, label=left:{\footnotesize $6+3N_L$}] (Nus) [above=5mm of N3s] {};

      \draw (Zk.east) -- (A1.west);
      \draw (A1.east) -- (A2.west);
      \draw (A2.east) -- (A3.west);
      \draw (A3.east) -- (A4.west);
      \draw (A4.east) -- (A5.west);
      \draw (A5.east) -- (N3.west);
      \draw (N3.east) -- (B4.west);
      \draw (B4.east) -- (B2.west);
      \draw (N3.north) -- (Nu.south); 

    \draw (Zk.west) -- (A1s.east);
      \draw (A1s.west) -- (A2s.east);
      \draw (A2s.west) -- (A3s.east);
      \draw (A3s.west) -- (A4s.east);
      \draw (A4s.west) -- (A5s.east);
      \draw (A5s.west) -- (N3s.east);
      \draw (N3s.west) -- (B4s.east);
      \draw (B4s.west) -- (B2s.east);
      \draw (N3s.north) -- (Nus.south); 
    \end{tikzpicture} }
  \end{gathered} $ \\ 
        \footnotesize$ \mathcal{K}_{N_L,N_R,3} \left( 3', 1+1+1\right)$ & $10+3(N_L+N_R-1)$ & $    \begin{gathered}
    \resizebox{0.75\textwidth}{!}{
    \begin{tikzpicture}
      \node[node, label=below:{\footnotesize $3$}] (Zk) [] {};
      \node[node, label={[below right=5pt and -9pt]:\rotatebox{-30}{\footnotesize $4+N_R$}}] (A1)  [right=6mm of Zk] {};
      \node[node, label={[below right=5pt and -9pt]:\rotatebox{-30}{\footnotesize $8+2N_R$}}] (A2) [right=6mm of A1] {};
      \node[node, label={[below right=5pt and -9pt]:\rotatebox{-30}{\footnotesize $12+3N_R$}}] (A3) [right=6mm of A2] {};
      \node[node, label={[below right=5pt and -9pt]:\rotatebox{-30}{\footnotesize $16+4N_R$}}] (A4) [right=6mm of A3] {};
      \node[node, label={[below right=5pt and -9pt]:\rotatebox{-30}{\footnotesize $20+5N_R$}}] (A5) [right=6mm of A4] {};
      \node[node, label={[below right=5pt and -9pt]:\rotatebox{-30}{\footnotesize $24+6N_R$}}] (N3) [right=6mm of A5] {};
      \node[node, label={[below right=5pt and -9pt]:\rotatebox{-30}{\footnotesize $16+4N_R$}}] (B4) [right=6mm of N3] {};
      \node[node, label={[below right=5pt and -9pt]:\rotatebox{-30}{\footnotesize $8+2N_R$}}] (B2) [right=6mm of B4] {};
      \node[node, label=right:{\footnotesize $12+3N_R$}] (Nu) [above=5mm of N3] {};
      \node[node, label=below left:\rotatebox{30}{\footnotesize $4+N_L$}] (A1s)  [left=6mm of Zk] {};
      \node[node, label=below left:\rotatebox{30}{\footnotesize $5+2N_L$}] (A2s) [left=6mm of A1s] {};
      \node[node, label=below left:\rotatebox{30}{\footnotesize $6+3N_L$}] (A3s) [left=6mm of A2s] {};
      \node[node, label=below left:\rotatebox{30}{\footnotesize $7+4N_L$}] (A4s) [left=6mm of A3s] {};
      \node[node, label=below left:\rotatebox{30}{\footnotesize $8+5N_L$}] (A5s) [left=6mm of A4s] {};
      \node[node, label=below left:\rotatebox{30}{\footnotesize $9+6N_L$}] (N3s) [left=6mm of A5s] {};
      \node[node, label=below left:\rotatebox{30}{\footnotesize $6+4N_L$}] (B4s) [left=6mm of N3s] {};
      \node[node, label=below left:\rotatebox{30}{\footnotesize $3+2N_L$}] (B2s) [left=6mm of B4s] {};
      \node[node, label=left:{\footnotesize $4+3N_L$}] (Nus) [above=5mm of N3s] {};

      \draw (Zk.east) -- (A1.west);
      \draw (A1.east) -- (A2.west);
      \draw (A2.east) -- (A3.west);
      \draw (A3.east) -- (A4.west);
      \draw (A4.east) -- (A5.west);
      \draw (A5.east) -- (N3.west);
      \draw (N3.east) -- (B4.west);
      \draw (B4.east) -- (B2.west);
      \draw (N3.north) -- (Nu.south); 

    \draw (Zk.west) -- (A1s.east);
      \draw (A1s.west) -- (A2s.east);
      \draw (A2s.west) -- (A3s.east);
      \draw (A3s.west) -- (A4s.east);
      \draw (A4s.west) -- (A5s.east);
      \draw (A5s.west) -- (N3s.east);
      \draw (N3s.west) -- (B4s.east);
      \draw (B4s.west) -- (B2s.east);
      \draw (N3s.north) -- (Nus.south); 
    \end{tikzpicture} }
  \end{gathered} $ \\ 
        \footnotesize$ \mathcal{K}_{N_L,N_R,3} \left( 2+1, 2+1\right)$ & $12+3(N_L+N_R-1)$ & $    \begin{gathered}
    \resizebox{0.75\textwidth}{!}{
    \begin{tikzpicture}
      \node[node, label=below:{\footnotesize $3$}] (Zk) [] {};
      \node[node, label={[below right=5pt and -9pt]:\rotatebox{-30}{\footnotesize $4+N_R$}}] (A1)  [right=6mm of Zk] {};
      \node[node, label={[below right=5pt and -9pt]:\rotatebox{-30}{\footnotesize $6+2N_R$}}] (A2) [right=6mm of A1] {};
      \node[node, label={[below right=5pt and -9pt]:\rotatebox{-30}{\footnotesize $9+3N_R$}}] (A3) [right=6mm of A2] {};
      \node[node, label={[below right=5pt and -9pt]:\rotatebox{-30}{\footnotesize $12+4N_R$}}] (A4) [right=6mm of A3] {};
      \node[node, label={[below right=5pt and -9pt]:\rotatebox{-30}{\footnotesize $15+5N_R$}}] (A5) [right=6mm of A4] {};
      \node[node, label={[below right=5pt and -9pt]:\rotatebox{-30}{\footnotesize $18+6N_R$}}] (N3) [right=6mm of A5] {};
      \node[node, label={[below right=5pt and -9pt]:\rotatebox{-30}{\footnotesize $12+4N_R$}}] (B4) [right=6mm of N3] {};
      \node[node, label={[below right=5pt and -9pt]:\rotatebox{-30}{\footnotesize $6+2N_R$}}] (B2) [right=6mm of B4] {};
      \node[node, label=right:{\footnotesize $9+3N_R$}] (Nu) [above=5mm of N3] {};
      \node[node, label=below left:\rotatebox{30}{\footnotesize $4+N_L$}] (A1s)  [left=6mm of Zk] {};
      \node[node, label=below left:\rotatebox{30}{\footnotesize $6+2N_L$}] (A2s) [left=6mm of A1s] {};
      \node[node, label=below left:\rotatebox{30}{\footnotesize $9+3N_L$}] (A3s) [left=6mm of A2s] {};
      \node[node, label=below left:\rotatebox{30}{\footnotesize $12+4N_L$}] (A4s) [left=6mm of A3s] {};
      \node[node, label=below left:\rotatebox{30}{\footnotesize $15+5N_L$}] (A5s) [left=6mm of A4s] {};
      \node[node, label=below left:\rotatebox{30}{\footnotesize $18+6N_L$}] (N3s) [left=6mm of A5s] {};
      \node[node, label=below left:\rotatebox{30}{\footnotesize $12+4N_L$}] (B4s) [left=6mm of N3s] {};
      \node[node, label=below left:\rotatebox{30}{\footnotesize $6+2N_L$}] (B2s) [left=6mm of B4s] {};
      \node[node, label=left:{\footnotesize $9+3N_L$}] (Nus) [above=5mm of N3s] {};

      \draw (Zk.east) -- (A1.west);
      \draw (A1.east) -- (A2.west);
      \draw (A2.east) -- (A3.west);
      \draw (A3.east) -- (A4.west);
      \draw (A4.east) -- (A5.west);
      \draw (A5.east) -- (N3.west);
      \draw (N3.east) -- (B4.west);
      \draw (B4.east) -- (B2.west);
      \draw (N3.north) -- (Nu.south); 

    \draw (Zk.west) -- (A1s.east);
      \draw (A1s.west) -- (A2s.east);
      \draw (A2s.west) -- (A3s.east);
      \draw (A3s.west) -- (A4s.east);
      \draw (A4s.west) -- (A5s.east);
      \draw (A5s.west) -- (N3s.east);
      \draw (N3s.west) -- (B4s.east);
      \draw (B4s.west) -- (B2s.east);
      \draw (N3s.north) -- (Nus.south); 
    \end{tikzpicture} }
  \end{gathered} $ \\ \bottomrule  
    \end{tabular}
    \caption{The $\mathcal{K}_{N_L,N_R,3} \left( \rho_L, \rho_R \right)$ LSTs. We provide the constant $\kappa_R$ and the associated magnetic quiver.}
    \label{tab:1E8LSTk=3}
\end{table}
\end{landscape}

\begin{landscape}
\begin{table}[p]
    \centering
    \begin{tabular}{c c c } \toprule
       \footnotesize$ \mathcal{K}_{N_L,N_R,3} \left( \rho_L,\rho_R\right)$  & $\kappa_R$  & Magnetic quiver\\ \hline
        \footnotesize$ \mathcal{K}_{N_L,N_R,3} \left( 2'+1, 2+1\right)$ & $11+3(N_L+N_R-1)$&$    \begin{gathered}
    \resizebox{0.85\textwidth}{!}{
    \begin{tikzpicture}
      \node[node, label=below:{\footnotesize $3$}] (Zk) [] {};
      \node[node, label={[below right=5pt and -9pt]:\rotatebox{-30}{\footnotesize $4+N_R$}}] (A1)  [right=6mm of Zk] {};
      \node[node, label={[below right=5pt and -9pt]:\rotatebox{-30}{\footnotesize $6+2N_R$}}] (A2) [right=6mm of A1] {};
      \node[node, label={[below right=5pt and -9pt]:\rotatebox{-30}{\footnotesize $9+3N_R$}}] (A3) [right=6mm of A2] {};
      \node[node, label={[below right=5pt and -9pt]:\rotatebox{-30}{\footnotesize $12+4N_R$}}] (A4) [right=6mm of A3] {};
      \node[node, label={[below right=5pt and -9pt]:\rotatebox{-30}{\footnotesize $15+5N_R$}}] (A5) [right=6mm of A4] {};
      \node[node, label={[below right=5pt and -9pt]:\rotatebox{-30}{\footnotesize $18+6N_R$}}] (N3) [right=6mm of A5] {};
      \node[node, label={[below right=5pt and -9pt]:\rotatebox{-30}{\footnotesize $12+4N_R$}}] (B4) [right=6mm of N3] {};
      \node[node, label={[below right=5pt and -9pt]:\rotatebox{-30}{\footnotesize $6+2N_R$}}] (B2) [right=6mm of B4] {};
      \node[node, label=right:{\footnotesize $9+3N_R$}] (Nu) [above=5mm of N3] {};
      \node[node, label=below left:\rotatebox{30}{\footnotesize $4+N_L$}] (A1s)  [left=6mm of Zk] {};
      \node[node, label=below left:\rotatebox{30}{\footnotesize $6+2N_L$}] (A2s) [left=6mm of A1s] {};
      \node[node, label=below left:\rotatebox{30}{\footnotesize $8+3N_L$}] (A3s) [left=6mm of A2s] {};
      \node[node, label=below left:\rotatebox{30}{\footnotesize $10+4N_L$}] (A4s) [left=6mm of A3s] {};
      \node[node, label=below left:\rotatebox{30}{\footnotesize $12+5N_L$}] (A5s) [left=6mm of A4s] {};
      \node[node, label=below left:\rotatebox{30}{\footnotesize $14+6N_L$}] (N3s) [left=6mm of A5s] {};
      \node[node, label=below left:\rotatebox{30}{\footnotesize $9+4N_L$}] (B4s) [left=6mm of N3s] {};
      \node[node, label=below left:\rotatebox{30}{\footnotesize $4+2N_L$}] (B2s) [left=6mm of B4s] {};
      \node[node, label=left:{\footnotesize $7+3N_L$}] (Nus) [above=5mm of N3s] {};

      \draw (Zk.east) -- (A1.west);
      \draw (A1.east) -- (A2.west);
      \draw (A2.east) -- (A3.west);
      \draw (A3.east) -- (A4.west);
      \draw (A4.east) -- (A5.west);
      \draw (A5.east) -- (N3.west);
      \draw (N3.east) -- (B4.west);
      \draw (B4.east) -- (B2.west);
      \draw (N3.north) -- (Nu.south); 

    \draw (Zk.west) -- (A1s.east);
      \draw (A1s.west) -- (A2s.east);
      \draw (A2s.west) -- (A3s.east);
      \draw (A3s.west) -- (A4s.east);
      \draw (A4s.west) -- (A5s.east);
      \draw (A5s.west) -- (N3s.east);
      \draw (N3s.west) -- (B4s.east);
      \draw (B4s.west) -- (B2s.east);
      \draw (N3s.north) -- (Nus.south); 
    \end{tikzpicture} }
  \end{gathered} $ \\ 
        \footnotesize$ \mathcal{K}_{N_L,N_R,3} \left( 3, 2+1\right)$ & $10+3(N_L+N_R-1)$ &$    \begin{gathered}
    \resizebox{0.85\textwidth}{!}{
    \begin{tikzpicture}
      \node[node, label=below:{\footnotesize $3$}] (Zk) [] {};
      \node[node, label={[below right=5pt and -9pt]:\rotatebox{-30}{\footnotesize $4+N_R$}}] (A1)  [right=6mm of Zk] {};
      \node[node, label={[below right=5pt and -9pt]:\rotatebox{-30}{\footnotesize $6+2N_R$}}] (A2) [right=6mm of A1] {};
      \node[node, label={[below right=5pt and -9pt]:\rotatebox{-30}{\footnotesize $9+3N_R$}}] (A3) [right=6mm of A2] {};
      \node[node, label={[below right=5pt and -9pt]:\rotatebox{-30}{\footnotesize $12+4N_R$}}] (A4) [right=6mm of A3] {};
      \node[node, label={[below right=5pt and -9pt]:\rotatebox{-30}{\footnotesize $15+5N_R$}}] (A5) [right=6mm of A4] {};
      \node[node, label={[below right=5pt and -9pt]:\rotatebox{-30}{\footnotesize $18+6N_R$}}] (N3) [right=6mm of A5] {};
      \node[node, label={[below right=5pt and -9pt]:\rotatebox{-30}{\footnotesize $12+4N_R$}}] (B4) [right=6mm of N3] {};
      \node[node, label={[below right=5pt and -9pt]:\rotatebox{-30}{\footnotesize $6+2N_R$}}] (B2) [right=6mm of B4] {};
      \node[node, label=right:{\footnotesize $9+3N_R$}] (Nu) [above=5mm of N3] {};
      \node[node, label=below left:\rotatebox{30}{\footnotesize $4+N_L$}] (A1s)  [left=6mm of Zk] {};
      \node[node, label=below left:\rotatebox{30}{\footnotesize $5+2N_L$}] (A2s) [left=6mm of A1s] {};
      \node[node, label=below left:\rotatebox{30}{\footnotesize $6+3N_L$}] (A3s) [left=6mm of A2s] {};
      \node[node, label=below left:\rotatebox{30}{\footnotesize $8+4N_L$}] (A4s) [left=6mm of A3s] {};
      \node[node, label=below left:\rotatebox{30}{\footnotesize $10+5N_L$}] (A5s) [left=6mm of A4s] {};
      \node[node, label=below left:\rotatebox{30}{\footnotesize $12+6N_L$}] (N3s) [left=6mm of A5s] {};
      \node[node, label=below left:\rotatebox{30}{\footnotesize $8+4N_L$}] (B4s) [left=6mm of N3s] {};
      \node[node, label=below left:\rotatebox{30}{\footnotesize $4+2N_L$}] (B2s) [left=6mm of B4s] {};
      \node[node, label=left:{\footnotesize $6+3N_L$}] (Nus) [above=5mm of N3s] {};

      \draw (Zk.east) -- (A1.west);
      \draw (A1.east) -- (A2.west);
      \draw (A2.east) -- (A3.west);
      \draw (A3.east) -- (A4.west);
      \draw (A4.east) -- (A5.west);
      \draw (A5.east) -- (N3.west);
      \draw (N3.east) -- (B4.west);
      \draw (B4.east) -- (B2.west);
      \draw (N3.north) -- (Nu.south); 

    \draw (Zk.west) -- (A1s.east);
      \draw (A1s.west) -- (A2s.east);
      \draw (A2s.west) -- (A3s.east);
      \draw (A3s.west) -- (A4s.east);
      \draw (A4s.west) -- (A5s.east);
      \draw (A5s.west) -- (N3s.east);
      \draw (N3s.west) -- (B4s.east);
      \draw (B4s.west) -- (B2s.east);
      \draw (N3s.north) -- (Nus.south); 
    \end{tikzpicture} }
  \end{gathered} $ \\ 
        \footnotesize$ \mathcal{K}_{N_L,N_R,3} \left( 3', 2+1\right)$ & $9+3(N_L+N_R-1)$ & $    \begin{gathered}
    \resizebox{0.85\textwidth}{!}{
    \begin{tikzpicture}
      \node[node, label=below:{\footnotesize $3$}] (Zk) [] {};
      \node[node, label={[below right=5pt and -9pt]:\rotatebox{-30}{\footnotesize $4+N_R$}}] (A1)  [right=6mm of Zk] {};
      \node[node, label={[below right=5pt and -9pt]:\rotatebox{-30}{\footnotesize $6+2N_R$}}] (A2) [right=6mm of A1] {};
      \node[node, label={[below right=5pt and -9pt]:\rotatebox{-30}{\footnotesize $9+3N_R$}}] (A3) [right=6mm of A2] {};
      \node[node, label={[below right=5pt and -9pt]:\rotatebox{-30}{\footnotesize $12+4N_R$}}] (A4) [right=6mm of A3] {};
      \node[node, label={[below right=5pt and -9pt]:\rotatebox{-30}{\footnotesize $15+5N_R$}}] (A5) [right=6mm of A4] {};
      \node[node, label={[below right=5pt and -9pt]:\rotatebox{-30}{\footnotesize $18+6N_R$}}] (N3) [right=6mm of A5] {};
      \node[node, label={[below right=5pt and -9pt]:\rotatebox{-30}{\footnotesize $12+4N_R$}}] (B4) [right=6mm of N3] {};
      \node[node, label={[below right=5pt and -9pt]:\rotatebox{-30}{\footnotesize $6+2N_R$}}] (B2) [right=6mm of B4] {};
      \node[node, label=right:{\footnotesize $9+3N_R$}] (Nu) [above=5mm of N3] {};
      \node[node, label=below left:\rotatebox{30}{\footnotesize $4+N_L$}] (A1s)  [left=6mm of Zk] {};
      \node[node, label=below left:\rotatebox{30}{\footnotesize $5+2N_L$}] (A2s) [left=6mm of A1s] {};
      \node[node, label=below left:\rotatebox{30}{\footnotesize $6+3N_L$}] (A3s) [left=6mm of A2s] {};
      \node[node, label=below left:\rotatebox{30}{\footnotesize $7+4N_L$}] (A4s) [left=6mm of A3s] {};
      \node[node, label=below left:\rotatebox{30}{\footnotesize $8+5N_L$}] (A5s) [left=6mm of A4s] {};
      \node[node, label=below left:\rotatebox{30}{\footnotesize $9+6N_L$}] (N3s) [left=6mm of A5s] {};
      \node[node, label=below left:\rotatebox{30}{\footnotesize $6+4N_L$}] (B4s) [left=6mm of N3s] {};
      \node[node, label=below left:\rotatebox{30}{\footnotesize $3+2N_L$}] (B2s) [left=6mm of B4s] {};
      \node[node, label=left:{\footnotesize $4+3N_L$}] (Nus) [above=5mm of N3s] {};

      \draw (Zk.east) -- (A1.west);
      \draw (A1.east) -- (A2.west);
      \draw (A2.east) -- (A3.west);
      \draw (A3.east) -- (A4.west);
      \draw (A4.east) -- (A5.west);
      \draw (A5.east) -- (N3.west);
      \draw (N3.east) -- (B4.west);
      \draw (B4.east) -- (B2.west);
      \draw (N3.north) -- (Nu.south); 

    \draw (Zk.west) -- (A1s.east);
      \draw (A1s.west) -- (A2s.east);
      \draw (A2s.west) -- (A3s.east);
      \draw (A3s.west) -- (A4s.east);
      \draw (A4s.west) -- (A5s.east);
      \draw (A5s.west) -- (N3s.east);
      \draw (N3s.west) -- (B4s.east);
      \draw (B4s.west) -- (B2s.east);
      \draw (N3s.north) -- (Nus.south); 
    \end{tikzpicture} }
  \end{gathered} $ \\ 
        \footnotesize$ \mathcal{K}_{N_L,N_R,3} \left( 2'+1, 2'+1\right)$ & $10+3(N_L+N_R-1)$ & $    \begin{gathered}
    \resizebox{0.85\textwidth}{!}{
    \begin{tikzpicture}
      \node[node, label=below:{\footnotesize $3$}] (Zk) [] {};
      \node[node, label={[below right=5pt and -9pt]:\rotatebox{-30}{\footnotesize $4+N_R$}}] (A1)  [right=6mm of Zk] {};
      \node[node, label={[below right=5pt and -9pt]:\rotatebox{-30}{\footnotesize $6+2N_R$}}] (A2) [right=6mm of A1] {};
      \node[node, label={[below right=5pt and -9pt]:\rotatebox{-30}{\footnotesize $8+3N_R$}}] (A3) [right=6mm of A2] {};
      \node[node, label={[below right=5pt and -9pt]:\rotatebox{-30}{\footnotesize $10+4N_R$}}] (A4) [right=6mm of A3] {};
      \node[node, label={[below right=5pt and -9pt]:\rotatebox{-30}{\footnotesize $12+5N_R$}}] (A5) [right=6mm of A4] {};
      \node[node, label={[below right=5pt and -9pt]:\rotatebox{-30}{\footnotesize $14+6N_R$}}] (N3) [right=6mm of A5] {};
      \node[node, label={[below right=5pt and -9pt]:\rotatebox{-30}{\footnotesize $9+4N_R$}}] (B4) [right=6mm of N3] {};
      \node[node, label={[below right=5pt and -9pt]:\rotatebox{-30}{\footnotesize $4+2N_R$}}] (B2) [right=6mm of B4] {};
      \node[node, label=right:{\footnotesize $7+3N_R$}] (Nu) [above=5mm of N3] {};
      \node[node, label=below left:\rotatebox{30}{\footnotesize $4+N_L$}] (A1s)  [left=6mm of Zk] {};
      \node[node, label=below left:\rotatebox{30}{\footnotesize $6+2N_L$}] (A2s) [left=6mm of A1s] {};
      \node[node, label=below left:\rotatebox{30}{\footnotesize $8+3N_L$}] (A3s) [left=6mm of A2s] {};
      \node[node, label=below left:\rotatebox{30}{\footnotesize $10+4N_L$}] (A4s) [left=6mm of A3s] {};
      \node[node, label=below left:\rotatebox{30}{\footnotesize $12+5N_L$}] (A5s) [left=6mm of A4s] {};
      \node[node, label=below left:\rotatebox{30}{\footnotesize $14+6N_L$}] (N3s) [left=6mm of A5s] {};
      \node[node, label=below left:\rotatebox{30}{\footnotesize $9+4N_L$}] (B4s) [left=6mm of N3s] {};
      \node[node, label=below left:\rotatebox{30}{\footnotesize $4+2N_L$}] (B2s) [left=6mm of B4s] {};
      \node[node, label=left:{\footnotesize $7+3N_L$}] (Nus) [above=5mm of N3s] {};

      \draw (Zk.east) -- (A1.west);
      \draw (A1.east) -- (A2.west);
      \draw (A2.east) -- (A3.west);
      \draw (A3.east) -- (A4.west);
      \draw (A4.east) -- (A5.west);
      \draw (A5.east) -- (N3.west);
      \draw (N3.east) -- (B4.west);
      \draw (B4.east) -- (B2.west);
      \draw (N3.north) -- (Nu.south); 

    \draw (Zk.west) -- (A1s.east);
      \draw (A1s.west) -- (A2s.east);
      \draw (A2s.west) -- (A3s.east);
      \draw (A3s.west) -- (A4s.east);
      \draw (A4s.west) -- (A5s.east);
      \draw (A5s.west) -- (N3s.east);
      \draw (N3s.west) -- (B4s.east);
      \draw (B4s.west) -- (B2s.east);
      \draw (N3s.north) -- (Nus.south); 
    \end{tikzpicture} }
  \end{gathered} $ \\ 
        \footnotesize$ \mathcal{K}_{N_L,N_R,3} \left( 3, 2'+1\right)$ & $9+3(N_L+N_R-1)$ &$    \begin{gathered}
    \resizebox{0.85\textwidth}{!}{
    \begin{tikzpicture}
      \node[node, label=below:{\footnotesize $3$}] (Zk) [] {};
      \node[node, label={[below right=5pt and -9pt]:\rotatebox{-30}{\footnotesize $4+N_R$}}] (A1)  [right=6mm of Zk] {};
      \node[node, label={[below right=5pt and -9pt]:\rotatebox{-30}{\footnotesize $6+2N_R$}}] (A2) [right=6mm of A1] {};
      \node[node, label={[below right=5pt and -9pt]:\rotatebox{-30}{\footnotesize $8+3N_R$}}] (A3) [right=6mm of A2] {};
      \node[node, label={[below right=5pt and -9pt]:\rotatebox{-30}{\footnotesize $10+4N_R$}}] (A4) [right=6mm of A3] {};
      \node[node, label={[below right=5pt and -9pt]:\rotatebox{-30}{\footnotesize $12+5N_R$}}] (A5) [right=6mm of A4] {};
      \node[node, label={[below right=5pt and -9pt]:\rotatebox{-30}{\footnotesize $14+6N_R$}}] (N3) [right=6mm of A5] {};
      \node[node, label={[below right=5pt and -9pt]:\rotatebox{-30}{\footnotesize $9+4N_R$}}] (B4) [right=6mm of N3] {};
      \node[node, label={[below right=5pt and -9pt]:\rotatebox{-30}{\footnotesize $4+2N_R$}}] (B2) [right=6mm of B4] {};
      \node[node, label=right:{\footnotesize $7+3N_R$}] (Nu) [above=5mm of N3] {};
      \node[node, label=below left:\rotatebox{30}{\footnotesize $4+N_L$}] (A1s)  [left=6mm of Zk] {};
      \node[node, label=below left:\rotatebox{30}{\footnotesize $5+2N_L$}] (A2s) [left=6mm of A1s] {};
      \node[node, label=below left:\rotatebox{30}{\footnotesize $6+3N_L$}] (A3s) [left=6mm of A2s] {};
      \node[node, label=below left:\rotatebox{30}{\footnotesize $8+4N_L$}] (A4s) [left=6mm of A3s] {};
      \node[node, label=below left:\rotatebox{30}{\footnotesize $10+5N_L$}] (A5s) [left=6mm of A4s] {};
      \node[node, label=below left:\rotatebox{30}{\footnotesize $12+6N_L$}] (N3s) [left=6mm of A5s] {};
      \node[node, label=below left:\rotatebox{30}{\footnotesize $8+4N_L$}] (B4s) [left=6mm of N3s] {};
      \node[node, label=below left:\rotatebox{30}{\footnotesize $4+2N_L$}] (B2s) [left=6mm of B4s] {};
      \node[node, label=left:{\footnotesize $6+3N_L$}] (Nus) [above=5mm of N3s] {};

      \draw (Zk.east) -- (A1.west);
      \draw (A1.east) -- (A2.west);
      \draw (A2.east) -- (A3.west);
      \draw (A3.east) -- (A4.west);
      \draw (A4.east) -- (A5.west);
      \draw (A5.east) -- (N3.west);
      \draw (N3.east) -- (B4.west);
      \draw (B4.east) -- (B2.west);
      \draw (N3.north) -- (Nu.south); 

    \draw (Zk.west) -- (A1s.east);
      \draw (A1s.west) -- (A2s.east);
      \draw (A2s.west) -- (A3s.east);
      \draw (A3s.west) -- (A4s.east);
      \draw (A4s.west) -- (A5s.east);
      \draw (A5s.west) -- (N3s.east);
      \draw (N3s.west) -- (B4s.east);
      \draw (B4s.west) -- (B2s.east);
      \draw (N3s.north) -- (Nus.south); 
    \end{tikzpicture} }
  \end{gathered} $ \\ 
        \footnotesize$ \mathcal{K}_{N_L,N_R,3} \left( 3', 2'+1\right)$ & $8+3(N_L+N_R-1)$ & $    \begin{gathered}
    \resizebox{0.85\textwidth}{!}{
    \begin{tikzpicture}
      \node[node, label=below:{\footnotesize $3$}] (Zk) [] {};
      \node[node, label={[below right=5pt and -9pt]:\rotatebox{-30}{\footnotesize $4+N_R$}}] (A1)  [right=6mm of Zk] {};
      \node[node, label={[below right=5pt and -9pt]:\rotatebox{-30}{\footnotesize $6+2N_R$}}] (A2) [right=6mm of A1] {};
      \node[node, label={[below right=5pt and -9pt]:\rotatebox{-30}{\footnotesize $8+3N_R$}}] (A3) [right=6mm of A2] {};
      \node[node, label={[below right=5pt and -9pt]:\rotatebox{-30}{\footnotesize $10+4N_R$}}] (A4) [right=6mm of A3] {};
      \node[node, label={[below right=5pt and -9pt]:\rotatebox{-30}{\footnotesize $12+5N_R$}}] (A5) [right=6mm of A4] {};
      \node[node, label={[below right=5pt and -9pt]:\rotatebox{-30}{\footnotesize $14+6N_R$}}] (N3) [right=6mm of A5] {};
      \node[node, label={[below right=5pt and -9pt]:\rotatebox{-30}{\footnotesize $9+4N_R$}}] (B4) [right=6mm of N3] {};
      \node[node, label={[below right=5pt and -9pt]:\rotatebox{-30}{\footnotesize $4+2N_R$}}] (B2) [right=6mm of B4] {};
      \node[node, label=right:{\footnotesize $7+3N_R$}] (Nu) [above=5mm of N3] {};
      \node[node, label=below left:\rotatebox{30}{\footnotesize $4+N_L$}] (A1s)  [left=6mm of Zk] {};
      \node[node, label=below left:\rotatebox{30}{\footnotesize $5+2N_L$}] (A2s) [left=6mm of A1s] {};
      \node[node, label=below left:\rotatebox{30}{\footnotesize $6+3N_L$}] (A3s) [left=6mm of A2s] {};
      \node[node, label=below left:\rotatebox{30}{\footnotesize $7+4N_L$}] (A4s) [left=6mm of A3s] {};
      \node[node, label=below left:\rotatebox{30}{\footnotesize $8+5N_L$}] (A5s) [left=6mm of A4s] {};
      \node[node, label=below left:\rotatebox{30}{\footnotesize $9+6N_L$}] (N3s) [left=6mm of A5s] {};
      \node[node, label=below left:\rotatebox{30}{\footnotesize $6+4N_L$}] (B4s) [left=6mm of N3s] {};
      \node[node, label=below left:\rotatebox{30}{\footnotesize $3+2N_L$}] (B2s) [left=6mm of B4s] {};
      \node[node, label=left:{\footnotesize $4+3N_L$}] (Nus) [above=5mm of N3s] {};

      \draw (Zk.east) -- (A1.west);
      \draw (A1.east) -- (A2.west);
      \draw (A2.east) -- (A3.west);
      \draw (A3.east) -- (A4.west);
      \draw (A4.east) -- (A5.west);
      \draw (A5.east) -- (N3.west);
      \draw (N3.east) -- (B4.west);
      \draw (B4.east) -- (B2.west);
      \draw (N3.north) -- (Nu.south); 

    \draw (Zk.west) -- (A1s.east);
      \draw (A1s.west) -- (A2s.east);
      \draw (A2s.west) -- (A3s.east);
      \draw (A3s.west) -- (A4s.east);
      \draw (A4s.west) -- (A5s.east);
      \draw (A5s.west) -- (N3s.east);
      \draw (N3s.west) -- (B4s.east);
      \draw (B4s.west) -- (B2s.east);
      \draw (N3s.north) -- (Nus.south); 
    \end{tikzpicture} }
  \end{gathered} $ \\ \bottomrule

    \end{tabular}
    \caption{The $\mathcal{K}_{N_L,N_R,3} \left( \rho_L, \rho_R \right)$ LSTs. We provide the constant $\kappa_R$ and the associated magnetic quiver.}
    \label{tab:2E8LSTk=3}
\end{table}
\end{landscape}

\begin{landscape}
\begin{table}[p]
    \centering
    \begin{tabular}{ c c c }\toprule
       \footnotesize$ \mathcal{K}_{N_L,N_R,3} \left( \rho_L,\rho_R\right)$  & $\kappa_R$  & Magnetic quiver\\ \hline
        
        \footnotesize$ \mathcal{K}_{N_L,N_R,3} \left( 3, 3\right)$ & $8+3(N_L+N_R-1)$ &$    \begin{gathered}
    \resizebox{0.85\textwidth}{!}{
    \begin{tikzpicture}
      \node[node, label=below:{\footnotesize $3$}] (Zk) [] {};
      \node[node, label={[below right=5pt and -9pt]:\rotatebox{-30}{\footnotesize $4+N_R$}}] (A1)  [right=6mm of Zk] {};
      \node[node, label={[below right=5pt and -9pt]:\rotatebox{-30}{\footnotesize $5+2N_R$}}] (A2) [right=6mm of A1] {};
      \node[node, label={[below right=5pt and -9pt]:\rotatebox{-30}{\footnotesize $6+3N_R$}}] (A3) [right=6mm of A2] {};
      \node[node, label={[below right=5pt and -9pt]:\rotatebox{-30}{\footnotesize $8+4N_R$}}] (A4) [right=6mm of A3] {};
      \node[node, label={[below right=5pt and -9pt]:\rotatebox{-30}{\footnotesize $10+5N_R$}}] (A5) [right=6mm of A4] {};
      \node[node, label={[below right=5pt and -9pt]:\rotatebox{-30}{\footnotesize $12+6N_R$}}] (N3) [right=6mm of A5] {};
      \node[node, label={[below right=5pt and -9pt]:\rotatebox{-30}{\footnotesize $8+4N_R$}}] (B4) [right=6mm of N3] {};
      \node[node, label={[below right=5pt and -9pt]:\rotatebox{-30}{\footnotesize $4+2N_R$}}] (B2) [right=6mm of B4] {};
      \node[node, label=right:{\footnotesize $6+3N_R$}] (Nu) [above=5mm of N3] {};
      \node[node, label=below left:\rotatebox{30}{\footnotesize $4+N_L$}] (A1s)  [left=6mm of Zk] {};
      \node[node, label=below left:\rotatebox{30}{\footnotesize $5+2N_L$}] (A2s) [left=6mm of A1s] {};
      \node[node, label=below left:\rotatebox{30}{\footnotesize $6+3N_L$}] (A3s) [left=6mm of A2s] {};
      \node[node, label=below left:\rotatebox{30}{\footnotesize $8+4N_L$}] (A4s) [left=6mm of A3s] {};
      \node[node, label=below left:\rotatebox{30}{\footnotesize $10+5N_L$}] (A5s) [left=6mm of A4s] {};
      \node[node, label=below left:\rotatebox{30}{\footnotesize $12+6N_L$}] (N3s) [left=6mm of A5s] {};
      \node[node, label=below left:\rotatebox{30}{\footnotesize $8+4N_L$}] (B4s) [left=6mm of N3s] {};
      \node[node, label=below left:\rotatebox{30}{\footnotesize $4+2N_L$}] (B2s) [left=6mm of B4s] {};
      \node[node, label=left:{\footnotesize $6+3N_L$}] (Nus) [above=5mm of N3s] {};

      \draw (Zk.east) -- (A1.west);
      \draw (A1.east) -- (A2.west);
      \draw (A2.east) -- (A3.west);
      \draw (A3.east) -- (A4.west);
      \draw (A4.east) -- (A5.west);
      \draw (A5.east) -- (N3.west);
      \draw (N3.east) -- (B4.west);
      \draw (B4.east) -- (B2.west);
      \draw (N3.north) -- (Nu.south); 

    \draw (Zk.west) -- (A1s.east);
      \draw (A1s.west) -- (A2s.east);
      \draw (A2s.west) -- (A3s.east);
      \draw (A3s.west) -- (A4s.east);
      \draw (A4s.west) -- (A5s.east);
      \draw (A5s.west) -- (N3s.east);
      \draw (N3s.west) -- (B4s.east);
      \draw (B4s.west) -- (B2s.east);
      \draw (N3s.north) -- (Nus.south); 
    \end{tikzpicture} }
  \end{gathered} $ \\ 
        \footnotesize$ \mathcal{K}_{N_L,N_R,3} \left( 3', 3\right)$ & $7+3(N_L+N_R-1)$ & $    \begin{gathered}
    \resizebox{0.85\textwidth}{!}{
    \begin{tikzpicture}
      \node[node, label=below:{\footnotesize $3$}] (Zk) [] {};
      \node[node, label={[below right=5pt and -9pt]:\rotatebox{-30}{\footnotesize $4+N_R$}}] (A1)  [right=6mm of Zk] {};
      \node[node, label={[below right=5pt and -9pt]:\rotatebox{-30}{\footnotesize $5+2N_R$}}] (A2) [right=6mm of A1] {};
      \node[node, label={[below right=5pt and -9pt]:\rotatebox{-30}{\footnotesize $6+3N_R$}}] (A3) [right=6mm of A2] {};
      \node[node, label={[below right=5pt and -9pt]:\rotatebox{-30}{\footnotesize $8+4N_R$}}] (A4) [right=6mm of A3] {};
      \node[node, label={[below right=5pt and -9pt]:\rotatebox{-30}{\footnotesize $10+5N_R$}}] (A5) [right=6mm of A4] {};
      \node[node, label={[below right=5pt and -9pt]:\rotatebox{-30}{\footnotesize $12+6N_R$}}] (N3) [right=6mm of A5] {};
      \node[node, label={[below right=5pt and -9pt]:\rotatebox{-30}{\footnotesize $8+4N_R$}}] (B4) [right=6mm of N3] {};
      \node[node, label={[below right=5pt and -9pt]:\rotatebox{-30}{\footnotesize $4+2N_R$}}] (B2) [right=6mm of B4] {};
      \node[node, label=right:{\footnotesize $6+3N_R$}] (Nu) [above=5mm of N3] {};
      \node[node, label=below left:\rotatebox{30}{\footnotesize $4+N_L$}] (A1s)  [left=6mm of Zk] {};
      \node[node, label=below left:\rotatebox{30}{\footnotesize $5+2N_L$}] (A2s) [left=6mm of A1s] {};
      \node[node, label=below left:\rotatebox{30}{\footnotesize $6+3N_L$}] (A3s) [left=6mm of A2s] {};
      \node[node, label=below left:\rotatebox{30}{\footnotesize $7+4N_L$}] (A4s) [left=6mm of A3s] {};
      \node[node, label=below left:\rotatebox{30}{\footnotesize $8+5N_L$}] (A5s) [left=6mm of A4s] {};
      \node[node, label=below left:\rotatebox{30}{\footnotesize $9+6N_L$}] (N3s) [left=6mm of A5s] {};
      \node[node, label=below left:\rotatebox{30}{\footnotesize $6+4N_L$}] (B4s) [left=6mm of N3s] {};
      \node[node, label=below left:\rotatebox{30}{\footnotesize $3+2N_L$}] (B2s) [left=6mm of B4s] {};
      \node[node, label=left:{\footnotesize $4+3N_L$}] (Nus) [above=5mm of N3s] {};

      \draw (Zk.east) -- (A1.west);
      \draw (A1.east) -- (A2.west);
      \draw (A2.east) -- (A3.west);
      \draw (A3.east) -- (A4.west);
      \draw (A4.east) -- (A5.west);
      \draw (A5.east) -- (N3.west);
      \draw (N3.east) -- (B4.west);
      \draw (B4.east) -- (B2.west);
      \draw (N3.north) -- (Nu.south); 

    \draw (Zk.west) -- (A1s.east);
      \draw (A1s.west) -- (A2s.east);
      \draw (A2s.west) -- (A3s.east);
      \draw (A3s.west) -- (A4s.east);
      \draw (A4s.west) -- (A5s.east);
      \draw (A5s.west) -- (N3s.east);
      \draw (N3s.west) -- (B4s.east);
      \draw (B4s.west) -- (B2s.east);
      \draw (N3s.north) -- (Nus.south); 
    \end{tikzpicture} }
  \end{gathered} $ \\ 
    
        \footnotesize$ \mathcal{K}_{N_L,N_R,3} \left( 3', 3'\right)$ & $6+3(N_L+N_R-1)$ & $    \begin{gathered}
    \resizebox{0.85\textwidth}{!}{
    \begin{tikzpicture}
      \node[node, label=below:{\footnotesize $3$}] (Zk) [] {};
      \node[node, label={[below right=5pt and -9pt]:\rotatebox{-30}{\footnotesize $4+N_R$}}] (A1)  [right=6mm of Zk] {};
      \node[node, label={[below right=5pt and -9pt]:\rotatebox{-30}{\footnotesize $5+2N_R$}}] (A2) [right=6mm of A1] {};
      \node[node, label={[below right=5pt and -9pt]:\rotatebox{-30}{\footnotesize $6+3N_R$}}] (A3) [right=6mm of A2] {};
      \node[node, label={[below right=5pt and -9pt]:\rotatebox{-30}{\footnotesize $7+4N_R$}}] (A4) [right=6mm of A3] {};
      \node[node, label={[below right=5pt and -9pt]:\rotatebox{-30}{\footnotesize $8+5N_R$}}] (A5) [right=6mm of A4] {};
      \node[node, label={[below right=5pt and -9pt]:\rotatebox{-30}{\footnotesize $9+6N_R$}}] (N3) [right=6mm of A5] {};
      \node[node, label={[below right=5pt and -9pt]:\rotatebox{-30}{\footnotesize $6+4N_R$}}] (B4) [right=6mm of N3] {};
      \node[node, label={[below right=5pt and -9pt]:\rotatebox{-30}{\footnotesize $3+2N_R$}}] (B2) [right=6mm of B4] {};
      \node[node, label=right:{\footnotesize $4+3N_R$}] (Nu) [above=5mm of N3] {};
      \node[node, label=below left:\rotatebox{30}{\footnotesize $4+N_L$}] (A1s)  [left=6mm of Zk] {};
      \node[node, label=below left:\rotatebox{30}{\footnotesize $5+2N_L$}] (A2s) [left=6mm of A1s] {};
      \node[node, label=below left:\rotatebox{30}{\footnotesize $6+3N_L$}] (A3s) [left=6mm of A2s] {};
      \node[node, label=below left:\rotatebox{30}{\footnotesize $7+4N_L$}] (A4s) [left=6mm of A3s] {};
      \node[node, label=below left:\rotatebox{30}{\footnotesize $8+5N_L$}] (A5s) [left=6mm of A4s] {};
      \node[node, label=below left:\rotatebox{30}{\footnotesize $9+6N_L$}] (N3s) [left=6mm of A5s] {};
      \node[node, label=below left:\rotatebox{30}{\footnotesize $6+4N_L$}] (B4s) [left=6mm of N3s] {};
      \node[node, label=below left:\rotatebox{30}{\footnotesize $3+2N_L$}] (B2s) [left=6mm of B4s] {};
      \node[node, label=left:{\footnotesize $4+3N_L$}] (Nus) [above=5mm of N3s] {};

      \draw (Zk.east) -- (A1.west);
      \draw (A1.east) -- (A2.west);
      \draw (A2.east) -- (A3.west);
      \draw (A3.east) -- (A4.west);
      \draw (A4.east) -- (A5.west);
      \draw (A5.east) -- (N3.west);
      \draw (N3.east) -- (B4.west);
      \draw (B4.east) -- (B2.west);
      \draw (N3.north) -- (Nu.south); 

    \draw (Zk.west) -- (A1s.east);
      \draw (A1s.west) -- (A2s.east);
      \draw (A2s.west) -- (A3s.east);
      \draw (A3s.west) -- (A4s.east);
      \draw (A4s.west) -- (A5s.east);
      \draw (A5s.west) -- (N3s.east);
      \draw (N3s.west) -- (B4s.east);
      \draw (B4s.west) -- (B2s.east);
      \draw (N3s.north) -- (Nus.south); 
    \end{tikzpicture} }
  \end{gathered} $ \\ \bottomrule
    \end{tabular}
    \caption{The $\mathcal{K}_{N_L,N_R,3} \left( \rho_L, \rho_R \right)$ LSTs. We provide the constant $\kappa_R$ and the associated magnetic quiver.}
    \label{tab:3E8LSTk=3}
\end{table}
\end{landscape}

Next, we turn to the exploration of the $\mathrm{Spin}(32)/\mathbb{Z}_2$ heterotic $\mathbb{C}^2/\mathbb{Z}_3$ orbifold LSTs. They have the following description at the generic point of the tensor branch\footnote{Note that the ranks of the gauge algebras are always positive in the cases that we are interested in, due to equation \eqref{eqn:sporticus}.}
\begin{equation}\label{eqn:KtildeNsu3}
     \widetilde{\mathcal{K}}_{N,3} \left(p, 16-p\right)= [SO(2p)]\stackon{$1$}{$\mathfrak{sp}_N$} \stackunder{\stackon{$1$}{$\mathfrak{su}_{2N+8-p}$}}{$[N_{\Lambda^2}=1]$}[SU(16-p)] \,,
\end{equation}
where $p$ in range $[0, 16]$ is an integer which describes the homomorphism $\mathbb{Z}_3 \rightarrow \mathrm{Spin}(32)/\mathbb{Z}_2$. 
For the $\mathrm{Spin}(32)/\mathbb{Z}_2$ LST with trivial homomorphism, i.e., $p = 16$, we can directly utilize the brane system as described in Section \ref{sec:Tdual} to understand the T-dual, which is the $E_8 \times E_8$ LST with trivial homomorphisms. We have
\begin{equation}
    \mathcal{K}_{N_L, N_R, K}(1+1+1, 1+1+1) \quad \xleftrightarrow{\, \text{ T-dual } \,} \quad \widetilde{\mathcal{K}}_{N,3} \left(16, 0\right) \,.
\end{equation}
As previously mentioned, care must be taken with the mapping of the instanton numbers, $N_L$, $N_R$, and $N$. In this case, we find that
\begin{equation}\label{eqn:sporticus}
    N = N_L + N_R + 6 \,,
\end{equation}
as observed in \cite{DelZotto:2022ohj}. We do not worry about which particular combination of the pair $(N_L, N_R)$ is T-dual to the $\mathrm{Spin}(32)/\mathbb{Z}_2$ model, as the quantities that we discuss in this section are insensitive to this information.

The magnetic quiver for the LSTs associated to the tensor branch geometries in equation \eqref{eqn:KtildeNsu3} can be straightforwardly worked out from the brane system. The magnetic quiver is:\footnote{For particular small and large values of $p$, one needs to be careful when reading off these magnetic quivers from the equations.}
\begin{itemize}
    \item For $0\le p \leq 8$ :
    \begin{equation}\label{eqn:robbie}
   \begin{gathered}
    \resizebox{0.8\textwidth}{!}{
    \begin{tikzpicture}
      \node[node, label=below right:\rotatebox{-30}{\footnotesize $2N+4$}] (Zk) [] {};
      \node[tnode] (A1) [right=6mm of Zk] {\footnotesize $\cdots$};
      \node[node, label={[below right=5pt and -9pt]:\rotatebox{-30}{\footnotesize $2N+10-p$}}] (A2)  [right=6mm of A1] {};
      \node[node, label={[below right=5pt and -9pt]:\rotatebox{-30}{\footnotesize $2N+10-p$}}] (A3) [right=6mm of A2] {};
      \node[tnode] (A4) [right=6mm of A3] {$\cdots$};
      \node[node, label={[below right=5pt and -9pt]:\rotatebox{-30}{\footnotesize $2N+10-p$}}] (A5) [right=6mm of A4] {};
      \node[node, label={[below right=5pt and -9pt]:\rotatebox{-30}{\footnotesize $2N+10-p$}}] (N3) [right=6mm of A5] {};
      \node[node, label={[below right=5pt and -9pt]:\rotatebox{-30}{\footnotesize $\lceil{N+5-\frac{p}{2}}\rceil$}}] (B4) [right=6mm of N3] {};
      \node[node, label={[below right=5pt and -9pt]:\rotatebox{-30}{\footnotesize $1$}}] (B2) [right=6mm of B4] {};
      \node[node, label=right:{\footnotesize $\floor{N+5-\frac{p}{2}}$}] (Nu) [above=5mm of N3] {};
      \node[node, label=below right:\rotatebox{-30}{\footnotesize $2N+3$}] (A1s)  [left=6mm of Zk] {};
      \node[node, label=below:{\footnotesize $2N+2$}] (A2s) [left=6mm of A1s] {};
      \node[node, label=below left:\rotatebox{30}{\footnotesize $2N+2$}] (A3s) [left=6mm of A2s] {};
      \node[tnode] (A4s) [left=6mm of A3s] {\footnotesize $\cdots$};
      \node[node, label=below left:\rotatebox{30}{\footnotesize $2N+2$}] (A5s) [left=6mm of A4s] {};
      \node[node, label=below left:\rotatebox{30}{\footnotesize $2N+2$}] (N3s) [left=6mm of A5s] {};
      \node[node, label=below left:\rotatebox{30}{\footnotesize $N+1$}] (B4s) [left=6mm of N3s] {};
      \node[node, label=left:{\footnotesize $N+1$}] (Nus) [above=5mm of N3s] {};
      \draw (Zk.east) -- (A1.west);
      \draw (A1.east) -- (A2.west);
      \draw (A2.east) -- (A3.west);
      \draw (A3.east) -- (A4.west);
      \draw (A4.east) -- (A5.west);
      \draw (A5.east) -- (N3.west);
      \draw (N3.east) -- (B4.west);
      \draw (B4.east) -- (B2.west);
      \draw (N3.north) -- (Nu.south); 
    \draw (Zk.west) -- (A1s.east);
      \draw (A1s.west) -- (A2s.east);
      \draw (A2s.west) -- (A3s.east);
      \draw (A3s.west) -- (A4s.east);
      \draw (A4s.west) -- (A5s.east);
      \draw (A5s.west) -- (N3s.east);
      \draw (N3s.west) -- (B4s.east);
      \draw (N3s.north) -- (Nus.south);
    \draw[decorate,decoration={brace,mirror}] (-6.5,-1)--(-2.5,-1) node[midway,below=2mm] () {\footnotesize $p$ Balanced nodes};
    \draw[decorate,decoration={brace,raise=2mm}] (-2,0)--(2.5,0) node[midway,above=2mm] () {\footnotesize $8-p$ nodes};
    \draw[decorate,decoration={brace,raise=2mm}] (2.7,0)--(5.5,0) node[midway,above=2mm] () {\footnotesize $6$  nodes};
    \end{tikzpicture} }
  \end{gathered}    \,. 
\end{equation}
\item For $8 < p \le 16$ :
    \begin{equation}\label{eqn:robbie2}
   \begin{gathered}
    \resizebox{0.8\textwidth}{!}{
    \begin{tikzpicture}
      \node[node, label=below right:\rotatebox{-30}{\footnotesize $2N$}] (Zk) [] {};
      \node[tnode] (A1) [right=6mm of Zk] {\footnotesize $\cdots$};
      \node[node, label={[below right=5pt and -9pt]:\rotatebox{-30}{\footnotesize $2N+10-p$}}] (A2)  [right=6mm of A1] {};
      \node[node, label={[below right=5pt and -9pt]:\rotatebox{-30}{\footnotesize $2N+10-p$}}] (A3) [right=6mm of A2] {};
      \node[tnode] (A4) [right=6mm of A3] {$\cdots$};
      \node[node, label={[below right=5pt and -9pt]:\rotatebox{-30}{\footnotesize $2N+10-p$}}] (A5) [right=6mm of A4] {};
      \node[node, label={[below right=5pt and -9pt]:\rotatebox{-30}{\footnotesize $2N+10-p$}}] (N3) [right=6mm of A5] {};
      \node[node, label={[below right=5pt and -9pt]:\rotatebox{-30}{\footnotesize $\lceil{N+5-\frac{p}{2}}\rceil$}}] (B4) [right=6mm of N3] {};
      \node[node, label={[below right=5pt and -9pt]:\rotatebox{-30}{\footnotesize $1$}}] (B2) [right=6mm of B4] {};
      \node[node, label=right:{\footnotesize $\floor{N+5-\frac{p}{2}}$}] (Nu) [above=5mm of N3] {};
      \node[node, label=below right:\rotatebox{-30}{\footnotesize $2N+1$}] (A1s)  [left=6mm of Zk] {};
      \node[node, label=below:{\footnotesize $2N+2$}] (A2s) [left=6mm of A1s] {};
      \node[node, label=below left:\rotatebox{30}{\footnotesize $2N+2$}] (A3s) [left=6mm of A2s] {};
      \node[tnode] (A4s) [left=6mm of A3s] {\footnotesize $\cdots$};
      \node[node, label=below left:\rotatebox{30}{\footnotesize $2N+2$}] (A5s) [left=6mm of A4s] {};
      \node[node, label=below left:\rotatebox{30}{\footnotesize $2N+2$}] (N3s) [left=6mm of A5s] {};
      \node[node, label=below left:\rotatebox{30}{\footnotesize $N+1$}] (B4s) [left=6mm of N3s] {};
      \node[node, label=left:{\footnotesize $N+1$}] (Nus) [above=5mm of N3s] {};
      \draw (Zk.east) -- (A1.west);
      \draw (A1.east) -- (A2.west);
      \draw (A2.east) -- (A3.west);
      \draw (A3.east) -- (A4.west);
      \draw (A4.east) -- (A5.west);
      \draw (A5.east) -- (N3.west);
      \draw (N3.east) -- (B4.west);
      \draw (B4.east) -- (B2.west);
      \draw (N3.north) -- (Nu.south); 
    \draw (Zk.west) -- (A1s.east);
      \draw (A1s.west) -- (A2s.east);
      \draw (A2s.west) -- (A3s.east);
      \draw (A3s.west) -- (A4s.east);
      \draw (A4s.west) -- (A5s.east);
      \draw (A5s.west) -- (N3s.east);
      \draw (N3s.west) -- (B4s.east);
      \draw (N3s.north) -- (Nus.south);
    \draw[decorate,decoration={brace,mirror}] (-6.5,-1)--(-2.5,-1) node[midway,below=2mm] () {\footnotesize $8$ Balanced nodes};
    \draw[decorate,decoration={brace,raise=2mm}] (-2,0)--(2.5,0) node[midway,above=2mm] () {\footnotesize $p-7$ nodes};
    \draw[decorate,decoration={brace,raise=2mm}] (2.7,0)--(5.5,0) node[midway,above=2mm] () {\footnotesize $13-p$  nodes};
    \end{tikzpicture} }
  \end{gathered}    \,. 
  \end{equation}
\end{itemize}

Here, we performed the $\mathfrak{d}_{16}$ transition described in Section \ref{sec:spin32} to move to infinite coupling. We can also determine the structure constant $\kappa_R$, which is
\begin{equation}
    \kappa_R=3N+9-p  \,.
\end{equation}

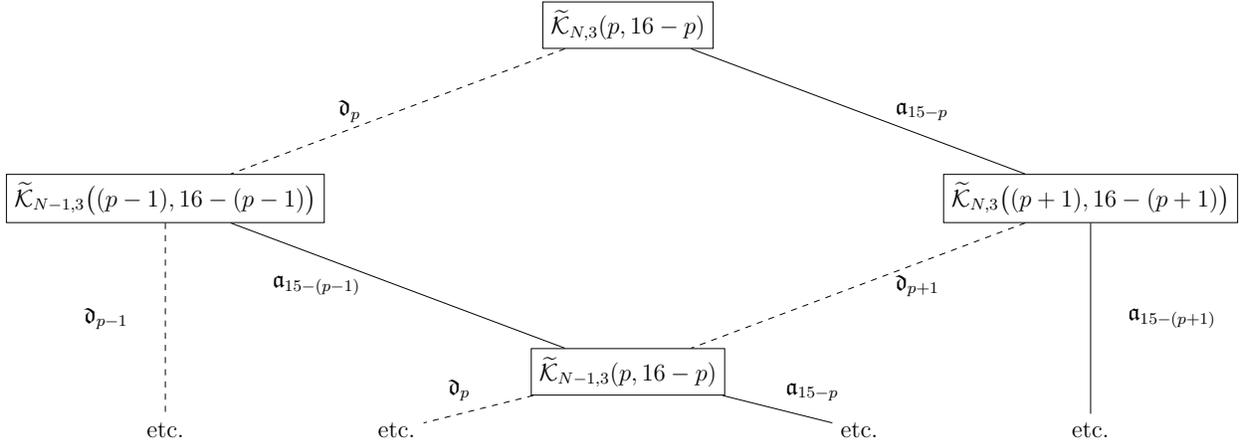
\begin{figure}[t]
    \centering
    \resizebox{\textwidth}{!}{
    \begin{tikzpicture}
        \node[draw] at (0,0) (nodo0) {$\widetilde{\mathcal{K}}_{N,3} (p,16-p)$};
        \node[draw] at (8,-3) (nodo1r) {$\widetilde{\mathcal{K}}_{N,3} \big( (p+1),16-(p+1) \big)$};
        \node[draw] at (-8,-3) (nodo1l) {$\widetilde{\mathcal{K}}_{N-1,3} \big( (p-1),16-(p-1) \big)$};
\draw[dashed] (nodo0)--(nodo1l) node[midway, left=0.5cm] () {$\mathfrak{d}_{p}$};
\draw (nodo0)--(nodo1r) node[midway, right=0.5cm] () {$\mathfrak{a}_{15-p}$};
\node[draw] at (0,-6) (nodo2) {$\widetilde{\mathcal{K}}_{N-1,3} (p,16-p)$};
        \node[] at (-8,-7) (nodo2l) {$\mathrm{etc.}$};
        \node[] at (-4,-7) (nodo2cl) {$\mathrm{etc.}$};
        \node[] at (4,-7) (nodo2cr) {$\mathrm{etc.}$};
        \node[] at (8,-7) (nodo2r) {$\mathrm{etc.}$};
\draw[dashed] (nodo1r)--(nodo2) node[midway, right=0.5cm] () {$\mathfrak{d}_{p+1}$};
\draw (nodo1l)--(nodo2) node[midway, left=0.5cm] () {$\mathfrak{a}_{15-(p-1)}$};      
\draw (nodo1r)--(nodo2r) node[midway, right=0.5cm] () {$\mathfrak{a}_{15-(p+1)}$};
\draw[dashed] (nodo1l)--(nodo2l) node[midway, left=0.5cm] () {$\mathfrak{d}_{p-1}$}; 
\draw[dashed] (nodo2)--(nodo2cl) node[midway, above left] () {$\mathfrak{d}_{p}$};
\draw (nodo2)--(nodo2cr) node[midway, above right] () {$\mathfrak{a}_{15-p}$};
    \end{tikzpicture}
    }
    \caption{Heterotic $\mathrm{Spin}\left(32\right)/ \mathbb{Z}_2$ $\widetilde{\mathcal{K}}_{N,3} (p,16-p)$ LST Hasse Diagram, the diagram is ordered by Higgs branch dimension thus the top of this Hasse diagram corresponds to the bottom point of the Hasse diagram engineered via quiver subtraction. Dashed slices connect theories with different values of $N$.}
    \label{fig:HasseLST3Spin32}
\end{figure}

We can now use the quiver subtraction algorithm to determine the Hasse diagram of the magnetic quiver in equation \eqref{eqn:robbie}. We draw a subdiagram of the Hasse diagram for the LST $\widetilde{\mathcal{K}}_{N,3} \left(16, 0\right)$ in Figure \ref{fig:HasseLST3Spin32}, where we draw only vertices that correspond to $K=3$ LSTs. We also focus only on Higgs branch RG flows that do not involve the Higgsing of matter in the second rank antisymmetric representation of the $\mathfrak{su}$ gauge algebra; such a Higgsing leads to a different model, which although in principle possible to study with these methods, we refrain from exploring here. Each elementary slice appearing in the Hasse diagram in Figure \ref{fig:HasseLST3Spin32} decreases the structure constant $\kappa_R$ and the Coulomb branch dimension in the same way as in equation \eqref{eqn:theexplorer}.

Thus, looking at the Hasse diagrams of each heterotic model starting from the T-dual pair $\mathcal{K}_{N_L,N_R,3} \left( 1+1+1, 1+1+1\right) \sim \widetilde{\mathcal{K}}_{N,3} \left(16, 0\right)$, where we kept implicit the equality $N=N_L+N_R+6$, as in Figures \ref{fig:HasseLST3E8} and \ref{fig:HasseLST3Spin32}, we can tabulate all the dual families. They appear in Table \ref{tbl:K3}. There are three families, corresponding to the value of $\kappa_R \, \text{mod} \, 3$, as has been pointed out in \cite{DelZotto:2022ohj}. Thus, we have generated a large collection of 6d $(1,0)$ LSTs that have the same T-duality-invariant properties. We leave a geometric study and verification of these putative T-dualities for future work. 

In this section, we have depicted the vertices on the Hasse diagram that corresponded to LSTs with $K=3$. However, when considering the quiver subtraction algorithm, we also find vertices which correspond to the $K=2$ heterotic orbifolds. The elementary slices, which are generically $A_m$ where $m$ has a dependence on $N_L + N_R$, typically involve a larger than unit change in $\kappa_R$ and $\operatorname{dim}(\mathcal{C})$, however, it is straightforward after drawing the Hasse diagram to identify T-dual pairs involving different values of $K$. An example of which consists of:
\begin{equation}
    \mathcal{K}_{N_L,N_R,3}(3,2'+1) \quad \xleftrightarrow{\, \text{ T-dual }\, \,\,} \quad \mathcal{K}_{\widetilde{N}_L,\widetilde{N}_R,2}(1+1,1+1) \,,
\end{equation}
when
\begin{equation}
    2(N_L+N_R)=3(\widetilde{N}_L+\widetilde{N}_R) \,.
\end{equation}

\section{Beyond \texorpdfstring{\boldmath{$\mathfrak{su}(K)$}}{su(K)}}\label{sec:beyond}

Throughout this paper, we have considered the LSTs corresponding to (Higgsed) $(N_L + N_R)$ $\mathbb{C}^2/\mathbb{Z}_K$ ALE instantons in $E_8 \times E_8$ heterotic string theory. Such LSTs can be obtained via the fusing of the $\mathfrak{su}(K)$ flavor symmetry of the (Higgsed) rank $N_L$ and rank $N_R$ $(\mathfrak{e}_8, \mathfrak{su}(K))$ orbi-instanton SCFTs. 

More generally, we can consider the rank $N$ $(\mathfrak{e}_8, \mathfrak{g})$ orbi-instanton SCFT, which is the worldvolume theory living on a stack of $N$ M5-branes, probing a $\mathbb{C}^2/\Gamma_\mathfrak{g}$ orbifold singularity, contained inside of an M9-brane \cite{DelZotto:2014hpa}. Such theories have an $\mathfrak{e}_8 \oplus \mathfrak{g}$ non-Abelian flavor symmetry, and are the parents under Higgs branch RG flow for a family of interacting SCFTs 
\begin{equation}
    \mathcal{O}_{N, \mathfrak{g}}(\rho, O) \,,
\end{equation}
where $\rho$ is a homomorphism $\Gamma_\mathfrak{g} \rightarrow E_8$ and $O$ is a nilpotent orbit of $\mathfrak{g}$. Let $O = 0$ be the nilpotent orbit of $\mathfrak{g}$ of trivial dimension, and then we can consider the little string theories obtained by fusing two such orbi-instantons along the $\mathfrak{g}$ flavor symmetry associated with the nilpotent orbits $O$:
\begin{equation}\label{eqn:snooty}
    \mathcal{K}_{N_L, N_R, \mathfrak{g}}(\rho_L, \rho_R) = \mathcal{O}_{N_L, \mathfrak{g}}(\rho_L, 0)  \,\,\text{---}\,\, \overset{\mathfrak{g}}{0} \,\,\text{---}\,\, \mathcal{O}_{N_R, \mathfrak{g}}(\rho_R, 0) \,.
\end{equation}
This LST is the theory of (Higgsed) $(N_L + N_R)$ $\mathbb{C}^2/\Gamma_\mathfrak{g}$ ALE instantons in $E_8 \times E_8$ heterotic string theory. 

From the geometric description in equation \eqref{eqn:snooty}, we can compute all of the numeric invariants in equation \eqref{eqn:invariants}, however, as usual, we see that they depend only on the linear combination $(N_L + N_R)$ and not the individual values of $N_L$ and $N_R$. For $\mathfrak{g} = \mathfrak{su}(K)$, we could go beyond this data and study the Hasse diagram of interacting fixed points on the Higgs branch of the LST, and this is sensitive to $N_L$ and $N_R$ individually. Such an analysis was possible as we could exploit a Type IIA brane system approach to determine the magnetic quiver describing the Higgs branch of $\mathfrak{g} = \mathfrak{su}(K)$ LSTs. The magnetic quiver approach is not available for generic $\mathfrak{g}$.\footnote{For $\mathfrak{g}=\mathfrak{so}(2K+8)$, a Type IIA brane system involving orientifold planes exists, which can, in some favorable cases, allow for the extraction of a magnetic quiver; see, e.g., \cite{Sperling:2021fcf}. For $\mathfrak{g}$ an exceptional Lie algebra, we must turn to alternative methods.} However, there is at least one case where we have access to (part of) the Higgs branch Hasse diagram for the heterotic $E_8 \times E_8$ LSTs with $\mathfrak{g} = \mathfrak{e}_6$. 

In Section \ref{sec:build}, the orbi-instanton building blocks involved an integer $N \geq 1$ of M5-branes inside of the M9-brane. In fact, there are natural 6d $(1,0)$ SCFTs associated to the $N = 0$ orbi-instantons. Consider the tensor branch effective field theories for the rank $N$ $(\mathfrak{e}_8, \mathfrak{g})$ orbi-instanton SCFT, for $\mathfrak{g} = \mathfrak{su}(K)$, $\mathfrak{so}(2K)$, $\mathfrak{e}_6$, $\mathfrak{e}_7$, and $\mathfrak{e}_8$, respectively:
\begin{equation}
    \begin{aligned}
        1\,2\overset{\mathfrak{su}_2}{2}\cdots\overset{\mathfrak{su}_K}{2} \, \underbrace{\, \overset{\mathfrak{su}_K}{2} \cdots \overset{\mathfrak{su}_K}{2} \,}_{N-1} &\,, \\
        1 \, 2 \overset{\mathfrak{su}_2}{2} \overset{\mathfrak{g}_2}{3} 1 \overset{\mathfrak{so}_9}{4} \overset{\mathfrak{sp}_1}{1} \overset{\mathfrak{so}_{11}}{4}
        \overset{\mathfrak{sp}_2}{1} \cdots \overset{\mathfrak{sp}_{K-4}}{1} \overset{\mathfrak{so}_{2K}}{4}\overset{\mathfrak{sp}_{K-4}}{1} \underbrace{\, \overset{\mathfrak{so}_{2K}}{4}\overset{\mathfrak{sp}_{K-4}}{1} \cdots \overset{\mathfrak{so}_{2K}}{4}\overset{\mathfrak{sp}_{K-4}}{1}\,}_{N-1} &\,, \\
        1\,2\overset{\mathfrak{su}_2}{2}\overset{\mathfrak{g}_2}{3}1\overset{\mathfrak{f}_4}{5}1\overset{\mathfrak{su}_3}{3}1\overset{\mathfrak{e}_6}{6}1\overset{\mathfrak{su}_3}{3}1 \, \underbrace{\, \overset{\mathfrak{e}_6}{6}1\overset{\mathfrak{su}_3}{3}1 \cdots \overset{\mathfrak{e}_6}{6}1\overset{\mathfrak{su}_3}{3}1 \,}_{N-1} &\,, \\
        1 \, 2 \overset{\mathfrak{su}_2}{2} \overset{\mathfrak{g}_2}{3} 1 \overset{\mathfrak{f}_4}{5} 1 \overset{\mathfrak{g}_2}{3} \overset{\mathfrak{su}_2}{2} 1 \overset{\mathfrak{e}_7}{8} 1 \overset{\mathfrak{su}_2}{2} \overset{\mathfrak{so}_7}{3} \overset{\mathfrak{su}_2}{2} 1 \underbrace{\, \overset{\mathfrak{e}_7}{8} 1 \overset{\mathfrak{su}_2}{2} \overset{\mathfrak{so}_7}{3} \overset{\mathfrak{su}_2}{2} 1 \cdots \overset{\mathfrak{e}_7}{8} 1 \overset{\mathfrak{su}_2}{2} \overset{\mathfrak{so}_7}{3} \overset{\mathfrak{su}_2}{2} 1 \,}_{N-1} &\,, \\
        1 \, 2 \overset{\mathfrak{su}_2}{2} \overset{\mathfrak{g}_2}{3} 1 \overset{\mathfrak{f}_4}{5} 1 \overset{\mathfrak{g}_2}{3} \overset{\mathfrak{su}_2}{2} 2 1 \overset{\displaystyle 1}{\overset{\mathfrak{e}_8}{(12)}} 1 \, 2 \overset{\mathfrak{su}_2}{2} \overset{\mathfrak{g}_2}{3} 1 \overset{\mathfrak{f}_4}{5} 1 \overset{\mathfrak{g}_2}{3} \overset{\mathfrak{su}_2}{2} 2 1 \underbrace{\, \overset{\mathfrak{e}_8}{(12)} 1 \, 2 \overset{\mathfrak{su}_2}{2} \overset{\mathfrak{g}_2}{3} 1 \overset{\mathfrak{f}_4}{5} 1 \overset{\mathfrak{g}_2}{3} \overset{\mathfrak{su}_2}{2} 2 1 \cdots \overset{\mathfrak{e}_8}{(12)} 1 \, 2 \overset{\mathfrak{su}_2}{2} \overset{\mathfrak{g}_2}{3} 1 \overset{\mathfrak{f}_4}{5} 1 \overset{\mathfrak{g}_2}{3} \overset{\mathfrak{su}_2}{2} 2 1 \,}_{N-1} &\,.
    \end{aligned}
\end{equation}
The continuation to $N = 0$ is then clear.\footnote{It is interesting to note that the rank $0$ $(\mathfrak{e}_8, \mathfrak{e}_8)$ orbi-instanton is a product of two interacting SCFTs: the rank one E-string and rank one $(\mathfrak{e}_8, \mathfrak{e}_8)$ conformal matter.} The $E_8$-homomorphism Higgsing, for $N \geq 1$, leads to a new interacting SCFT for which the tensor branch effective field theory is only modified along the ramp, and the part on the right, which is repeated $N-1$ times, is unmodified. For $N = 0$, some $E_8$-homomorphisms will not trigger a flow to interacting SCFT fixed points.

Conveniently, the rank $0$ $(\mathfrak{e}_8, \mathfrak{e}_6)$ orbi-instanton SCFT has the following description at the generic point of its tensor branch:
\begin{equation}
    \begin{gathered} 1\,2\overset{\mathfrak{su}_2}{2}\overset{\mathfrak{g}_2}{3}1\overset{\mathfrak{f}_4}{5}1\overset{\mathfrak{su}_3}{3}1 \end{gathered} \,.
\end{equation}
This is exactly the same tensor branch as the rank $1$ $(\mathfrak{e}_8, \mathfrak{e}_8)$ conformal matter theory, Higgsed on the left and the right by the pair of $\mathfrak{e}_8$ nilpotent orbits $0$ and $A_2$,\footnote{We use the standard Bala--Carter notation \cite{MR417306,MR417307} to label the nilpotent orbits of exceptional Lie algebras.} respectively.\footnote{The Higgsed conformal matter theories are obtained from conformal matter via giving nilpotent vacuum expectation values to the moment maps of the two flavor symmetry factors; these SCFTs are usually denoted $\mathcal{T}_{\mathfrak{g}, N}(O_L, O_R)$, where $O_L$ and $O_R$ are nilpotent orbits of $\mathfrak{g}$.} The interacting fixed points on the Higgs branch of this SCFT have been extensively studied in \cite{Baume:2021qho}. We consider further Higgsing to the theories $\mathcal{T}_{\mathfrak{e}_8, 1}(O, A_2)$ for $O$ any nilpotent orbit of $\mathfrak{e}_8$ in Table \ref{tbl:e6nilpoi}. From the tensor branch descriptions of each of these Higgsed theories, it is straightforward to see that, for each nilpotent orbit $O$ in Table \ref{tbl:e6nilpoi}, there is a homomorphism $\rho: \Gamma_{\mathfrak{e}_6} \rightarrow E_8$, such that the $E_8$-homomorphism Higgsing of the rank zero orbi-instanton gives the same SCFT as the nilpotent Higgsing of the conformal matter theory.\footnote{The tensor branch descriptions associated with the homomorphisms $\Gamma_{\mathfrak{e}_6} \rightarrow E_8$ were proposed in \cite{Frey:2018vpw}. The tensor branch descriptions associated with a pair of $\mathfrak{e}_8$ nilpotent orbits (used to Higgs rank one $(\mathfrak{e}_8, \mathfrak{e}_8)$ conformal matter) were given in \cite{Baume:2021qho}. Here, we point out that some $E_8$-homomorphisms and some pairs of $\mathfrak{e}_8$ nilpotent orbits lead to the same tensor branch, and thus the same SCFT; we highlight that these different perspectives are useful in extracting the physical properties.} Thus, we can label these $E_8$-homomorphisms by their corresponding nilpotent orbit.

\begin{table}[p]
    \centering
    \begin{threeparttable}
        \begin{tabular}{crccccc}
         \toprule
         BC Label & Tensor Branch & Flavor & $\kappa_F$ & $\delta\kappa_R$ & $d_C$ & $d_H$ \\\midrule
         $0$ & $\begin{gathered} 1\,2\overset{\mathfrak{su}_2}{2}\overset{\mathfrak{g}_2}{3}1\overset{\mathfrak{f}_4}{5}1\overset{\mathfrak{su}_3}{3}1 \end{gathered}$ & $\mathfrak{e}_8$ & $1$ & $31$ & $18$ & $192$\\
         $A_1$ & $\begin{gathered} 1\overset{\mathfrak{su}_2}{2}\overset{\mathfrak{g}_2}{3}1\overset{\mathfrak{f}_4}{5}1\overset{\mathfrak{su}_3}{3}1 \end{gathered}$ & $\mathfrak{e}_7$ & $1$ & $30$ & $17$ & $163$\\
         $2A_1$ & $\begin{gathered} \overset{\mathfrak{su}_2}{1}\overset{\mathfrak{g}_2}{3}1\overset{\mathfrak{f}_4}{5}1\overset{\mathfrak{su}_3}{3}1 \end{gathered}$ & $\mathfrak{so}_{13}$ & $1$ & $29$ & $16$ & $146$\\
         $3A_1$ & $\begin{gathered} 1\overset{\mathfrak{g}_2}{3}1\overset{\mathfrak{f}_4}{5}1\overset{\mathfrak{su}_3}{3}1 \end{gathered}$ & $\mathfrak{f}_4 \oplus \mathfrak{su}_2$ & $1$, $1$ & $28$ & $15$ & $136$\\
         $A_2$ & $\begin{gathered} 1\overset{\mathfrak{su}_3}{3}1\overset{\mathfrak{f}_4}{5}1\overset{\mathfrak{su}_3}{3}1 \end{gathered}$ & $\mathfrak{e}_6$ & $1$ & $27$ & $15$ & $135$\\
         $4A_1$ & $\begin{gathered} \overset{\mathfrak{g}_2}{2}1\overset{\mathfrak{f}_4}{5}1\overset{\mathfrak{su}_3}{3}1 \end{gathered}$ & $\mathfrak{sp}_4$ & $1$& $27$ & $14$ & $148$\\
         $A_2 + A_1$ & $\begin{gathered} \overset{\mathfrak{su}_3}{2}1\overset{\mathfrak{f}_4}{5}1\overset{\mathfrak{su}_3}{3}1 \end{gathered}$ & $\mathfrak{su}_6$ & $1$ & $26$ & $14$ & $144$ \\
         $A_2 + 2A_1$ & $\begin{gathered} \overset{\mathfrak{su}_2}{2}1\overset{\mathfrak{f}_4}{5}1\overset{\mathfrak{su}_3}{3}1 \end{gathered}$ & $\mathfrak{so}_7 \oplus \mathfrak{su}_2$ & $1,3( 2)$ & $25$ & $13$ & $119$\\
         $A_2 + 3A_1$ & $\begin{gathered} 2\,1\overset{\mathfrak{f}_4}{5}1\overset{\mathfrak{su}_3}{3}1 \end{gathered}$ & $\mathfrak{g}_2 \oplus \mathfrak{su}_2$ & $1( 2)$, $1$ & $24$ & $12$ & $115$ \\
         $2A_2$ & $\begin{gathered} 1\underset{\displaystyle 1}{\overset{\mathfrak{f}_4}{5}}1\overset{\mathfrak{su}_3}{3}1 \end{gathered}$ & $\mathfrak{g}_2 \oplus \mathfrak{g}_2$ & $1$, $1$ & $23$ & $12$ & $114$ \\
         $A_3$ & $\begin{gathered} \overset{\mathfrak{su}_2}{1}\overset{\mathfrak{so}_9}{4}1\overset{\mathfrak{su}_3}{3}1 \end{gathered}$ & $\mathfrak{so}_{11} \oplus \mathfrak{u}_1$ & $1$ & $21$ & $12$ & $118$\\
         $2A_2 + A_1$ & $\begin{gathered} 1\overset{\mathfrak{f}_4}{4}1\overset{\mathfrak{su}_3}{3}1 \end{gathered}$ & $\mathfrak{g}_2 \oplus \mathfrak{sp}_1$ & $1,3$ & $22$ & $11$ & $111$\\
         $A_3 + A_1$ & $\begin{gathered} 1\overset{\mathfrak{so}_9}{4}1\overset{\mathfrak{su}_3}{3}1 \end{gathered}$ & $\mathfrak{so}_7 \oplus \mathfrak{sp}_1  \oplus \mathfrak{u}_1$ & $1,1$ & $20$ & $11$ & $110$ \\
         $D_4(a_1)$ & $\begin{gathered} 1\overset{\mathfrak{so}_8}{4}1\overset{\mathfrak{su}_3}{3}1 \end{gathered}$ & $\mathfrak{so}_8  \oplus \mathfrak{u}_1  \oplus \mathfrak{u}_1$ & $1$ & $19$ & $11$ & $109$\\
         $2A_2 + 2A_1$ & $\begin{gathered} \overset{\mathfrak{f}_4}{3}1\overset{\mathfrak{su}_3}{3}1 \end{gathered}$ & $\mathfrak{sp}_2$ & $3$ & $21$ & $10$ & $108$\\
         $A_3 + 2A_1$ & $\begin{gathered} \overset{\mathfrak{so}_9}{3}1\overset{\mathfrak{su}_3}{3}1 \end{gathered}$ & $\mathfrak{sp}_2 \oplus \mathfrak{su}_2  \oplus \mathfrak{u}_1$ & $1,1$ & $19$ & $10$ & $106$\\
         $D_4(a_1) + A_1$ & $\begin{gathered} \overset{\mathfrak{so}_8}{3}1\overset{\mathfrak{su}_3}{3}1 \end{gathered}$ & $\mathfrak{sp}_1 \oplus \mathfrak{sp}_1 \oplus \mathfrak{sp}_1  \oplus \mathfrak{u}_1  \oplus \mathfrak{u}_1$ & $1,1,1$ & $18$ & $10$ & $104$\\
         $A_3 + A_2$ & $\begin{gathered} \overset{\mathfrak{so}_7}{3}1\overset{\mathfrak{su}_3}{3}1 \end{gathered}$ & $\mathfrak{sp}_2 \oplus \mathfrak{su}_2 \oplus \mathfrak{u}_1$ & $1,2( 3)$ & $17$ & $9$ & $103$\\
         $A_3 + A_2 + A_1$ & $\begin{gathered} \overset{\mathfrak{g}_2}{3}1\overset{\mathfrak{su}_3}{3}1 \end{gathered}$ & $\mathfrak{su}_3 \oplus \mathfrak{sp}_1$ & $2( 3),1$ & $16$ & $8$ & $101$\\
         $D_4(a_1) + A_2$ & $\begin{gathered} \overset{\mathfrak{su}_3}{3}1\overset{\mathfrak{su}_3}{3}1 \end{gathered}$ & $\mathfrak{su}_3 \oplus \mathfrak{su}_3$ & $1( 3),1( 3)$ & $15$ & $8$ & $100$ \\
         $2A_3$ & $\begin{gathered} \overset{\mathfrak{su}_2}{2}\overset{\mathfrak{su}_3}{2}1 \end{gathered}$ & $\mathfrak{su}_4 \oplus \mathfrak{u}_1$ & $1( 2)$ & $11$ & $6$ & $96$\\
         \bottomrule
        \end{tabular}
    \end{threeparttable}
    \caption{Properties of the 6d $(1,0)$ SCFTs $\mathcal{T}_{\mathfrak{e}_8, 1}(O, A_2)$, where $O$ is given in the Bala--Carter label column. In the flavor column, we do not write the universal $\mathfrak{e}_6$ factor. In the $\kappa_F$ column, the structure constants are in the same order as the simple algebras in the flavor column, and the factor in parentheses is a non-unit weight in the null eigenvector.}
    \label{tbl:e6nilpoi}
\end{table}

Therefore, we can use the duality between the rank zero orbi-instanton theory and the conformal matter theory to consider the following $E_8 \times E_8$ heterotic instantons on $\mathbb{C}^2/\Gamma_{\mathfrak{e}_6}$:
\begin{equation}\label{eqn:SPECIALLSTs}
    \mathcal{K}_{N_L=0, N_R=0, \mathfrak{e}_6}(O_L, O_R) = \mathcal{T}_{\mathfrak{e}_8, 1}(O_L, A_2) \,\,\text{---}\,\, \overset{\mathfrak{e}_6}{0} \,\,\text{---}\,\, \mathcal{T}_{\mathfrak{e}_8, 1}(O_R, A_2) \,.
\end{equation}
Here, we have fused along the $\mathfrak{e}_6$ flavor symmetry associated with the $A_2$ nilpotent orbit of each conformal matter building block. At the generic point of the tensor branch of such a little string theory, the effective field theory description can be read off from Table \ref{tbl:e6nilpoi}. In this case, since $N_L = N_R = 0$, the contraction map is unique. We also determine the contributions to the relevant invariants of the LSTs from the building blocks and list them in the table.\footnote{In fact, it is straightforward to determine the numerical invariants of $\mathcal{K}_{N_L, N_R, \mathfrak{e}_6}(O_L, O_R)$, again using the correspondence between (a subset of the) $E_8$-homomorphisms and (a subset of the) $\mathfrak{e}_8$ nilpotent orbits. However, as we will see, only for $N_L = N_R = 0$ do we have a reliable alternative understanding of the Higgs branch structure, due to the duality to conformal matter.}

The class of LSTs in equation \eqref{eqn:SPECIALLSTs} is particularly interesting as the rank one $(\mathfrak{g}, \mathfrak{g})$ conformal matter theories, compactified on a torus, have known alternative descriptions in the 4d $\mathcal{N}=2$ construction known as class $\mathcal{S}$ \cite{Ohmori:2015pua,Ohmori:2015pia,Baume:2021qho}. Class $\mathcal{S}$ theories have known 3d mirrors \cite{Benini:2010uu}, and therefore we have a description of the Higgs branch of this orbi-instanton SCFT:
\begin{equation}
    \mathcal{T}_{\mathfrak{e}_8, 1}(O, A_2) \qquad \xlongrightarrow{\text{3d mirror}} \qquad \begin{gathered}
    \begin{tikzpicture}
    \node[tnode] (T1) {$T_{O}(E_8)$};
    \node[node, label=below:{\footnotesize $\mathfrak{e}_8$}] (N3) [right=9mm of T1] {};
    \node[tnode] (T2) [right=9mm of N3] {$T_{A_2}(E_8)$};
    \node[tnode] (T4) [above=6mm of N3] {$T_{E_8(a_1)}(E_8)$};
    
    \draw (T1.east) -- (N3.west);
    \draw (N3.east) -- (T2.west);    
    \draw (N3.north) -- (T4.south);
\end{tikzpicture}
    \end{gathered} \,.
\end{equation}
Here, $T_O(G)$ are the 3d $\mathcal{N}=4$ theories studied in \cite{Gaiotto:2008ak}. In Section \ref{sec:LSTs}, we argued that the magnetic quiver describing the Higgs branch of the $\mathcal{K}_{N_L, N_R, \mathfrak{su}(K)}(\rho_L, \rho_R)$ LSTs as obtained from the Type IIA brane system was identical to the putative magnetic quiver obtained from $\mathfrak{su}(K)$ Coulomb gauging of the 3d mirrors of the two orbi-instanton building blocks. Via the same Coulomb gauging logic, we expect that the 3d mirror describing the Higgs branch of the $\mathcal{K}_{N_L=0, N_R=0, \mathfrak{e}_6}(O_L, O_R)$ LSTs is given by the following theory:
\begin{equation}
\begin{gathered}
    \begin{tikzpicture}
    \node[tnode] (T1sx) {$T_{O_L}(E_8)$};
    \node[node, label=below:{\footnotesize $\mathfrak{e}_8$}] (N3sx) [right=9mm of T1] {};
    \node[tnode] (T2sx) [right=9mm of N3sx] {$T_{A_2}(E_8)$};
    \node[tnode] (T4sx) [above=6mm of N3sx] {$T_{E_8(a_1)}(E_8)$};

    \node[tnode] (T1dx) [right=9mm of T2sx] {$T_{A_2}(E_8)$};
    \node[node, label=below:{\footnotesize $\mathfrak{e}_8$}] (N3dx) [right=9mm of T1dx] {};
    \node[tnode] (T2dx) [right=9mm of N3dx] {$T_{O_R}(E_8)$};
    \node[tnode] (T4dx) [above=6mm of N3dx] {$T_{E_8(a_1)}(E_8)$};

    \draw[dashed,red] (3.5,-0.5)to[in=-90,out=180](3.0,0.0)to[in=-180,out=90](3.5,0.5)--(6.85,0.5)to[in=90,out=0](7.35,0)to[in=0,out=-90](6.85,-0.5);
    \draw[dashed,red] (3.5,-0.5)--(6.85,-0.5) node[midway,below] () {\footnotesize $E_6$ Coulomb Gauging};
    
    \draw (T1sx.east) -- (N3sx.west);
    \draw (N3sx.east) -- (T2sx.west);    
    \draw (N3sx.north) -- (T4sx.south);
    \draw (T1dx.east) -- (N3dx.west);
    \draw (N3dx.east) -- (T2dx.west);    
    \draw (N3dx.north) -- (T4dx.south);
\end{tikzpicture}
    \end{gathered} \,.
\end{equation}
Unfortunately, the quiver subtraction algorithm that we have elucidated in Algorithm \ref{alg:QS} is not formulated for such non-Lagrangian quivers, and thus we cannot determine the Higgs branch Hasse diagram from that perspective. 

Alternatively, we can explore a subdiagram of the Hasse diagram via different methods. Instead of $E_8$-homomorphisms, we are studying theories that are associated with nilpotent orbits, and nilpotent orbits have a known partial ordering, which is also related to the partial ordering on the foliation of the symplectic singularity, see, e.g., \cite{Collingwood_1993,Chacaltana:2012zy,Slodowy_1980}. The partial ordering on nilpotent orbits is related to the Higgs branch renormalization group flows of the conformal matter theories, and thus to the rank $0$ orbi-instanton and LSTs under consideration here. The Hasse diagram of $\mathcal{K}_{0,0,\mathfrak{e}_6}(O_L, O_R)$ is a truncation, dropping all vertices where there is a nilpotent orbit not appearing in Table \ref{tbl:e6nilpoi}, of the Hasse diagram that appears in Figures 7.3 and 7.4 of \cite{Baume:2021qho}; we do not redraw it here. Studying this Hasse diagram allows for the straightforward extraction of LSTs with the same T-duality-invariant properties; we list these potential T-dual families in Appendix \ref{app:bigtables}.

In this way, we have determined a variety of potential T-dualities among the heterotic $E_8 \times E_8$ $\mathbb{C}^2/\Gamma_{\mathfrak{e}_6}$ orbifold LSTs. We can then use the known T-duality between the $E_8 \times E_8$ and $\mathrm{Spin}(32)/\mathbb{Z}_2$ models with trivial homomorphisms
\begin{equation}
    \mathcal{K}_{0,0, \mathfrak{e}_6}(0, 0) \quad \xleftrightarrow{\, \text{ T-dual } \,\,} \quad \widetilde{\mathcal{K}}_{0, \mathfrak{e}_6}(\varnothing) \,,
\end{equation}
to expand the families listed in Appendix \ref{app:bigtables} via the incorporation of the $\mathrm{Spin}(32)/\mathbb{Z}_2$ LSTs.\footnote{We have followed \cite{DelZotto:2022ohj} in defining $\widetilde{\mathcal{K}}_{N, \mathfrak{e}_6}(\varnothing)$ as the LST associated to the tensor branch configuration:
\begin{equation}
    \overset{\mathfrak{sp}_{10+N}}{1}
    \overset{\mathfrak{so}_{24+4N}}{4}
    \overset{\mathfrak{sp}_{6+3N}}{1}
    \overset{\mathfrak{su}_{8+4N}}{2}
    \overset{\mathfrak{su}_{4+2N}}{2}\,.
\end{equation}} 

It is worthwhile to point out a subtlety with the study of $\kappa_F$. Let us consider a particular potential T-dual pair from Appendix \ref{app:bigtables}. The LSTs with $O_L = A_3 + A_2$ and $O_R = D_4(a_1) + A_1$ has the following flavor symmetries and structure constants
\begin{equation}
    \mathfrak{sp}(2) \oplus \mathfrak{su}(2) \oplus \mathfrak{sp}(1) \oplus \mathfrak{sp}(1) \oplus \mathfrak{sp}(1) \oplus \mathfrak{u}(1)^{\oplus 3} \,, \qquad \kappa_F = (1, 6, 1, 1, 1) \,.
\end{equation}
One of the putative T-dual theories has $O_L = A_3 + A_1$ and $O_R = D_4(a_1) + A_2$. For this theory, the flavor algebras and associated structure constants are
\begin{equation}
    \mathfrak{so}(7) \oplus \mathfrak{sp}(1) \oplus \mathfrak{su}(3) \oplus \mathfrak{su}(3) \oplus \mathfrak{u}(1) \,, \qquad \kappa_F = (1, 1, 3, 3) \,.
\end{equation}
We can see that, if these theories are to be T-dual, then the Wilson lines turned on along the $S^1$ must be such that the rank-preserving decompositions
\begin{equation}
    \mathfrak{sp}(2) \oplus \mathfrak{su}(2) \oplus \mathfrak{sp}(1) \oplus \mathfrak{sp}(1) \oplus \mathfrak{sp}(1) \oplus \mathfrak{u}(1)^{\oplus 3} \rightarrow \mathfrak{f}_\text{5d} \,,
\end{equation}
and
\begin{equation}
    \mathfrak{so}(7) \oplus \mathfrak{sp}(1) \oplus \mathfrak{su}(3) \oplus \mathfrak{su}(3) \oplus \mathfrak{u}(1) \rightarrow \mathfrak{f}_\text{5d} \,,
\end{equation}
must have the correct embedding indices to match the structure constants in 5d. A careful analysis of the branching rules may allow us to rule out T-dual pairs by arguing that there is no set of Wilson lines compatible with the matching of $\kappa_F$. We leave this for future work.

\section{Outlook}\label{sec:conc}

In this paper, we constructed magnetic quivers describing the Higgs branch of the 6d $(1,0)$ LSTs that we referred to as
\begin{equation}
    \mathcal{K}_{N_L, N_R, K}(\rho_L, \rho_R) \,.
\end{equation}
These are the theories living on the worldvolume of NS5-branes in $E_8 \times E_8$ heterotic string theory, and probing a $\mathbb{C}^2/\mathbb{Z}_K$ orbifold singularity. In the Ho\v{r}ava--Witten heterotic M-theory picture, there are $N_L$ M5-branes inside of one of the M9-branes, and $N_R$ M5-branes inside of the other. Unlike the dimensions of the branches of moduli space, the rank of the flavor algebra, and the structure constants for the generalized global symmetries, the magnetic quiver is sensitive to $N_L$ and $N_R$, not just their sum.

From the magnetic quiver describing the Higgs branch of this class of LSTs, we determine the Hasse diagram between LSTs obtained via Higgs branch renormalization group flow from the theory with trivial boundary conditions $\rho_L$ and $\rho_R$. We track the structure constants for the generalized global symmetries along the RG flow and demonstrate monotonicity properties. Using the structure of the Higgs branch Hasse diagram, we identify LSTs with the same invariants, and thus which are candidate T-dual theories. A similar analysis for the Higgs branch Hasse diagram of the LSTs living on the worldvolume of NS5-branes in $\mathrm{Spin}(32)/\mathbb{Z}_2$ heterotic string theory probing a $\mathbb{C}^2/\mathbb{Z}_K$ singularity, yields the $\mathrm{Spin}(32)/\mathbb{Z}_2$ LSTs that belong to the same T-duality-families as the $E_8 \times E_8$ models. Despite the absence of a magnetic quiver, using a novel duality between particular orbi-instanton and conformal matter theories, we are also able to explore the Higgs branch Hasse diagram of some $\mathbb{C}^2/\Gamma_{\mathfrak{e}_6}$ orbifold LSTs, and again extract candidate T-dual families.

\paragraph{\uline{T-dualities and geometry:}} One of the consequences of the extraction of the Hasse diagram of the Higgs branch in this paper is the generation of an extensive landscape of LSTs which fall into large T-duality families. These families are obtained by the matching of invariant properties of the 5d theories after $S^1$-compactification, however, this provides only a necessary, but not sufficient check of T-duality. To prove the existence of T-duality, or a more general $n$-ality among a large family of LSTs, we should observe that the compactification spaces engineering the LSTs are, in fact, identical, up to a choice of inequivalent fibration structures. This geometric analysis of T-duality amongst heterotic ALE instantons has recently been performed for certain families in \cite{DelZotto:2022xrh}, building on the classic work \cite{Aspinwall:1997ye}. We leave a careful analysis of T-dualities for future work \cite{DFLM}. 

\paragraph{\uline{MQs for more LST models:}} The Higgs branch of little string theories has been understood to be a powerful portal on the search for possible T-dual models via the matching of duality invariant quantities. Hence an interesting research direction comprises the study of all the LST models that allow the formulation of a magnetic quiver. In this spirit, a first step has been taken in \cite{ZeroMarcus} analysing the simplest low-rank LSTs obtained via an $I_N$ fibration of a $\mathbb{P}^2$ surface: 
\begin{equation}
    [N_f=16] \stackon{$0$}{$\mathfrak{su}_N$} [N_{\Lambda^2}=2] \,, \quad [N_f=16] \stackon{$0$}{$\mathfrak{sp}_N$} [N_{\Lambda^2}=1] \,.
\end{equation}
The simplicity of the model allows for a Type IIA construction from which one can extract magnetic quivers and run Higgs branch RG flows. A further continuation of this program of studying the Higgs branches of LSTs via magnetic quivers is to analyze brane constructions for a variety of models ranging from loop-like configurations to other non-heterotic theories. 

\paragraph{\uline{Compact models:}} Herein, we considered F-theory compactifications that engineer little string theories; thus the base of the associated elliptic fibration is non-compact, which decouples the 6d $(1,0)$ supergravity degrees of freedom. Via the duality to heterotic string theory, we used this perspective to study the worldvolume theories on NS5-branes in $E_8 \times E_8$ heterotic string theory compactified on ADE orbifolds $\mathbb{C}^2/\Gamma$. Alternatively, we can ask about the compactification of the heterotic string on a \emph{compact} K3-manifold -- in particular, one with a singularity that looks locally like $\mathbb{C}^2/\Gamma$. The resulting 6d theory is no longer an LST, but an LST coupled to a 6d supergravity theory. Which LSTs can be coupled in such a way is highly constrained by the restrictions on which compact K3-manifolds can exist. A detailed analysis of the non-minimal fibers that exist over the compact finite-volume curve engineering the LST sector in a compact K3 compactification (in fact, more generally in compact Calabi--Yau threefold compactifications of F-theory) has been discussed in great detail in \cite{Alvarez-Garcia:2023gdd,Alvarez-Garcia:2023qqj}. 

\section*{Acknowledgements}

We thank Rafael \'Alvarez-Garc\'ia, Florent Baume, Michele Del Zotto, Marco Fazzi, Jonathan Heckman, and Marcus Sperling for interesting discussions. 
The authors acknowledge support from DESY (Hamburg, Germany), a member of the Helmholtz Association HGF. This work was partially supported by the Deutsche Forschungsgemeinschaft under Germany's Excellence Strategy - EXC 2121 ``Quantum Universe'' - 390833306. 

\appendix

\section{T-dual Pairs for \texorpdfstring{\boldmath{$\mathfrak{e}_6$}}{e6} Orbifolds}\label{app:bigtables}

In Section \ref{sec:beyond}, we studied the heterotic $E_8 \times E_8$ LSTs associated to the orbifold $\mathbb{C}^2/\Gamma_{\mathfrak{e}_6}$, and with $N_L = N_R = 0$. We were able to elucidate the Hasse diagram of the Higgs branch by using a duality between the rank zero $(\mathfrak{e}_8, \mathfrak{e}_6)$ orbi-instanton and the Higgsed rank one $(\mathfrak{e}_8, \mathfrak{e}_8)$ conformal matter theory. From this Hasse diagram, it is elementary to extract families of LSTs with the same T-duality-invariant properties.

We have attached to the {\tt arXiv} submission of this article a {\tt Mathematica} notebook that searches for T-dual pairs amongst the heterotic $E_8 \times E_8$ LSTs that we denoted
\begin{equation}
    \mathcal{K}_{0,0,\mathfrak{e}_6}(O_L, O_R) \,,
\end{equation}
where $O_L$ and $O_R$ are chosen from the nilpotent orbits of $\mathfrak{e}_8$. The resulting families are written here in Table \ref{tbl:nicetable}.

The heterotic $\mathrm{Spin}(32)/\mathbb{Z}_2$ LSTs associated with the $\mathbb{C}^2/\Gamma_{\mathfrak{e}_6}$ orbifold have a tensor branch description of the following form:
\begin{equation}\label{eqn:e6e6e6}
    \overset{\mathfrak{sp}_{v_1}}{1}
    \overset{\mathfrak{so}_{2v_2}}{4}
    \overset{\mathfrak{sp}_{v_3}}{1}
    \overset{\mathfrak{su}_{v_4}}{2}
    \overset{\mathfrak{su}_{v_5}}{2}\,.
\end{equation}
The {\tt Mathematica} notebook can also take any pair $(\kappa_R, \operatorname{dim}(\mathcal{C}))$ and determine all possible $v_i$ for which the LST in equation \eqref{eqn:e6e6e6} is consistent and has the same pair of invariants. In this way, each family in Table \ref{tbl:nicetable} can be augmented by the potentially T-dual $\mathrm{Spin}(32)/\mathbb{Z}_2$ LSTs. We have chosen to leave this information in the {\tt Mathematica} notebook, rather than reproduce it here.

\begin{table}[H]
    \centering
    \scriptsize
        \caption{Putative families of heterotic $E_8 \times E_8$ $\mathcal{K}_{0,0,\mathfrak{e}_6}(O_L, O_R)$ LST T-dual theories based on the matching of the invariant quantities $\kappa_R$ and $\operatorname{dim}(\mathcal{C})$.}
    \begin{threeparttable}
    \begin{tabular}{ccl}
        \toprule
        $\kappa_R$ & $\operatorname{dim}(\mathcal{C})$ & Pairs $(O_L, O_R)$ \\\midrule
        $72$ & $40$ & $(A_1, A_1)$, $(0, 2A_1)$ \\
        $71$ & $39$ & $(A_1, 2A_1)$, $(0, 3A_1)$ \\
        $70$ & $38$ & $(A_1, 3A_1)$, $(2A_1, 2A_1)$, $(0, 4A_1)$ \\
        $69$ & $38$ & $(A_1, A_2)$, $(0, A_2 + A_1)$ \\
        $69$ & $37$ & $(A_1, 4A_1)$, $(2A_1, 3A_1)$ \\
        $68$ & $37$ & $(A_1, A_2 + A_1)$, $(2A_1, A_2)$, $(0, A_2 + 2A_1)$ \\
        $68$ & $36$ & $(3A_1, 3A_1)$, $(2A_1, 4A_1)$ \\
        $67$ & $36$ & $(A_1, A_2 + 2A_1)$, $(3A_1, A_2)$, $(2A_1, A_2 + A_1)$, $(3A_1, 4A_1)$, $(0, A_2 + 3A_1)$ \\
        $66$ & $36$ & $(A_2, A_2)$, $(0, 2A_2)$ \\
        $66$ & $35$ & $(A_1, A_2 + 3A_1)$, $(4A_1, A_2)$, $(3A_1, A_2 + A_1)$, $(2A_1, A_2 + 2A_1)$ \\
        $65$ & $35$ & $(A_2, A_2 + A_1)$, $(A_1, 2A_2)$, $(0, 2A_2 + A_1)$ \\
        $65$ & $34$ & $(2A_1, A_2 + 3A_1)$, $(3A_1, A_2 + 2A_1)$, $(4A_1, A_2 + A_1)$ \\
        $64$ & $34$ & $(0, 2A_2 + 2A_1)$, $(A_2 + A_1, A_2 + A_1)$, $(2A_1, 2A_2)$, $(A_1, 2A_2 + A_1)$, $(A_2, A_2 + 2A_1)$ \\
        $64$ & $33$ & $(3A_1, A_2 + 3A_1)$, $(4A_1, A_2 + 2A_1)$ \\
        $63$ & $35$ & $(0, A_3 + A_1)$, $(A_1, A_3)$ \\
        $63$ & $33$ & $(2A_1, 2A_2+A_1)$, $(3A_1, 2A_2)$, $(A_1, 2A_2 + 2A_1)$, $(A_2, A_2+3A_1)$, $(A_2 + A_1, A_2 + 2A_1)$ \\
        $62$ & $34$ & $(0, A_3 + 2A_1)$, $(2A_1, A_3)$, $(A_1, A_3 + A_1)$ \\
        $62$ & $32$ & $(A_2 + A_1, A_2 + 3A_1)$, $(3A_1, 2A_2 + A_1)$, $(4A_1, 2A_2)$, $(A_2 + 2A_1, A_2 + 2A_1)$, $(2A_1, 2A_2 + 2A_1)$ \\
        $61$ & $34$ & $(0, D_4(a_1) + A_1$, $(A_1, D_4(a_1))$ \\
        $61$ & $33$ & $(A_1, A_3 + 2A_1)$, $(3A_1, A_3)$, $(2A_1, A_3 + A_1)$ \\
        $61$ & $32$ & $(A_2, 2A_2 + A_1)$, $(A_2 + A_1, 2A_2)$ \\
        $61$ & $31$ & $(3A_1, 2A_2 + 2A_1)$, $(4A_1, 2A_2 + A_1)$, $(A_2 + 2A_1, A_2 + 3A_1)$ \\ 
        $60$ & $33$ & $(A_2, A_3)$, $(A_1, D_4(a_1) + A_1)$, $(2A_1, D_4(a_1))$, $(0, A_3 + A_2)$ \\
        $60$ & $32$ & $(2A_1, A_3 + 2A_1)$, $(3A_1, A_3 + A_1)$, $(4A_1, A_3)$ \\
        $60$ & $31$ & $(A_2 + A_1, 2A_2 + A_1)$, $(A_2, 2A_2 + 2A_1)$, $(A+2 + 2A_1, 2A_2)$ \\
        $60$ & $30$ & $(A_2 + 3A_1, A_2 + 3A_1)$, $(4A_1, 2A_2 + 2A_1)$ \\
        $59$ & $32$ & $(0, A_3 + A_2 + A_1)$, $(2A_1, D_4(a_1) + A_1)$, $(3A_1, D_4(a_1))$, $(A_1, A_3 + A_2)$, $(A_2, A_3 + A_1)$, $(A_2 + A_1, A_3)$ \\
        $59$ & $31$ & $(3A_1, A_3 + 2A_1)$, $(4A_1, A_3 + A_1)$ \\
        $59$ & $30$ & $(A_2 + 3A_1, 2A_2)$, $(A_2 + A_1, 2A_2 + 2A_1)$, $(A_2 + 2A_1, 2A_2 + A_1)$ \\
        $58$ & $32$ & $(0, D_4(a_1) + A_2)$, $(A_2, D_4(a_1))$ \\
        \multirow{2}{*}{$58$} & \multirow{2}{*}{$31$} & $(A_2 + A_1, A_3 + A_1)$, $(A_2 + 2A_1, A_3)$, $(A_2, A_3 + 2A_1)$, $(A_1, A_3 + A_2 + A_1)$, \\ & & $\qquad (4A_1, D_4(a_1))$, $(3A_1, D_4(a_1) + A_1)$, $(2A_1, A_3 + A_2)$ \\
        $58$ & $30$ & $(2A_2, 2A_2)$, $(4A_1, A_3 + 2A_1)$ \\
        $58$ & $29$ & $(A_2 + 3A_1, 2A_2 + A_1)$, $(A_2 + 2A_1, 2A_2 + 2A_1)$ \\
        $57$ & $31$ & $(A_2 + A_1, D_4(a_1))$, $(A_2, D_4(a_1) + A_1)$, $(A_1, D_4(a_1) + A_2)$ \\
        \multirow{2}{*}{$57$} & \multirow{2}{*}{$30$} & $(2A_1 , A_3 + A_2 + A_1 ),(3A_1 , A_3 + A_2),(4A_1 , D_4(a_1) + A_1)$, \\ & & $\qquad (A_2 + 2A_1 , A_3 + A_1),(A_2 + 3A_1, A_3),(A_2 + A_1 , A_3 + 2A_1)$ \\
        $56$ & $30$ & $(2A_1 , D_4(a_1) + A_2),(2A_2 , A_3),(A_2 , A_3 + A_2),(A_2 + 2A_1 , D_4(a_1)),(A_2 + A_1 , D_4(a_1) + A_1)$\\
        $56$ & $29$ & $(3A_1 , A_3 + A_2 + A_1),(4A_1 , A_3 + A_2),(A_2 + 2A_1 , A_3 + 2A_1),(A_2 + 3A_1 , A_3 + A_1)$ \\
        $56$ & $28$ & $(2A_2 + A_1 ,2A_2 + A_1),(2A_2 , 2A_2 + 2A_1)$ \\
        \multirow{2}{*}{$55$} & \multirow{2}{*}{$29$} & $(2A_2 , A_3 + A_1),(2A_2 + A_1 , A_3),(3A_1 , D_4(a_1) + A_2 ),(A_2 , A_3 + A_2 + A_1),(A_2 + 2A_1 , D_4(a_1) + A_1 )$, \\ & & $\qquad(A_2 + 3A_1 , D_4(a_1)),(A_2 + A_1 , A_3 + A_2)$ \\
        $55$ & $28$ & $(A_2 + 3A_1 , A_3 + 2A_1),(4A_1 , A_3 + A_2 + A_1)$ \\
        $54$ & $30$ & $(A_3 , A_3),(0 , 2A_3)$ \\
        $54$ & $29$ & $(A_2 , D_4(a_1) + A_2),(2A_2 , D_4(a_1))$ \\
        \multirow{2}{*}{$54$} & \multirow{2}{*}{$28$} & $(2A_2 , A_3 + 2A_1),(2A_2 + 2A_1 , A_3),(2A_2 + A_1 , A_3 + A_1),(4A_1 , D_4(a_1) + A_2),(A_2 + 2A_1 , A_3 + A_2)$, \\ & & $\qquad(A_2 + 3A_1 , D_4(a_1) + A_1),(A_2 + A_1 , A_3 + A_2 + A_1)$ \\
        $53$ & $29$ & $(2A_3 , A_1),(A_3 , A_3 + A_1)$ \\
        $53$ & $28$ & $(2A_2 , D_4(a_1) + A_1),(2A_2 + A_1 , D_4(a_1)),(A_2 + A_1 , D_4(a_1) + A_2 )$ \\
        $53$ & $27$ & $(2A_2 + 2A_1 , A_3 + A_1),(2A_2 + A_1 , A_3 + 2A_1),(A_2 + 2A_1 , A_3 + A_2 + A_1),(A_2 + 3A_1 , A_3 + A_2)$ \\ 
        $52$ & $28$ & $(2A_1 , 2A_3),(A_3 , A_3 + 2A_1),(A_3 + A_1 , A_3 + A_1)$ \\
        $52$ & $27$ & $(2A_2 , A_3 + A_2 ),(2A_2 + 2A_1 , D_4(a_1)),(2A_2 + A_1 , D_4(a_1) + A_1 ),(A_2 + 2A_1 , D_4(a_1) + A_2$\\
        $52$ & $26$ & $(2A_2 + 2A_1 , A_3 + 2A_1),(A_2 + 3A_1 , A_3 + A_2 + A_1)$\\
            \end{tabular}
    \end{threeparttable}
    \label{tbl:nicetable}
\end{table}

\begin{table}[t]
    \centering
    \scriptsize
    \begin{threeparttable}
    \begin{tabular}{ccl}
        \toprule
        $\kappa_R$ & $\operatorname{dim}(\mathcal{C})$ & Pairs $(O_L, O_R)$ \\\midrule
        $51$ & $28$ & $(A_3 , D_4(a_1) + A_1),(A_3 + A_1 , D_4(a_1))$\\
        $51$ & $27$ & $(2A_3, 3A_1),(A_3 + 2A_1 , A_3 + A_1)$\\
        $51$ & $26$ & $(2A_2 , A_3 + A_2 + A_1 ),(2A_2 + 2A_1 , D_4(a_1) + A_1 ),(2A_2 + A_1 , A_3 + A_2),(A_2 + 3A_1 , D_4(a_1) + A_2)$ \\
        $50$ & $27$ & $(2A_3,A_2),(A_3, A_3 + A_2),(A_3 + 2A_1 , D_4(a_1)),(A_3 + A_1 , D_4(a_1) + A_1)$ \\
        $50$ & $26$ & $(A_3 + 2A_1,A_3 + 2A_1),(2A_2,D_4(a_1) + A_2),(2A_3 ,4A_1)$ \\
        $50$ & $25$ & $(2A_2 + 2A_1,A_3 + A_2),(2A_2 + A_1,A_3 + A_2 + A_1)$\\
        $49$ & $26$ & $(2A_3, A_2 + A_1)$, $(A_3, A_3 + A_2 + A_1)$, $(A_3 + 2A_1, D_4(a_1) + A_1)$, $(A_3 + A_1, A_3 + A_2)$ \\
        $48$ & $26$ & $(A_3, D_4(a_1) + A_2)$, $(D_4(a_1) + A_1, D_4(a_1) + A_1)$ \\
        $48$ & $25$ & $(2A_3, A_2 + 2A_1)$, $(A_3 + 2A_1, A_3 + A_2)$, $(A_3 + A_1, A_3 + A_2 + A_1)$ \\
        $47$ & $25$ & $(A_3 + A_1, D_4(a_1) + A_2)$, $(A_3 + A_2, D_4(a_1) + A_1)$, $(A_3 + A_2 + A_1, D_4(a_1))$ \\
        $47$ & $24$ & $(2A_3, A_2 + 3A_1)$, $(A_3 + 2A_1, A_3 + A_2 + A_1)$ \\
        $46$ & $24$ & $(2A_2, 2A_3)$, $(A_3 + 2A_1, D_4(a_1) + A_2)$, $(A_3 + A_2 + A_1, D_4(a_1) + A_1)$, $(A_3 + A_2, A_3 + A_2)$ \\
        $45$ & $23$ & $(2A_2 + A_1, 2A_3)$, $(A_3 + A_2, A_3 + A_2 + A_1)$ \\
        $44$ & $22$ & $(2A_2 + 2A_1, 2A_3)$, $(A_3 + A_2 + A_1, A_3 + A_2 + A_1)$ \\
        $42$ & $22$ & $(D_4(a_1) + A_2, D_4(a_1) + A_2)$, $(2A_3, A_3 + 2A_1)$ \\
         & 
    \end{tabular}
    \end{threeparttable}
\end{table}

\bibliography{bibliography}{}
\bibliographystyle{sortedbutpretty}

\end{document}